\documentclass[prd,twocolumn,superscriptaddress,showpacs,amsmath,amssymb]{revtex4}
\usepackage{comment}
\usepackage{multirow}
\usepackage{dcolumn}
\usepackage{bm}

\excludecomment{printrobustnotes}
\excludecomment{printallnotes}

\usepackage{ifpdf}
\ifpdf
  \usepackage[pdftex]{graphicx}
  \usepackage{epstopdf}
\else
  \usepackage[dvips]{graphicx}
\fi

\setkeys{Gin}{width=.99\columnwidth, height=.75\textheight, keepaspectratio}
\usepackage{dcolumn}% Align table columns on decimal point
\usepackage{bm}% bold math
\usepackage{url}
\usepackage{xspace}

\newenvironment{cfigure}[1][tbp]{\begin{figure}[#1]\centering}{\end{figure}}
\newenvironment{cfigure1c}[1][tbp]{\begin{figure*}[#1]\centering}{\end{figure*}}

\newcommand{\fig}[1]{Fig.~\ref{#1}}
\newcommand{\tab}[1]{Table~\ref{#1}}
\newcommand{\secref}[1]{Sec.~\ref{#1}}
\newcommand{\eqnref}[1]{Eq.~\eqref{#1}}
\newcommand{\refref}[1]{Ref.~\cite{#1}}
\newcommand{\Fig}[1]{Figure~\ref{#1}}
\newcommand{\Tab}[1]{Table~\ref{#1}}
\newcommand{\Secref}[1]{Section~\ref{#1}}

\newcommand{\note}[1]{\xspace}
\newcommand{\robustnote}[1]{\xspace}
\begin{printrobustnotes}
\renewcommand{\robustnote}[1]{\textbf{\textit{#1}}\xspace}
\end{printrobustnotes}
\begin{printallnotes}
\renewcommand{\note}[1]{\textbf{\textit{#1}}\xspace}
\renewcommand{\robustnote}[1]{\textbf{\textit{#1}}\xspace}
\end{printallnotes}
\newcommand{\abs}[1]{\left| #1 \right|}

\newcommand{\twotag}  {2-tag\xspace}
\newcommand{\onetag}  {1-tag\xspace}

\newcommand{\zerotag} {0-tag\xspace}

\newcommand{\ppbar}  {\ensuremath{p\bar{p}}\xspace}
\newcommand{\ttbar}  {\ensuremath{t\bar{t}}\xspace}

\newcommand{\bbbar}  {\ensuremath{b\bar{b}}\xspace}
\newcommand{\ccbar}  {\ensuremath{c\bar{c}}\xspace}
\newcommand{\qqbar}  {\ensuremath{q\bar{q}}\xspace}

\newcommand{\etadet} {\ensuremath{\eta_{det}}\xspace}

\newcommand{\Ht}     {\ensuremath{H_{T}}\xspace}
\newcommand{\et}     {\ensuremath{E_{T}}\xspace}

\newcommand{\pt}     {\ensuremath{p_{T}}\xspace}

\newcommand{\met}    {\ensuremath{\text{\raisebox{.3ex}{$\not$}}\et}\xspace}

\newcommand{\ljets}  {lepton + jets\xspace}
\newcommand{\dil}    {dilepton\xspace}
\newcommand{\Ljets}  {Lepton + jets\xspace}
\newcommand{\Dil}    {Dilepton\xspace}
\newcommand{\wjets}  {\ensuremath{W+\text{jets}}\xspace}
\newcommand{\chisq}  {\ensuremath{\chi^{2}}\xspace}

\newcommand{\mtop}    {\ensuremath{\mathrm{M}_{\text{top}}}\xspace}

\newcommand{\mreco}   {\ensuremath{m_{t}^{\text{reco}}}\xspace}
\newcommand{\mtr}     {\ensuremath{m_{t}^{\text{reco}}}\xspace}

\newcommand{\mtnwa}   {\ensuremath{m_{t}^{\text{NWA}}}\xspace}
\newcommand{\mnwa}    {\ensuremath{m_{t}^{\text{NWA}}}\xspace}

\newcommand{\runii}  {run~II\xspace}

\newcommand{\ns}    {\ensuremath{n_{s}}\xspace}
\newcommand{\nb}    {\ensuremath{n_{b}}\xspace}
\newcommand{\nbzero}{\ensuremath{n_{b_{0}}}\xspace}

\newcommand{\mjj}    {\ensuremath{m_{jj}}\xspace}

\newcommand{\jes}    {\ensuremath{\mathrm{JES}}\xspace}
\newcommand{\djes}   {\ensuremath{\Delta_{\mathrm{JES}}}\xspace}
\newcommand{\djesf}  {\ensuremath{\Delta_{\mathrm{JES}_f}}\xspace}

\newcommand{\genunit}[2]{\ensuremath{#1~\mathrm{#2}}\xspace}

\newcommand{\cm}[1]     {\genunit{#1}{cm}}

\newcommand{\tev}[1]    {\genunit{#1}{TeV}}
\newcommand{\gev}[1]    {\genunit{#1}{GeV}}
\newcommand{\pb}[1]     {\genunit{#1}{pb}}
\newcommand{\sigunit}[1]{\genunit{#1}{\sigma}}

\newcommand{\gevc}[1]   {\ensuremath{#1~\mathrm{GeV}/c}}
\newcommand{\gevcc}[1]  {\ensuremath{#1~\mathrm{GeV}/c^{2}}}

\newcommand{\invfb}[1]  {\ensuremath{#1~\mathrm{fb}^{-1}}}
\newcommand{\degrees}[1]{\ensuremath{#1^{\mathrm{o}}}\xspace}
\newcommand{\sigcunit}[1]{\ensuremath{#1~\mathrm{\sigma_{c}}}\xspace}

\newcommand{\gevnoarg}  {\ensuremath{\mathrm{GeV}}}

\newcommand{\sigcnoarg} {\ensuremath{\mathrm{\sigma_{c}}}\xspace}

\newcommand{\measErr}[2]{\ensuremath{#1 \pm #2}\xspace}
\newcommand{\measAErr}[3]{\ensuremath{#1~^{+#2}_{-#3}}\xspace}

\newcommand{\measStatJES}[2]{\ensuremath{#1 \pm #2~(\textrm{stat.}+\jes)}\xspace}
\newcommand{\measAStat}[3]{\ensuremath{#1~^{+#2}_{-#3}~(\textrm{stat.})}\xspace}

\newcommand{\measStatSyst}[3]{\ensuremath{#1 \pm #2~(\textrm{stat.}) \pm #3~(\textrm{syst.})}\xspace}
\newcommand{\measStatJESSyst}[3]{\ensuremath{#1 \pm #2~(\textrm{stat.}+\jes) \pm #3~(\textrm{other syst.})}\xspace}
\newcommand{\measStatJESSystALIGNED}[3]{\ensuremath{#1 &\pm #2~(\textrm{stat.}+\jes)\\ &\pm #3~(\textrm{other syst.})}\xspace}
\newcommand{\measAStatSystALIGNED}[4]{\ensuremath{#1&~^{+#2}_{-#3}~(\textrm{stat.})\\ &\pm #4~(\textrm{syst.})}\xspace}

\newcommand{\measStatASyst}[4]{\ensuremath{#1 \pm #2~\textrm{(stat.)}~^{+#3}_{-#4}~\textrm{(syst.)}}\xspace}
\newcommand{\measStatMT}[3]{\ensuremath{#1 \pm #2~\mathrm{(stat.+\mtop)~#3}}}

\newcommand{\lshape}{\ensuremath{{\mathcal L}_{\text{shape}}}\xspace}
\newcommand{\lsample}{\ensuremath{{\mathcal L}_{\text{sample}}}\xspace}

\newcommand{\lbg}{\ensuremath{{\mathcal L}_{\text{bg}}}\xspace}

\newcommand{\lljonetag}{\ensuremath{{\mathcal L}_{\text{lepton+jets,1-tag}}}\xspace}
\newcommand{\lljtwotag}{\ensuremath{{\mathcal L}_{\text{lepton+jets,2-tag}}}\xspace}
\newcommand{\ldilnontag}{\ensuremath{{\mathcal L}_{\text{dilepton,non-tagged}}}\xspace}
\newcommand{\ldiltag}{\ensuremath{{\mathcal L}_{\text{dilepton,tagged}}}\xspace}

\newcommand{\bv}[1]{\mbox{\boldmath$#1$}\xspace}

\begin{document}

%\title{Simultaneous measurement of the top quark mass in the lepton+jets and dilepton channels at CDF II}
\title{The first measurement of the top quark mass at CDF II in the lepton+jets and dilepton channels simultaneously}
\affiliation{Institute of Physics, Academia Sinica, Taipei, Taiwan 11529, Republic of China} 
\affiliation{Argonne National Laboratory, Argonne, Illinois 60439} 
\affiliation{University of Athens, 157 71 Athens, Greece} 
\affiliation{Institut de Fisica d'Altes Energies, Universitat Autonoma de Barcelona, E-08193, Bellaterra (Barcelona), Spain} 
\affiliation{Baylor University, Waco, Texas  76798} 
\affiliation{Istituto Nazionale di Fisica Nucleare Bologna, $^w$University of Bologna, I-40127 Bologna, Italy} 
\affiliation{Brandeis University, Waltham, Massachusetts 02254} 
\affiliation{University of California, Davis, Davis, California  95616} 
\affiliation{University of California, Los Angeles, Los Angeles, California  90024} 
\affiliation{University of California, San Diego, La Jolla, California  92093} 
\affiliation{University of California, Santa Barbara, Santa Barbara, California 93106} 
\affiliation{Instituto de Fisica de Cantabria, CSIC-University of Cantabria, 39005 Santander, Spain} 
\affiliation{Carnegie Mellon University, Pittsburgh, PA  15213} 
\affiliation{Enrico Fermi Institute, University of Chicago, Chicago, Illinois 60637} 
\affiliation{Comenius University, 842 48 Bratislava, Slovakia; Institute of Experimental Physics, 040 01 Kosice, Slovakia} 
\affiliation{Joint Institute for Nuclear Research, RU-141980 Dubna, Russia} 
\affiliation{Duke University, Durham, North Carolina  27708} 
\affiliation{Fermi National Accelerator Laboratory, Batavia, Illinois 60510} 
\affiliation{University of Florida, Gainesville, Florida  32611} 
\affiliation{Laboratori Nazionali di Frascati, Istituto Nazionale di Fisica Nucleare, I-00044 Frascati, Italy} 
\affiliation{University of Geneva, CH-1211 Geneva 4, Switzerland} 
\affiliation{Glasgow University, Glasgow G12 8QQ, United Kingdom} 
\affiliation{Harvard University, Cambridge, Massachusetts 02138} 
\affiliation{Division of High Energy Physics, Department of Physics, University of Helsinki and Helsinki Institute of Physics, FIN-00014, Helsinki, Finland} 
\affiliation{University of Illinois, Urbana, Illinois 61801} 
\affiliation{The Johns Hopkins University, Baltimore, Maryland 21218} 
\affiliation{Institut f\"{u}r Experimentelle Kernphysik, Universit\"{a}t Karlsruhe, 76128 Karlsruhe, Germany} 
\affiliation{Center for High Energy Physics: Kyungpook National University, Daegu 702-701, Korea; Seoul National University, Seoul 151-742, Korea; Sungkyunkwan University, Suwon 440-746, Korea; Korea Institute of Science and Technology Information, Daejeon, 305-806, Korea; Chonnam National University, Gwangju, 500-757, Korea} 
\affiliation{Ernest Orlando Lawrence Berkeley National Laboratory, Berkeley, California 94720} 
\affiliation{University of Liverpool, Liverpool L69 7ZE, United Kingdom} 
\affiliation{University College London, London WC1E 6BT, United Kingdom} 
\affiliation{Centro de Investigaciones Energeticas Medioambientales y Tecnologicas, E-28040 Madrid, Spain} 
\affiliation{Massachusetts Institute of Technology, Cambridge, Massachusetts  02139} 
\affiliation{Institute of Particle Physics: McGill University, Montr\'{e}al, Canada H3A~2T8; and University of Toronto, Toronto, Canada M5S~1A7} 
\affiliation{University of Michigan, Ann Arbor, Michigan 48109} 
\affiliation{Michigan State University, East Lansing, Michigan  48824}
\affiliation{Institution for Theoretical and Experimental Physics, ITEP, Moscow 117259, Russia} 
\affiliation{University of New Mexico, Albuquerque, New Mexico 87131} 
\affiliation{Northwestern University, Evanston, Illinois  60208} 
\affiliation{The Ohio State University, Columbus, Ohio  43210} 
\affiliation{Okayama University, Okayama 700-8530, Japan} 
\affiliation{Osaka City University, Osaka 588, Japan} 
\affiliation{University of Oxford, Oxford OX1 3RH, United Kingdom} 
\affiliation{Istituto Nazionale di Fisica Nucleare, Sezione di Padova-Trento, $^x$University of Padova, I-35131 Padova, Italy} 
\affiliation{LPNHE, Universite Pierre et Marie Curie/IN2P3-CNRS, UMR7585, Paris, F-75252 France} 
\affiliation{University of Pennsylvania, Philadelphia, Pennsylvania 19104}
\affiliation{Istituto Nazionale di Fisica Nucleare Pisa, $^y$University of Pisa, $^z$University of Siena and $^{aa}$Scuola Normale Superiore, I-56127 Pisa, Italy} 
\affiliation{University of Pittsburgh, Pittsburgh, Pennsylvania 15260} 
\affiliation{Purdue University, West Lafayette, Indiana 47907} 
\affiliation{University of Rochester, Rochester, New York 14627} 
\affiliation{The Rockefeller University, New York, New York 10021} 
\affiliation{Istituto Nazionale di Fisica Nucleare, Sezione di Roma 1, $^{bb}$Sapienza Universit\`{a} di Roma, I-00185 Roma, Italy} 

\affiliation{Rutgers University, Piscataway, New Jersey 08855} 
\affiliation{Texas A\&M University, College Station, Texas 77843} 
\affiliation{Istituto Nazionale di Fisica Nucleare Trieste/Udine, $^{cc}$University of Trieste/Udine, Italy} 
\affiliation{University of Tsukuba, Tsukuba, Ibaraki 305, Japan} 
\affiliation{Tufts University, Medford, Massachusetts 02155} 
\affiliation{Waseda University, Tokyo 169, Japan} 
\affiliation{Wayne State University, Detroit, Michigan  48201} 
\affiliation{University of Wisconsin, Madison, Wisconsin 53706} 
\affiliation{Yale University, New Haven, Connecticut 06520} 
\author{T.~Aaltonen}
\affiliation{Division of High Energy Physics, Department of Physics, University of Helsinki and Helsinki Institute of Physics, FIN-00014, Helsinki, Finland}
\author{J.~Adelman}
\affiliation{Enrico Fermi Institute, University of Chicago, Chicago, Illinois 60637}
\author{T.~Akimoto}
\affiliation{University of Tsukuba, Tsukuba, Ibaraki 305, Japan}
\author{M.G.~Albrow}
\affiliation{Fermi National Accelerator Laboratory, Batavia, Illinois 60510}
\author{B.~\'{A}lvarez~Gonz\'{a}lez}
\affiliation{Instituto de Fisica de Cantabria, CSIC-University of Cantabria, 39005 Santander, Spain}
\author{S.~Amerio$^x$}
\affiliation{Istituto Nazionale di Fisica Nucleare, Sezione di Padova-Trento, $^x$University of Padova, I-35131 Padova, Italy} 

\author{D.~Amidei}
\affiliation{University of Michigan, Ann Arbor, Michigan 48109}
\author{A.~Anastassov}
\affiliation{Northwestern University, Evanston, Illinois  60208}
\author{A.~Annovi}
\affiliation{Laboratori Nazionali di Frascati, Istituto Nazionale di Fisica Nucleare, I-00044 Frascati, Italy}
\author{J.~Antos}
\affiliation{Comenius University, 842 48 Bratislava, Slovakia; Institute of Experimental Physics, 040 01 Kosice, Slovakia}
% ...

\collaboration{CDF Collaboration\footnote{With visitors from $^a$University of Massachusetts Amherst, Amherst, Massachusetts 01003,
$^b$Universiteit Antwerpen, B-2610 Antwerp, Belgium, 
$^c$University of Bristol, Bristol BS8 1TL, United Kingdom,
$^d$Chinese Academy of Sciences, Beijing 100864, China, 
$^e$Istituto Nazionale di Fisica Nucleare, Sezione di Cagliari, 09042 Monserrato (Cagliari), Italy,
$^f$University of California Irvine, Irvine, CA  92697, 
$^g$University of California Santa Cruz, Santa Cruz, CA  95064, 
$^h$Cornell University, Ithaca, NY  14853, 
$^i$University of Cyprus, Nicosia CY-1678, Cyprus, 
$^j$University College Dublin, Dublin 4, Ireland,
$^k$Royal Society of Edinburgh/Scottish Executive Support Research Fellow,
$^l$University of Edinburgh, Edinburgh EH9 3JZ, United Kingdom, 
$^m$Universidad Iberoamericana, Mexico D.F., Mexico,
$^n$Queen Mary, University of London, London, E1 4NS, England,
$^o$University of Manchester, Manchester M13 9PL, England, 
$^p$Nagasaki Institute of Applied Science, Nagasaki, Japan, 
$^q$University of Notre Dame, Notre Dame, IN 46556,
$^r$University de Oviedo, E-33007 Oviedo, Spain, 
$^s$Simon Fraser University, Vancouver, British Columbia, Canada V6B 5K3,
$^t$Texas Tech University, Lubbock, TX  79409, 
$^u$IFIC(CSIC-Universitat de Valencia), 46071 Valencia, Spain,
$^v$University of Virginia, Charlottesville, VA  22904,
$^{dd}$On leave from J.~Stefan Institute, Ljubljana, Slovenia, 
}}
\noaffiliation


%\collaboration{CDF Collaboration}
%\date{\today}

\begin{abstract}
We present a measurement of the mass of the top quark
using data corresponding to an integrated luminosity 
of \invfb{1.9} of \ppbar collisions collected at
$\sqrt{s}=\tev{1.96}$ with the CDF II detector at Fermilab's Tevatron.
This is the first measurement of the top quark mass using top-antitop
pair candidate events in the \ljets and \dil decay channels
simultaneously. We reconstruct two observables in each channel and
use a non-parametric kernel density estimation technique to derive
two-dimensional probability density functions from simulated signal and background
samples.  The observables are the top quark mass and the invariant
mass of two jets from the $W$ decay in the \ljets channel, and the
top quark mass and the scalar sum of transverse energy of the event in
the \dil channel. We perform a simultaneous fit for the top quark
mass and the jet energy scale, which is constrained \emph{in situ} by
the hadronic $W$ boson mass. Using 332 \ljets candidate events and 144
\dil candidate events, we measure the top quark mass to be $\mtop =
\gevcc{\measStatJESSyst{171.9}{1.7}{1.1}} = \gevcc{\measErr{171.9}{2.0}}$.
%
%We present a template-based measurement of the mass of the top quark
%(\mtop) using \invfb{1.9} of \ppbar collision data collected at
%$\sqrt{s}=\tev{1.96}$ with the CDF II detector at Fermilab's Tevatron.
%This is the first measurement of \mtop using top-antitop pair (\ttbar)
%events from both the \ljets and \dil topologies.  Two observables are
%reconstructed in each event, and a non-parametric kernel density
%estimation technique is used to derive two-dimensional templates from
%large samples of simulated signal and background events.  A
%simultaneous fit is performed for \mtop and \djes; the latter is a
%measure of the CDF jet energy scale, and constraining it reduces the
%largest systematic uncertainty in the measurement.  Using 332 \ljets
%candidate events and 144 \dil candidate events, the measured top quark
%mass is $\mtop = \gevcc{\measStatSyst{171.9}{1.7}{1.1}} = \gevcc{\measErr{171.9}{2.0}}$.
\end{abstract}

% activate the following line for publication

\maketitle

\section{Introduction}
\label{sec:intro}

With a mass of approximately \gevcc{172}~\cite{PDBook}, the top quark
($t$) is by far the most massive fundamental object observed to date
in nature, some 40 times as massive as its isospin partner, the bottom
quark ($b$). This large mass leads to an important role for the top
quark in theoretical predictions from the standard model (SM) of
particle physics. In particular, electroweak radiative corrections to the $W$
boson mass, due to loops containing top quarks, play an important role
in constraining the mass of the Higgs boson, which also contributes to
radiative corrections. If the Higgs boson is discovered, a
precise measurement of the mass of the top quark will help provide an
important test of the SM, and would confirm that the newly observed
object is the SM Higgs boson and not some other scalar particle or source of
new physics. Independent of the Higgs boson, the large mass of the top
quark may make precision measurements throughout the top quark sector
necessary to help disentangle models of new physics~\cite{newphysics}.

The CDF and D\O\ collaborations jointly announced discovery of the top
quark in 1995~\cite{r_run1topdiscCDF, r_run1topdiscD0}, but it was not
until the availability of large datasets from Run II at the Tevatron
that precision measurements of the top quark mass have been possible. The
2007 world average of published top quark mass measurements, $\mtop =
\gevcc{\measStatSyst{172.5}{1.5}{2.3}}$~\cite{PDBook}, compares to the
CDF measurement of $\mtop = \gevcc{\measStatASyst{174}{10}{13}{12}}$
upon finding first evidence for the top quark~\cite{PhysRevLett.73.225}.
According to the SM, top quarks at the Tevatron are produced mainly 
in \ttbar pairs resulting from \qqbar annihilation (85\%) and gluon-gluon 
fusion (15\%). A top quark decays more than 99\% of the time to
a $W$ boson and a $b$ quark. The topology of \ttbar events
depends on the subsequent decay of each of the two $W$ bosons. Each
$W$ boson can decay hadronically, to a pair of quarks, or
leptonically, to a charged lepton and a neutrino. Due to the difficulty
of reconstructing $\tau$ leptons when they decay, for the
purposes of this analysis only electrons ($e$) and muons ($\mu$) are
considered as charged lepton candidates. The semileptonically decaying
$\tau$ leptons can enter the data set when they are reconstructed as electrons
or muons.

\Dil \ttbar events are those in which both $W$ bosons decay
leptonically. The \dil channel has a small branching ratio of $\sim
5\%$, and suffers from underconstrained kinematics resulting from the
presence of two neutrinos in each event that escape undetected. The
advantages of the \dil channel are a low background rate and simple
combinatorics with only two quarks in the final state.  \Ljets events are the
roughly $30\%$ of \ttbar events in which one $W$ boson decays
hadronically and the other decays leptonically. The \ljets channel has
four quarks, one lepton, and one neutrino in the final state, and 
sufficient amount of information measured in the 
detector to constrain the kinematics
of the \ttbar decay. 

This analysis describes a measurement of
the top quark mass in both the \ljets and \dil decay channels using
data collected at the Tevatron with the CDF II detector in
 a \invfb{1.9} integrated luminosity run.
This is the first analysis to combine likelihoods from
multiple measurements of the top quark mass in different decay
topologies into a single joint likelihood with a robust treatment of
the correlations in systematic uncertainties between the two channels
channels. 

In the analysis described in this article we follow the template
strategy~\cite{Abe:1994st}. In the \ljets channel we
determine the kinematics of the decay by fitting for a
reconstructed top quark mass \mtr~\cite{Abulencia:2005aj}. In the
\dil channel, due to the two undetected neutrinos, there is not
enough information to constrain the four-vectors of the top
quarks. Instead we use the neutrino weighting algorithm (NWA),
in which we scan over top quark masses, performing an
integration over the polar angles of the neutrinos at each mass to
calculate a weight based on the agreement with the measured momentum
imbalance in the event~\cite{PhysRevD.60.052001, abulencia:112006}. We select as the observable in an event the top quark
mass (\mtnwa) that yields the highest weight.

Uncertainties in jet modeling and in the calorimeter response result
in a systematic uncertainty in the jet energy calibration, which in
turn induces the largest systematic uncertainty in top quark mass
measurements. In the \ljets channel, the hadronically decaying $W$
boson provides an \textit{in situ} calibration sample for these
effects. The invariant mass of the hadronically decaying $W$ bosons
(\mjj) is used in the likelihood fit to measure and constrain the jet
energy calibration. This procedure reduces the combined statistical
and systematic uncertainty on the top quark mass measurement. The
calibration obtained from the \ljets channel is for the first time 
applied to both
channels by performing a simultaneous likelihood fit.  To improve the
precision of the \dil measurement we also use a second observable,
\Ht~\cite{Abe:1995ar}, which is a scalar sum of the \ttbar decay
product transverse energies.

Conventional procedures for combining values of the top quark mass
measured in different decay channels require correlations in
the systematic effects between the channels as inputs~\cite{Lyons:1988rp}.
In addition, combinations
of existing top quark mass results must also assume the functional form of the
likelihood in each measurement. Typically, the likelihoods are
assumed to be Gaussian, and
likelihood asymmetries or other departures from Gaussian
behavior are not taken into account. The analysis we describe here is
free of these two assumptions. The systematic uncertainties are evaluated
using the simultaneous fit, and the measurement uses one likelihood
function that depends on the data in the two channels. As a
cross-check, we also measure the top quark mass separately in the
\ljets and \dil channels, including a full evaluation of
systematic uncertainties.

\section{Experimental Setup}
\label{sec:experiment}
\subsection{CDF II Detector}

The Collider Detector at Fermilab is located at one of two
collision points along the ring of the Tevatron accelerator, which
collides bunches of protons and anti-protons at a center-of-mass energy of
\tev{1.96}. 
The detector has an approximate cylindrical geometry around the Tevatron beamline and is described in a Cartesian or in a polar coordinate system. 
%The detector has a cylindrical geometry with the origin of
%the coordinate system located at the center of the detector.
%An elevation view of the detector is shown in \fig{ref:elev}.
In Cartesian coordinates the $z$ axis is located along the beam axis with positive $z$ in the
direction of the proton beam, the $x$ axis pointing outward in the
plane of the Tevatron ring, and the $y$ coordinate pointing up. It is
often more convenient to use polar coordinates: the azimuthal angle
$\phi$ is the angle from the $x$ axis in the plane transverse to the
beamline; the polar angle $\theta$ is the angle from the proton beam
direction. The pseudorapidity $\eta \equiv -\ln(\tan\frac{\theta}{2})$
is a quantity numerically close to rapidity for highly relativistic
particles; differences in pseudorapidity are therefore nearly
invariant with respect to boosts along the $z$ axis.  Collisions occur
along the beamline and are distributed about the center of detector
with a spread of about \cm{30}. We distinguish between $\eta$ defined
with respect to $z=0$ (\etadet) and $\eta$ defined with respect to the
event collision point. It is common to reference a region in a cone of
$\Delta R$ around an object. This refers to the nearby region in
$\eta$-$\phi$ space: $\Delta R \equiv \sqrt{(\Delta\eta)^2 +
(\Delta\phi)^2}$. The transverse momentum \pt refers to the 
momentum in the plane transverse to the beamline.  The transverse
energy of an object is defined as $\et \equiv E\sin\theta$.  

A detailed description of the CDF II detector is provided in
\refref{r_cdfDetector}. In this section we briefly introduce the
detector subsystems relevant to this analysis, starting with the
detectors closest to the interaction region.

%  \begin{cfigure}
%  \includegraphics[width=\columnwidth]{plots_sec_experiment/cdfelev.epsi}
%  \caption{Elevation view of the CDF II detector}
%  \label{ref:elev}
%  \end{cfigure}

Charged particles are observed in the silicon tracking detectors. The
innermost silicon detector, layer 00~\cite{SVXII}, is a single-sided
silicon strip detector mounted directly onto the beryllium beampipe,
providing axial tracking information at a radius of 1.6 cm. The silicon
vertex detector (SVX II) consists of five double-sided silicon strip
detectors locating at radii from 2.5 cm to 10.6 cm from the beamline
and 90 cm in length,
providing axial and stereo information. Tracking of charged particles in
the central region $(\abs{\etadet} < 1.0)$ is provided by a 310 cm long
cylindrical open cell drift chamber, the central outer tracker (COT),
located at radii between 43 and 132 cm. The tracking
detectors are immersed in a 1.4 T solenoidal magnetic field, allowing
for charge determination and momentum measurements of charged
particles~\cite{COT}.

The calorimeter system measures the energy and position of particles
passing through and interacting with dense material. CDF uses
lead-scintillator and steel-scintillator sampling devices for the
electromagnetic and hadronic calorimetry, respectively. The calorimeter system is
comprised of the central electromagnetic (CEM)~\cite{CEM}, central
(CHA) and wall (WHA) hadronic calorimeters~\cite{CHAWHA} covering
$\abs{\etadet} < 1.0$, and the plug electromagnetic (PEM)~\cite{PEM}
and hadronic (PHA) calorimeters covering $1.1 < \abs{\etadet} <
3.6$. Shower maximum detectors are embedded in the central (CES) and
plug (PES) electromagnetic calorimeters at approximately six radiation
lengths from the collision point to provide the transverse shape of
the shower~\cite{CEM, PES}. 

Muon detectors are
located beyond the calorimeters.  Directly
outside the CHA is the central muon detector
(CMU), which covers $\abs{\etadet} < 0.6$. 
Located behind further 60 cm of steel shielding is the central muon upgrade
(CMP) detector. The central muon extension (CMX) covers the
region $0.6 < \abs{\etadet} < 1.0$. Muons pass through
the calorimeter and the shielding and leave
behind a series of hits (stubs) in the muon detectors,
which consist of four layers of single-wire drift cells.

CDF employs a three-level trigger system to select potentially
interesting events, reducing the interaction rate from the \genunit{1.7}{MHz} average bunch crossing rate
to a more manageable 75--\genunit{150}{Hz}.
The \ttbar candidate events used in this analysis are collected by
triggers that identify at least one high-\pt lepton candidate.
For central electron events~(CEM events),
the first-level trigger requires a cluster in the electromagnetic calorimeter
with $\et > \gev{8}$, a matching
track in the COT with $\pt > \gevc{8}$, and a ratio of
energy deposited in the hadronic to electromagnetic calorimeters
less than 1:8. At the second trigger level, the cluster energy
requirement is tightened to $\et > \gev{16}$, and the third trigger level 
makes basic
electron identification cuts and further tightens the energy
requirement to $\et > \gev{18}$. 
%These events are referred to as CEM events.
For muons in the central region, the first-level
trigger requires stubs in both the
CMU and CMP detectors (CMUP events) or the CMX detector (CMX events)
% consistent with muons with $\pt > \gevc{6}$
and a matching track in the drift chamber with $\pt > \gevc{4}$ for CMUP events and with $\pt > \gevc{8}$ for CMX events.
At the second trigger level, 
the $\pt$ requirement is increased to $\gevc{15}$, and at 
the third trigger level to $\gevc{18}$. 

\section{Event Selection and Background Estimation}
\label{sec:eventsel}

The energetic, charged leptons and missing transverse energy from at
least one leptonic $W$ boson decay help distinguish \ljets and \dil
 \ttbar events from the QCD multijet background.
%distinguish \ljets and \dil \ttbar
%events from the QCD multijet background.  
Further rejection of
background events containing real leptons is achieved by requiring
high-\pt jets, and in some cases identifying one or more of those jets
as arising from a $b$ quark.  We briefly describe here the
reconstruction of physics objects in the detector, as well as event
selection, background estimation, and Monte Carlo (MC) simulation.

\subsection{Selection of leptons}
\label{sec:sel_primitives}

Lepton identification is similar in the \ljets and \dil channels.
Events in both channels require a clean lepton in the central region
of the detector.  The second lepton in \dil events can be from less
pure categories such as forward or non-isolated leptons.  The major
distinctions between the two channels are noted in the following
descriptions.

A small cluster of towers in the CEM containing $\et>\gev{20}$ with a
COT track that extrapolates to the face of a tower in the cluster is
identified as an electron candidate. Electrons deposit most of their energy in
the electromagnetic calorimeter; therefore we require the ratio of
hadronic energy to electromagnetic energy to be less than 
$0.055+0.00045E$, where $E$ is the total energy of the electron. 
To reject
backgrounds with energetic $\pi^0$s plus a track, we require the ratio
of energy in the cluster to the track momentum to be not more than 2.0 for
electrons with $\et<\gev{100}$. The lateral shower development
measured in the calorimeter and the CES is required to 
match the electron shower shape as measured on a test beam.
We also require that the COT track
extrapolated to the depth of the shower maximum detector
 matches a CES cluster
in the $r$-$z$ and $r$-$\phi$ planes~\cite{wzcrosssection}.

Full COT tracking information is available only for $\abs{\etadet} <
1.0$. To reconstruct forward (PHX) electron candidates with
$\abs{\etadet} > 1.0$, we use clusters in the PES, the energy
measurement in the PEM, and knowledge of the \ppbar interaction
vertex. Silicon detector hits are then added to form a
track~\cite{PHX}. We require that forward electrons have $\et >
\gev{20}$ and a ratio of hadronic to electromagnetic energy less than
0.05. We also require that shower profiles measured in the PEM and PES
match the electron shower shape as measured on a test beam~\cite{wzcrosssection}. 
While PHX electrons are accepted in the dilepton sample, they are used in the lepton + jet sample only to veto \dil events.

Muon candidates are required to have a track with $\pt>\gevc{20}$
matched to a calorimeter tower with electromagnetic energy less than
\gev{2} and hadronic energy less than \gev{6}. The energy of the tower
is required to be low since muons are minimum ionizing in the
calorimeter material. Both requirements are slightly looser if the
track momentum is greater than \gevc{100}. If the track extrapolates 
to a fiducial region of the muon chambers, we require that it is matched
to a stub inside these detectors. We categorize muons based on the
detector chamber that the muon traverses. Muons fiducial to both the CMU
and CMP detectors are called CMUP muons. If a track does not
extrapolate to any muon chambers but all other quality criteria are
satisfied, we accept this object as a ``stubless'' muon (CMIO). Only
CMUP and CMX muons enter the \ljets dataset. All muon categories
are allowed into the \dil dataset for one of the two leptons to increase the 
statistics of the sample.

Electrons and muons produced in $W$ boson decays will in general be
well separated from other objects in the event. For electrons we
define isolation as the ratio of energy deposited within $\Delta R <
0.4$ around and excluding the electron cluster to the electron cluster
energy itself. For muons, isolation is defined as the ratio of the
transverse energy within $\Delta R < 0.4$ from the tower crossed by
the muon to the muon \pt. The tower traversed by the muon is excluded from 
transverse energy sum.
We categorize
leptons as isolated if their isolation variable is less than 0.1. The \ljets channel 
uses only
isolated leptons. 
In the \dil channel, to increase the sample statistics we allow
one non-isolated lepton of any type
except for PHX electrons and CMIO muons. 

\subsection{Jet corrections and systematic uncertainties}
\label{sec:Jcorr}

The property of quark confinement ~\cite{qcdasymp1, qcdasymp2}
ensures that bare quarks are not directly observable after QCD processes
take effect. Quarks and
gluons (``partons'') instead manifest themselves in the detector as jets of
particles flowing in the direction of the original gluon or quark
(parton). The sum of energies of particles within a cone around the
direction of the fragmenting parton is strongly correlated with its
energy. Jets for this analysis are reconstructed with the cone-based
clustering algorithm {\sc jetclu}~\cite{r_jetclu}, using a cone in
the azimuth-pseudorapidity space $\Delta R = \sqrt{\eta^{2}+\phi^{2}}=0.4$.
 The four-vector of a jet is constructed
based on energies and locations of calorimeter towers belonging to it.
A detailed explanation of CDF calibration of jet energy, the corrections
applied to it and
the associated systematic uncertainties can be found in
\refref{r_NIMJES}. We briefly summarize these corrections below.

From the raw jet energy several stages of corrections are applied that
attempt to remove effects masking the initial parton energy. In the
first stage, the relative corrections normalize the detector response
as a function of $\etadet$ so that jets at all $\etadet$ have the same $\et$ response as jets in the well 
understood central region of the detector, $0.2 <
\abs{\etadet} < 0.6$. The response varies across $\etadet$ due to
uninstrumented regions of the detector and different amounts of
material in front of the calorimeters. The next correction accounts
for the average energy in a jet cone due to additional
\ppbar inelastic interactions occurring during the same bunch
crossing. After accounting for energy due to extra \ppbar
interactions, corrections are applied for calorimeter non-linear
response to hadron jets and energy loss in uninstrumented regions.
  At this stage, jet
energies should be independent of the CDF detector, and correspond to
the energies of ``particle jets'', which are defined as all long lived 
particles from the primary \ppbar collision within the jet cone.
Energy from spectator partons
in the hard collision process that
breaks up the proton and anti-proton to form the \ttbar system is
accounted for by the underlying energy correction. The final
correction accounts for out-of-cone effects, in which some of the
original parton energy lies outside the jet cone.

Modeling of each of the effects described above is a potential source
of uncertainty on the measurement of jet energies. The combined
fractional uncertainty on the jet energy calibration~(\sigcnoarg) is shown in
\fig{f_jestot}. The overall jet energy calibration is referred
to as the jet energy scale (JES). We measure the difference \djes 
between the JES effects in simulation and data 
in units of \sigcnoarg. The \textit{a priori} \djes estimate at CDF
is thus by definition \sigcunit{\measErr{0}{1}}. 
For the \ljets channel, jets with
$\abs{\etadet} < 2.0$ and $\et > \gev{20}$ after applying the
relative, multiple interactions, and hadron jet response linearity corrections
are referred to as ``tight jets.'' Jets not passing the tight cuts but
having $\et > \gev{12}$ and $\abs{\etadet} < 2.4$ are referred to as
``loose jets.'' The \dil channel uses jets with $|\etadet|<2.5$ and
$\et>\gev{15}$ after corrections to the particle jet level.

\begin{cfigure}
\includegraphics[width=\columnwidth]{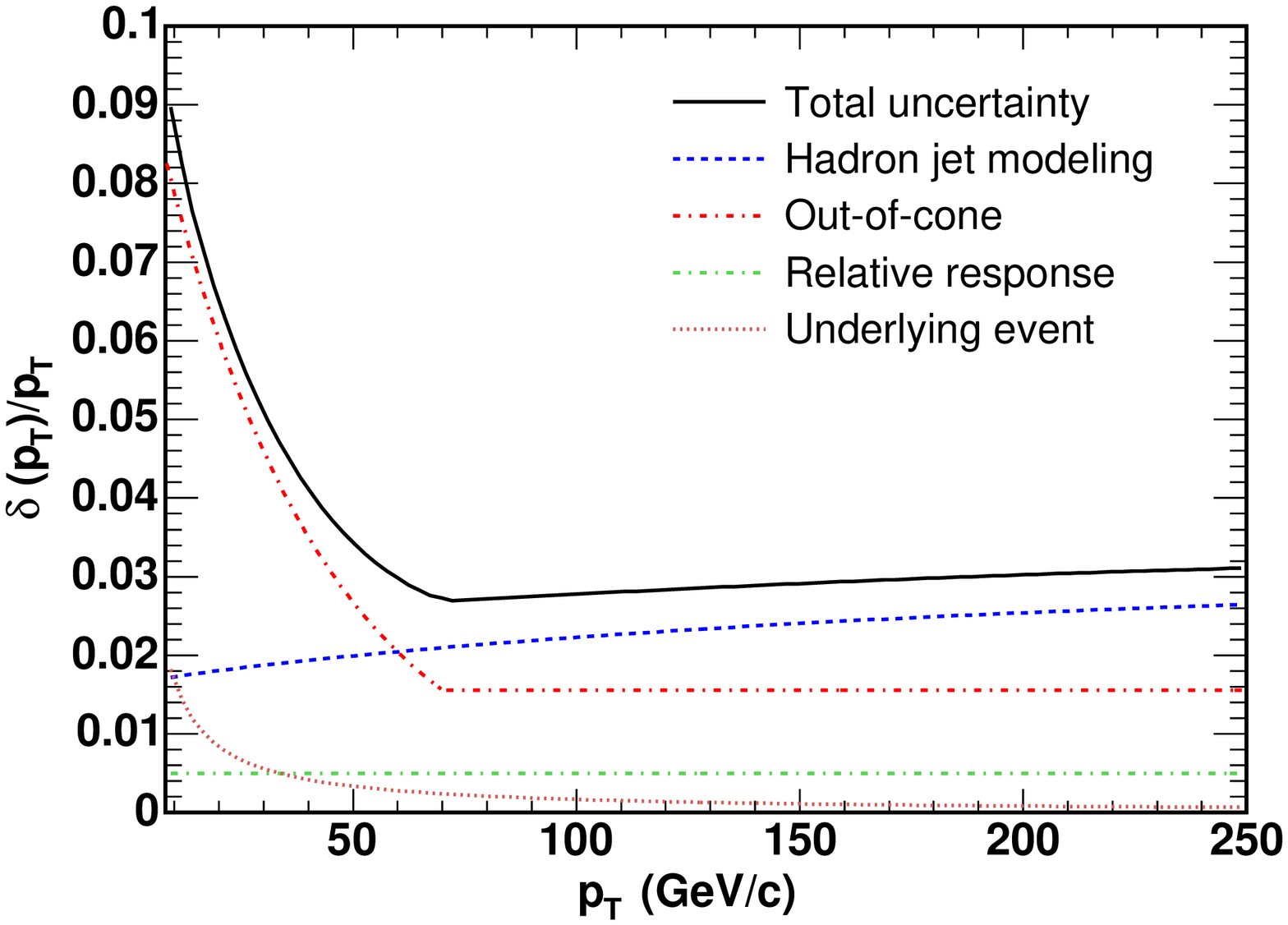}
\caption[Combined jet energy scale uncertainty]
{Fractional uncertainty on corrected jet \pt as a function of corrected jet \pt.}
\label{f_jestot}
\end{cfigure}

%\subsection{Missing $\et$}
\subsection{Neutrinos}
\label{c_eventselection_s_met}
Neutrinos from leptonic W decays escape the detector undetected, leading to an
imbalance of energy in the transverse plane of the detector. This
missing transverse energy ($\met$) is defined with respect to the
event's interaction vertex:

\begin{equation}
\label{e_met}
\met = \left| \sum_i -\vec{n^i} \et^i\right|
\end{equation}
\noindent
where the sum on $i$ runs over all calorimeter towers, $\vec{n^i}$ is
a unit vector in the transverse plane 
pointing from the beamline to the face of the $i$th calorimeter
tower, and $\et^i$ is the $\et$ in the $i$th tower, calculated using
the reconstructed event vertex. In events with
muons, the measured $\pt$ of the muon track is used in the $\met$
calculation, and not the energy deposited by the muon in the
calorimeters. Since jets are measured with better energy resolution than the
raw calorimeter towers, calorimeter tower energies are replaced
with jet energies for the towers clustered into jets.

% The jets are corrected 
%to particle level but without the
%application of the multiple interactions correction, since these
%interactions deposit energy outside the jets that cannot be removed.
%\note{Too much detail?  I would remove the sentence.}

\subsection{Identification of $b$ jets}
\label{c_eventselection_s_btagging}
The $B$ hadrons produced in the fragmentation of $b$ quarks
have an average lifetime on the order of 1.5 ps. Given typical boosts
of $b$ jets in \ttbar events at the Tevatron, this implies that $B$ hadrons
on average travel several millimeters in the detector before decaying,
leading to the identifiable signature of a displaced vertex.
This analysis uses the {\sc secvtx} algorithm~\cite{r_run1topxsec}
to find jets containing secondary vertices
consistent with the decay of a $B$ hadron; when such a displaced
vertex is found, the jet is identified (``tagged'') as a $b$-jet.
In the \ljets
(\dil) channel, only the four (two) jets with largest \et are
checked for $b$ tags.
%tags are considered only on the four (two) jets with largest \et. 
Only tight jets are allowed to have $b$ tags in the
\ljets channel. The tagging efficiency of $b$ quark jet depends
on the \et and \etadet of the jet, but is roughly 40\% within
the fiducial acceptance of the silicon detector, with a roughly 1\%
per-jet probability to incorrectly tag a light flavor jet. Jets
arising from charm quarks have finite lifetimes as well, and have a
per-jet tagging efficiency of roughly 8\%.

\subsection{\Ljets selection and background estimation}
\label{sec:ljets_sel_bgest}
\label{c_eventselection_s_samplediv}

The \ljets decay channel has four quarks in the final state, so we
require at least four jets in the detector. Only events with at least
one $b$ tag are considered; events with no $b$ tags have a
substantial background from production of $W$ bosons with four or more
jets, and are also particularly sensitive to systematic uncertainties.
We divide \ljets events into two exclusive subsamples based on the number
of $b$ tags.
For events with a single $b$
tag (\onetag), exactly four tight jets are required. Events with additional
tight jets and events with only three tight jets, but one or more loose
jets were predicted by MC simulation to contribute very little in terms of 
measurement precision.  
Events with
two or more tags (\twotag) have extremely small background contamination, and
also have fewer ways to assign jets to the quarks from the \ttbar
decay, so we allow events with three tight jets and one or more loose
jets, as well as events with four or more tight jets. For each \ljets
subsample, we also require a single, high-\pt, isolated CEM, CMUP or
CMX lepton and \met higher than \gev{20}, indicating the presence of an
escaping neutrino. Events with multiple leptons are vetoed.  The
\ljets event selection is summarized in \tab{table_eventselection_lj}.

\begin{table}
\begin{center}
\caption{\Ljets event selection summary.
Jets are corrected to the particle jet level.}
\begin{ruledtabular}
\begin{tabular}{ccc}
\label{table_eventselection_lj}
&\onetag&\twotag\\
\hline
$b$-tags (Leading 4 jets)&$=1$&$>1$\\
Lepton \pt (GeV/c), \et (GeV)&\multicolumn{2}{c}{$>20$}\\
\met (GeV)&\multicolumn{2}{c}{$>20$}\\
Leading 3 jets \et (GeV)&\multicolumn{2}{c}{$>20$}\\
4th jet \et (GeV)&$>$20&$>$12\\
Extra jets \et (GeV)&$<$20&Any\\
\end{tabular}
\end{ruledtabular}
\end{center}
\end{table}

Background estimates for the \ljets events are derived from a hybrid
of data- and MC-based measurements, similarly to previous dedicated
analysis~\cite{r_run2topxsec}. Data are used whenever
possible, and MC information is used to fill in any remaining
gaps in knowledge. 
In particular, the overall rate of events with real $W$
bosons and additional jets (\wjets), which dominate the background
sample, is determined using the data.  The fractions of these events
with one or two charm quarks (\emph{Wc} and $W\ccbar$ events) and two
bottom quarks ($W\bbbar$ events) are determined from MC samples. 
Overall normalizations of \wjets events come from the data after
subtracting off an estimate for the fraction of $W$ events coming from
QCD processes (non-$W$ events), and separating out a MC based estimate
for other
processes with real $W$'s (\emph{WW/WZ/ZZ}, $\ttbar$ and single-top
production), where the lepton trigger efficiencies and MC-data differences
have been taken into account.
The number of background events also depends on the rate
to mistag light-quark jets; this fake rate is determined using data samples
triggered by presence of jets.  
Fake tags come mostly from mismeasured tracks, through
interactions with material in the detector and real decays of
long-lived light-flavor particles such as $K_s$ and $\Lambda$ also
contribute.

Energetic charged leptons from $W$ boson decay can be faked by QCD
events via conversions (electrons) or misidentified pions and kaons
(muons), as well as from semi-leptonic heavy-flavor decays. In such
events, the \met requirement can also be passed when jets are
mismeasured or fall into uninstrumented regions of the detector. %Non-$W$
The events which do not include a $W$ boson, typically do fail the
 \met requirement, however, so the entire \met 
distribution, including the low-\met region, is used to fit for the
number of events with fake $W$ bosons and high \met. 
The QCD background is modeled in the data by events passing all
cuts where instead of a lepton an electron-like 
object is required. These pass all the kinematic cuts that are imposed
on the electrons,
but fail shower development or track quality cuts. 
In an alternative, high statistics model for the QCD background 
data events are used where the isolation cut on the charged lepton is removed
and an anti-isolation
cut is imposed, requiring the isolation variable defined earlier to be
greater than 0.2. The non-isolated model is not used to obtain the QCD
background normalization, however it is used at later stages in the analysis.

%requiring an additional energy of more than 20\% of the lepton 
%energy in a cone of $\Delta R = 0.4$ around the lepton.

\Tab{tablefinalbackgroundlj} shows the expected number
of background events in the \onetag and \twotag samples after all cuts,
the expected number of signal events based on the theoretical
cross-section (\pb{6.7})~\cite{theory_csection} at $\mtop = \gevcc{175}$, and the observed number
of events. Uncertainties on the event yield expectations are due to integrated luminosity, statistics of the MC samples as well as uncertain \met spectrum in the QCD model.

\begin{table}
\begin{center}
\caption{Expected event yield for the \ljets selection after all cuts. Uncertainties quoted are due to the uncertainty on the integrated luminosity, statistics of MC samples and uncertainty in QCD modeling.}
\begin{ruledtabular}
\begin{tabular}{ccc}
\label{tablefinalbackgroundlj}
&\onetag&\twotag\\
\hline
$W\bbbar$		&9.1$\pm$3.7	&2.1$\pm$0.9\\
$W\ccbar\mbox{, }Wc$	&8.3$\pm$3.4	&0.5$\pm$0.3\\
$W$ (mistags)		&10.4$\pm$2.3	&0.2$\pm$0.1\\
Single top		&2.0$\pm$0.1	&0.7$\pm$0.1\\
Diboson			&2.4$\pm$0.2	&0.21$\pm$0.02\\
QCD			&10.4$\pm$8.7	&0.3$\pm$1.6\\
\hline
Total Background & \measErr{42.7}{12.5} & \measErr{4.2}{1.9} \\
\hline
\ttbar (\pb{6.7}) & 156.7$\pm$21.1& 76.6$\pm$12.0 \\
\hline
Observed & 233 & 99\\
\end{tabular}
\end{ruledtabular}
\end{center}
\end{table}

%$W\bbbar$&9.1&2.1\\
%$W\ccbar$&5.0&0.4\\
%$Wc$&3.3&0.1\\
%$W$ (mistags)&10.4&0.2\\
%single top&2.0&0.7\\
%diboson&2.4&0.2\\
%QCD&10.4&0.3\\

\subsection{\Dil selection and background estimation}
\label{sec:dil_sel_bgest}

Signal events in the \dil channel contain two $b$ quarks, two
neutrinos and two oppositely charged leptons. We require that at least
one of the leptons is isolated and categorized as a CEM, CMUP or CMX 
lepton. The second lepton can belong to any category. If the
second lepton is a PHX electron or a CMIO muon, it must be
isolated. All other types of second leptons can be non-isolated to increase statistics of the sample.
We require a minimum of two jets in each \dil event.

To account for the two neutrinos, we
require $\met>\gev{25}$. We increase the requirement to
$\met>\gev{50}$ if the angle in the $r$-$\phi$ plane between
the \met vector and any jet or lepton is less than \degrees{20} in order to
reduce backgrounds such as Drell-Yan production of
$\tau$ pairs and QCD events where jets fall into
uninstrumented regions of the calorimeter.

We require that the $\Ht$ - the scalar sum of
transverse energies of jets, leptons and \met be greater than
\gev{200}. This has a small effect on signal acceptance, 
as the sum of energies of the
\ttbar decay products must be equal to or greater than twice the top
quark mass.

In events where the charged leptons are the same flavor, backgrounds
with \emph{Z} bosons are removed by requiring
the invariant mass of the \dil pair to be smaller than \gevcc{76}
or larger than \gevcc{106}. We impose this requirement only if the \met 
significance ($s_{\met}$) is smaller than 4.0$\gevnoarg^{1/2}$:
\begin{equation}
s_{\met}\equiv\frac{\met}{\sqrt{E_{T}^{sum}}} < \gev{4.0}^{1/2}
\label{eq:metsig}
\end{equation}
\noindent
where $E_{T}^{sum}$ is defined as the scalar sum of the transverse
energies of all calorimeter towers, with muons and jets corrected as
in the \met calculation.

Backgrounds for the \dil channel include Drell-Yan processes,
diboson production and QCD multijet production. We estimate the
contribution from Drell-Yan production of $\tau$ pairs and
diboson production using MC samples normalized to the theoretical cross
sections. We apply
trigger efficiencies as well as corrections accounting for differences
between the data and MC simulations in lepton
identification efficiencies and jet multiplicity distributions.

We employ a combined data-MC sample technique to estimate the
contamination from Drell-Yan $ee$ and $\mu\mu$. This background contains 
two components in the signal region: events outside the $Z$ 
boson window (\gevcc{76-106}), 
and events inside the window and passing the \met significance cut.
We first count the number of events in data with a 
dilepton invariant mass within the $Z$ window.
After subtracting expected contributions from other sources,
 we multiply this estimate by the ratio 
of the number of events outside the window to the number of
events inside the window, as measured in the MC samples. This gives the number of Drell-Yan
events outside the $Z$ window. We estimate the contribution from events
inside the $Z$ window with high $s_{\met}$ by multiplying the number
of events in data inside the window by the ratio
of events passing and failing the \met significance cut, again obtained
in MC samples.

The data are used to estimate the contribution of events where a real 
lepton is produced in association with multiple
jets and one of the jets is misidentified as a second lepton.
Data samples triggered on presence of jets are used to obtain
the probability for a jet to 
fake a charged lepton. These probabilities depend on lepton category 
and the jet \et. We apply these probabilities to the single-lepton 
data to obtain an estimate for the fake background contribution.

\note{We are not consistent with how we refer to \dil tagged (aka
\onetag) and non-tagged (aka untagged, \zerotag) subsamples.  Partly
my fault.}
We divide the \dil sample into non-tagged and tagged subsamples,
which have very different purity. Since the fake background is modeled
directly from data, the probabilities for fake leptons to be
reconstructed are summed separately in events with and without a $b$ tag. 
All other
backgrounds are modeled using MC samples. For all MC events, we
calculate the probability for each jet to be tagged, accounting for
the probabilities for light flavor jets to be mistagged. Given the tag
probabilities for the two leading jets, we calculate the probability
for each event to enter the non-tagged and tagged subsamples. Signal
and background estimates for the \dil channel are summarized in
\tab{tablefinalbackgrounddil}. Uncertainties are due to integrated luminosity,
MC sample statistics and fake rates. 

\begin{table}
\begin{center}
\caption{Expected event yield for the \dil selection after all cuts.
Uncertainties quoted capture the uncertainty on integrated luminosity,
statistics of the MC samples and uncertainties on the fake rates.}
\begin{ruledtabular}
\begin{tabular}{ccc}
\label{tablefinalbackgrounddil}
		&non-tagged	&tagged		\\
\hline
Diboson			&9.1	$\pm$2.2	&0.3	$\pm$0.1\\
Drell-Yan		&16.0	$\pm$2.5	&0.9	$\pm$0.1\\
Fakes			&19.3	$\pm$5.6	&2.7	$\pm$1.0\\
\hline
Total Background	&44.3	$\pm$7.0	&3.9	$\pm$1.0\\
\hline
\ttbar (\pb{6.7})	&40.1	$\pm$3.1	&55.8	$\pm$4.2\\
\hline
Observed & 83 & 61 \\
\end{tabular}
\end{ruledtabular}
\end{center}
\end{table}

\subsection{MC simulation}
\label{sec:MC}
The signal (\ttbar) MC simulation is modeled by {\sc pythia} version
6.216~\cite{Sjostrand:2000wi}, with {\sc herwig} version
6.510~\cite{r_herwig} used as a cross-check. Most background kinematics
are estimated from MC samples. The diboson backgrounds are modeled with 
{\sc pythia} version
6.216 and the \wjets and Drell-Yan+jets
backgrounds are modeled by {\sc alpgen}
version $2.10^{\prime}$~\cite{r_alpgen}, with jet fragmentation modeled by
{\sc pythia} version 6.325~\cite{Sjostrand:2000wi}.
A matching scheme~\cite{mlm1}
is used to ensure that there is no double-counting of phase space in
background events, as it is otherwise possible for events with hard
hadronic shower evolution to give states already described by events
at the matrix element level. The \wjets and Drell-Yan+jets background MC is
divided into $n$-parton samples, where $n$ refers to the total number
of partons (quarks or gluons), including heavy flavor. The samples are combined
according to the cross sections reported by {\sc alpgen}, accounting
for possibly different efficiencies for the samples to pass event
selection. Similarly, events with heavy flavor after fragmentation are
checked to ensure no double-counting of phase space across samples
with different flavor types at the matrix element level. Double-counting can
occur if samples generated at the matrix element level with
light-flavor partons produce charm or bottom quark pairs in the parton shower.
We remove such events unless both heavy flavor partons are within the same jet.
In addition the heavy flavor quark pairs generated at the matrix element
level 
can enter the same jet effectively reducing heavy jet multiplicity, therefore
we also remove events of this type. Depending on the multiplicity of light
and heavy flavor final state partons we remove a fraction of events between
a few percent and approximately 20\%.

% To account for this effect we remove events where
%one of the $b$ or $c$ quark generated by showering is found outside 
%from
%the MC samples generated with only light quarks in the final state
%at the matrix element level.

%,
%or if .
%In the MC samples generated with only light quarks in the final state
%at the matrix element level (e.g. $W$ + 2 partons) we remove events where a
%$b$ quark or a $c$ quark generated by showering is found flowing outside
%of a jet

%We therefore remove events where a hard emmission of a $b$ or $c$ is found. 
%, or if events generated at the matrix element level with charm quarks
%produce bottom quarks during fragmentation. //HMM the code doesn't
%distinguish between bb and cc samples -WTF

%% OK, this may be too much detail, so let's leave it out?

%In the MC samples generated with only light quarks in the final state
%at the matrix element level (e.g. $W$ + 2 partons, ),  we remove events
%where hard emmission of $b$ or a $c$ quark is found in MC generator information. 
%We allow events in which a soft emmission
%of a \bbbar or \ccbar quark pair occurs, provided that each
%heavy flavor quark is contained within a cone of $\Delta R < 0.4$ from
%a jet. 
%In the samples generated with heavy flavor quarks at the matrix
%element level, we remove events where a hard production of \bbbar
%or \ccbar occurs with both quarks falling within $\Delta R < 0.4$
%of a jet.
%Heavy flavor quarks generated at fragmentation must have $\pt$ and
%$\eta$ outside the cutoffs imposed during event generation, or else
%the events are rejected, ensuring that no double-counting occurs.

Electroweak production of single top quarks in both the $s$- and $t$-channels
contributes very few events to our sample.  These events are
treated as background, and are modeled using a fixed mass of
$\mtop = \gevcc{175.0}$.
Single top quark events are generated by
{\sc madevent}~\cite{maltoni-2003-0302}; fragmentation is modeled
with {\sc pythia} version 6.409~\cite{pythia6.4}.

To model multiple proton-antiproton interactions occurring in a single
bunch crossing, we add interactions where no partons with high
transverse momenta are produced to the events simulated for each
process. Those minimum bias collisions are simulated with {\sc
pythia} version
6.216 . The number of minimum bias interactions added to a given
event is equal to the expected number of \ppbar interactions, which depends on 
the instantaneous luminosity profile of the data run
of the event. The instantaneous luminosity profile is matched between
MC samples and data only for the first \invfb{1.2} of
integrated luminosity.  
This incorrect model is a source of bias of
\gevcc{0.4} for the \dil-only fit. No bias is present in the fitted
top quark mass in the combined and \ljets only fits, however a bias of
\sigcunit{0.04} is present in the fitted \djes in both
measurements. The bias in the \dil measurement
is higher due to the  particular choice of observables used to make
the measurement
and also due to the fact that the \textit{in situ} calibration absorbs
the bias on mass and converts it into bias on \djes.
These biases were found in studies of MC samples 
with increased number of \ppbar interactions and are corrected for in 
quoted results.

\section{Event reconstruction}
\label{sec:massreco}

After selection and reconstruction of event parameters of physics interest,
the data are processed to form
estimators for the top quark mass. Simply forming invariant masses is
not possible, as there are two top quarks per event. Events in the
lepton+jets channel have many possible assignments of jets to the
quarks, each of which give different reconstructed top quark masses. Each
dilepton event has two undetected neutrinos, resulting in
underconstrained kinematics. Both channels must account for the
possibility that, due to radiation effects, jets in the detector may
not correspond to quarks from the hard scattering. Both topologies also
contain non-negligible backgrounds. We approach these problems 
by constructing quantities strongly correlated to the top quark
mass and comparing the data to MC predictions that include all of the
above effects. For the lepton+jets channel, we additionally
account for the unique
and dense environment of \ttbar events by applying jet
corrections specific to the \ttbar events.

A set of generated distributions for a particular top quark mass
is referred to
as a template. A template is then a probability density function for a set of 
observables.
  Our measurement of the top quark mass is then a
determination of the most likely parent template for the data. Further
complicating the analysis, however, is the strong correlation between
the \djes in the detector and quantities sensitive to the top quark
mass, including the top quark mass estimators. 
%When a top quark decays
%to a $W$ boson and a $b$ quark, the narrow width and large mass of the
%$W$ ensure that most of the mass information is carried by the $b$
%quark. 
As explained in \secref{sec:Jcorr}, scaling measured jet
energies back to original parton energies is a difficult task, and any
uncertainty on the JES directly translates to a systematic uncertainty
on \mtop.  To reduce this effect, we introduce a second template in
the lepton+jets channel that uses the hadronic decay of the $W$ boson
to make an \emph{in situ} measurement of \djes that can be applied to
all jets in the event sample, including those from $b$ quarks and those in
the dilepton channel. The narrow width of the $W$ makes its dijet
mass (\mjj) a good estimator for \djes.

\subsection{Top-quark specific corrections}
\label{sec:ev_reco_tscorrs}

The jet corrections described in \secref{sec:Jcorr}
are generic algorithms derived for
application in all high-energy CDF analyses. As such, they miss out on
several key features of $\ttbar$ events in the lepton+jets
channel. The generic jet corrections assume flat $\pt$ spectra for all jets. 
The bias for the $\pt$ spectra expected from the physics process under
consideration may be corrected specifically. Top-antitop
events have two different, non-flat $\pt$ spectra for the $W$ decay jets
and the $b$ jets.  The generic jet corrections also do
not account for differences between jets coming from $b$ quarks
and jets coming from light-flavor quarks. To account for
all these effects, we derive jet corrections specific to the lepton+jets
$\ttbar$ environment from MC simulations. After event selection,
jets are corrected to the particle jet level. We separate the corrected 
jets by flavor (whether they came from
a $b$ quark or a light quark), and then into different $\eta$ and
$\pt$ bins. The top-specific corrections are derived such that the \pt of a jet
corresponds to the most probable value of the quark producing the jet. 

The top-specific corrections can be over 50\% for low-\pt jets in the central
region, and slightly negative for high-\pt jets.
We apply the corrections to the $\pt$ of the jet. We assume the
direction of the jet to be well measured, and do not apply 
correction to the jet angles ($\eta$ and $\phi$).
To obtain the jet energy,
the mass of jets assumed to come from $b$ quarks is fixed to
$\gevcc{5.0}$ and the mass of jets assumed to come from light quarks
is fixed to $\gevcc{0.5}$, though the mass effects are small compared
to typical jet energies in $\ttbar$ events.
The top-specific
corrections also provide the resolution on jet energy, once again separately
for the two flavors and as a function of $\pt$ and $\eta$. The
resolution is worse than 20\% for low-\pt jets, and better than 10\%
for high-\pt jets.

\subsection{Lepton+jets reconstruction}
\label{sec:ev_reco_ljets}

The lepton+jets decay channel gives overconstrained kinematics for the
\ttbar system. Detailed information on the lepton+jets kinematic
fitter can be found in \refref{Abulencia:2005aj}.  The minimization package
{\sc minuit}~\cite{minuit} is used
to minimize a $\chi^2$-like function for the overconstrained kinematic
system:
\begin{eqnarray}
\label{eq_chi2}
\chisq &=&
\sum_{i = \ell,\text{4 jets}}
\frac{(p_T^{i,\text{fit}} - p_T^{i,\text{meas}})^2}{\sigma_i^2} +
\sum_{j = x,y}
\frac{(U_j^{\text{fit}} - U_j^{\text{meas}})^2}{\sigma_j^2}
\nonumber \\
&+& \frac{(M_{jj} - M_W)^2}{\Gamma_W^2}
+ \frac{(M_{\ell\nu} - M_W)^2}{\Gamma_W^2}
\nonumber \\
&+& \frac{(M_{bjj} - \mtr)^2}{\Gamma_t^2}
+ \frac{(M_{b\ell\nu} -\mtr)^2}{\Gamma_t^2}
\end{eqnarray}

The first term constrains the $\pt$ of the lepton and the 4 jets in
the event to their measured values, within their uncertainties
$\sigma_i$. The unclustered energy ($U$) is the energy in the
calorimeter not associated with the primary lepton or one of the four
leading jets. The second term constrains the $x$ and $y$ components of
the unclustered energy ($U_x \equiv U\sin\theta\cos\phi$, $U_y \equiv
U\sin\theta\sin\phi$) in the detector close to their measured values
within uncertainties $\sigma_j$.
The third term in the \chisq expression constrains the dijet mass of
the two jets assigned as $W$ decay daughters to the well measured W
mass within the $W$ boson decay width. The fourth term similarly
constrains the invariant mass of the leptonic $W$ decay daughters.
The last two terms constrain the invariant masses of the three-body
top decay daughters to be consistent within the top quark decay
width of \gev{1.5}. The value of \mtr is a free parameter in the fit, and is taken
as the reconstructed mass used in the templates.

The neutrino transverse momentum is not a direct parameter in the
$\chi^2$ minimization, but is instead related to the unclustered
energy, and is calculated at every stage of the minimization process:
\begin{equation}
\label{eq_uncl}
p_{x,y} (\nu) = -\left(\sum_{\text{jets}} p_{x,y} (\text{jet}) +
p_{x,y} (\text{lepton}) + U_{x,y} \right)
\end{equation}
The longitudinal component of the neutrino momentum is a free
parameter that is effectively determined by the constraint on the
invariant mass of the leptonic $W$.

With the assumption that the leading (most energetic) four jets in the
detector come from the four final quarks at the hard scatter level,
there are 12 possible assignments of jets to quarks. The minimization
is performed for each assignment, with $\mtr$ taken from the
assignment that yields the lowest \chisq.  Events with the lowest
$\chisq>9.0$ are removed from the sample to reject poorly
reconstructed events not fitting the \ttbar hypothesis. The 
cut was optimized for expected statistical precision however we find 
that there is no strong dependence of the expected precision on the
value of the cut.
Identifying
$b$ jets reduces the number of combinations since tagged jets are assigned
only to final state $b$ quarks. 
%the combinatorics---any leading jet that is tagged is
%allowed assignment only to a $b$ quark. 
In rare events with more than
two tags among the leading four jets, only two tags of highest $E_{T}$ are 
assigned to the $b$ quarks and the additional tags are ignored. 

The calculation of the dijet mass \mjj is independent of the above
minimization procedure to derive \mreco.  Given a pair of jets, a simple invariant mass
is calculated from the jet four-vectors; in particular, the $W$ mass
constraint of the kinematic fitter is not applied.  There are multiple
ways to choose two jets among the four or more jets in $\ttbar$
events. Tagged jets are assumed to come from final state $b$ quarks.
Additionally, the two jets from the hadronic $W$ decay daughters are
assumed to be among the leading four jets. For two-tag events, there is
only one choice for the jet pair to be associated to the $W$ boson.
For \onetag events, there are three
possible dijet masses to be made from the 3 non-tagged leading jets;
we pick the single dijet mass closest to the well known $W$ mass. This
sculpts the distribution, but is the choice most likely to be correct in
selecting the two jets from the $W$ decay daughters,
and was found to give the best
sensitivity to \djes.  We correct the jets using the light quark
top-specific corrections.

\subsection{Lepton+jets template results}
\label{sec:ev_reco_ljets_templates}

We process MC samples with different values of $\mtop$ 
%MC samples with different values of $\mtop$ are processed
with full detector simulation~\cite{Gerchtein:2003ba} and event
selection. The kinematic
fitter is applied to each event, giving the \mreco templates shown in
\fig{f_ljtemplates} (a) and (c). Though the peak of the templates depends
strongly on \mtop, the reconstruction is not perfect, and \mreco only
gives an estimate for \mtop. The large tails in the templates are a
result of incorrect jet-quark assignments. The \twotag subsample, with
fewer jet-parton assignments, has narrower templates, and therefore
has more sensitivity to the top quark mass. Templates for \mjj masses
for three different values of \djes are shown in \fig{f_ljtemplates}
(b) and (d). The \onetag \mjj templates are narrower than the \twotag
templates due to sculpting of the distributions\note{(explain!)}.
The sculpting also yields smaller shifts in the \onetag template then in the \twotag template, as the \djes varies.

\begin{cfigure1c}
\begin{tabular}{cc}
\includegraphics[width=0.49\textwidth]{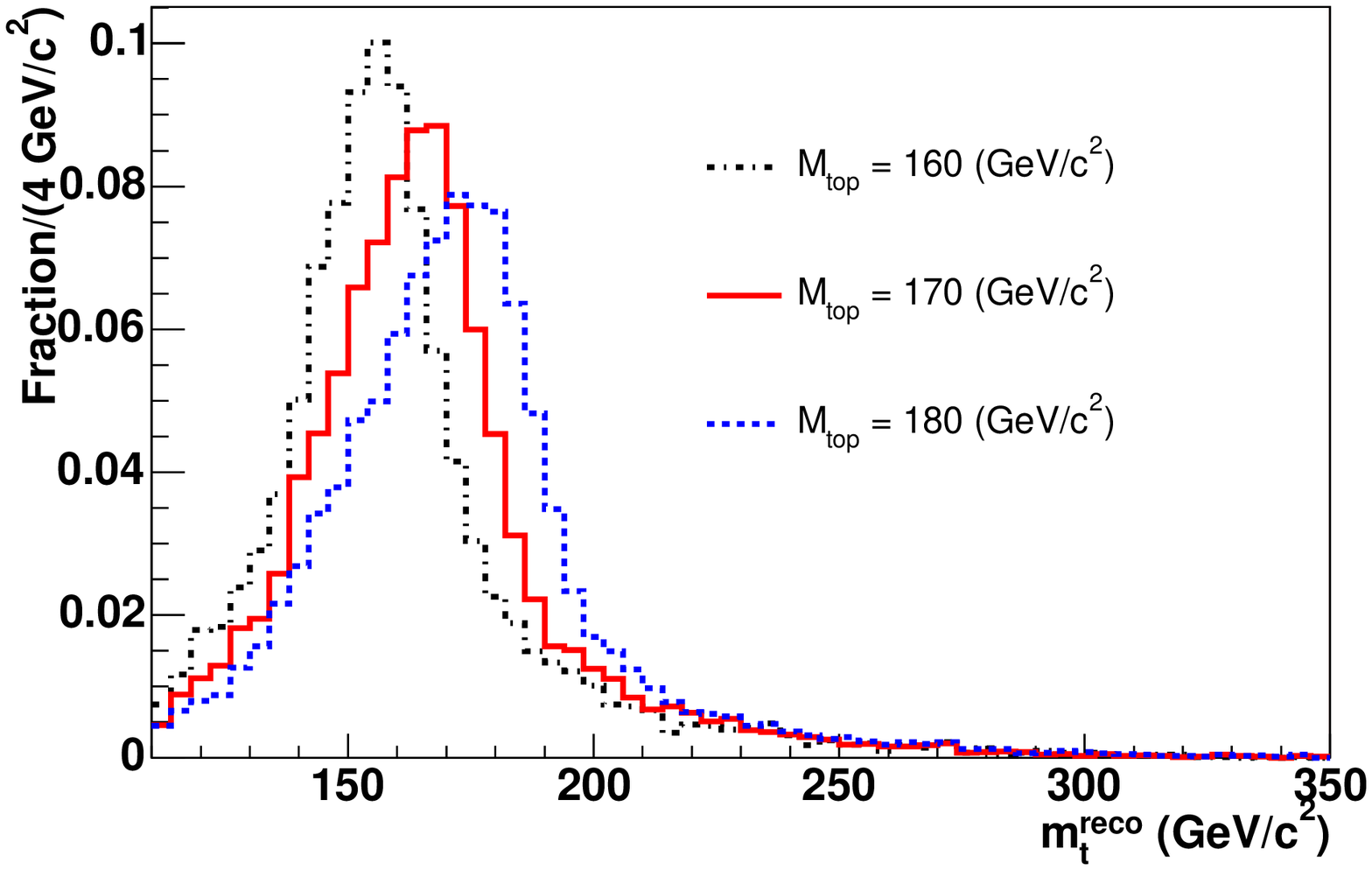} & 
\includegraphics[width=0.49\textwidth]{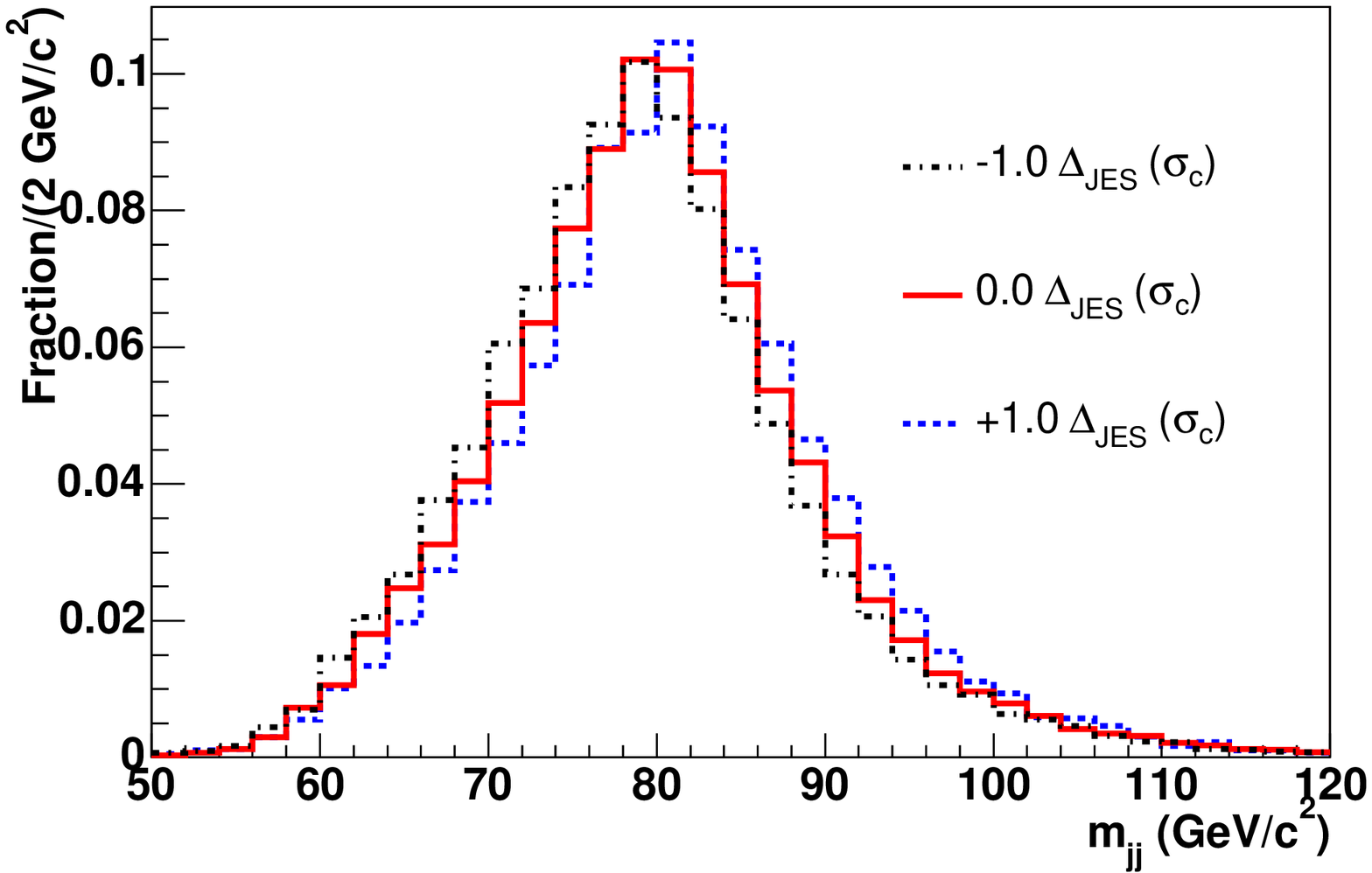}
\\
(a) \onetag \mreco & (b) \onetag \mjj \\
\includegraphics[width=0.49\textwidth]{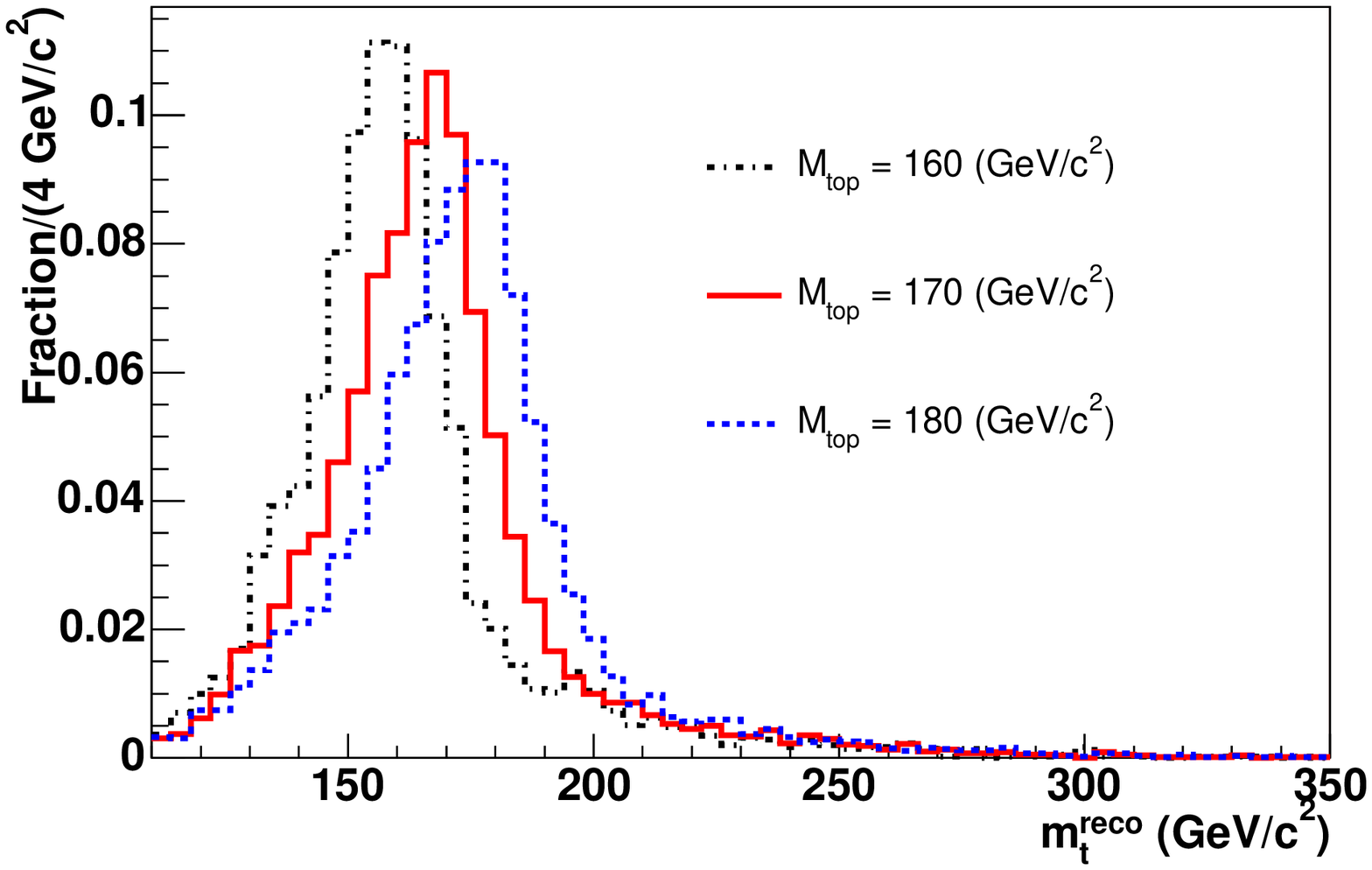} 
 &
\includegraphics[width=0.49\textwidth]{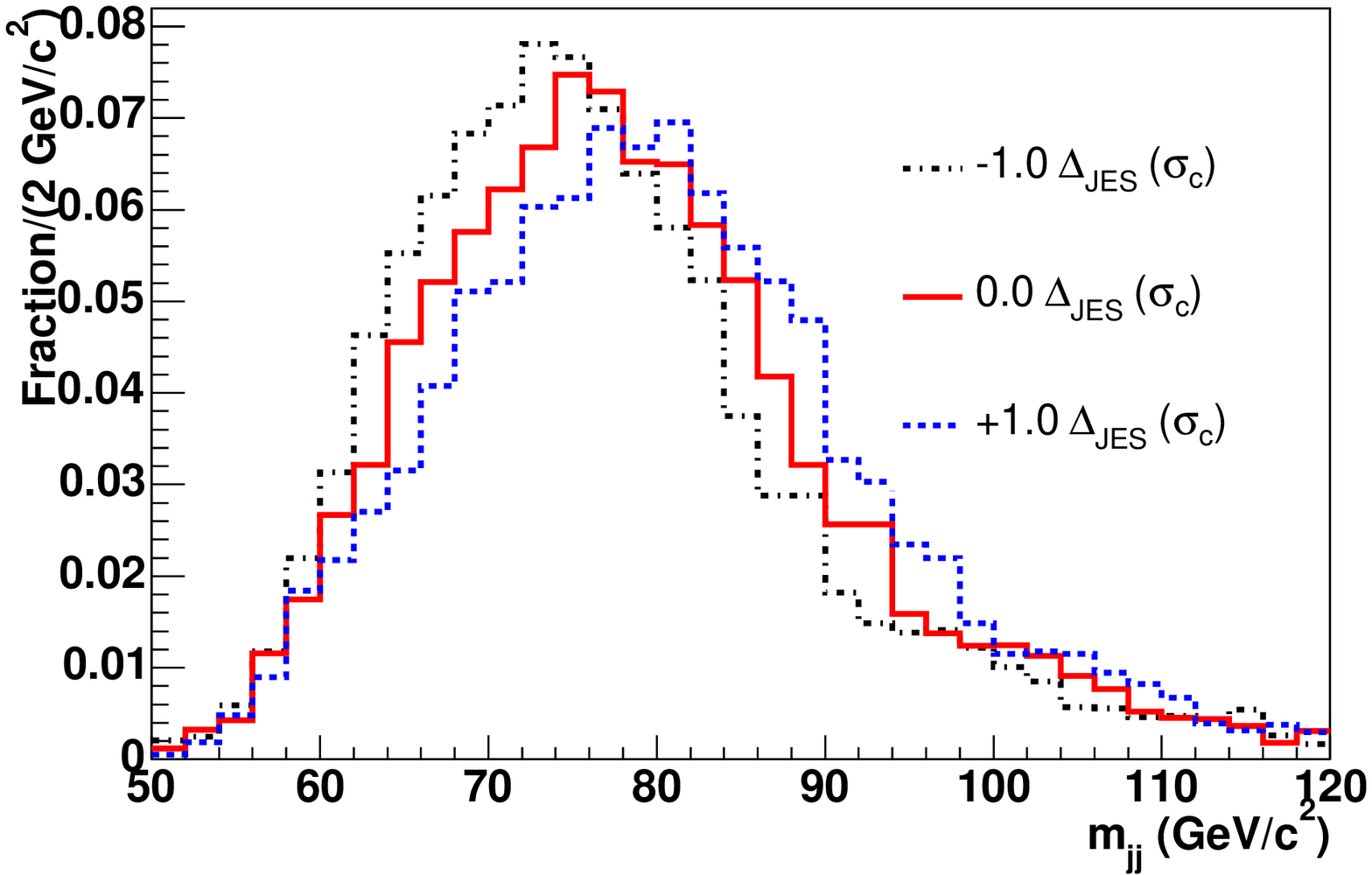} \\
(c) \twotag \mreco & (d) \twotag \mjj 
\end{tabular}
\caption
{Template distributions for MC events passing the \ljets selection.  Shown are the \onetag \mreco (a) and \mjj
(b) distributions, and the \twotag \mreco (c) and
\mjj (d) distributions. The \mreco distributions are plotted using events with three values of \mtop and with the nominal \sigcunit{\djes=0.0}. The \mjj distributions are plotted using events with three values of \djes and with \gevcc{\mtop=170}.}
\label{f_ljtemplates}
\end{cfigure1c}

\subsection{Dilepton templates}
\label{sec:ev_reco_dil}

In the dilepton channel, the measured quantities and assumptions on
the masses of particles in the decay cascade do not provide enough
constraints to reconstruct the four-vectors of the top
quarks. Instead, we form a reconstructed top quark mass (\mtnwa) for
each dilepton candidate using the Neutrino Weighting Algorithm
~\cite{PhysRevD.60.052001, abulencia:112006}. The
algorithm assigns a weight to the event as a function of top quark
mass. A top quark mass scan is performed in the range 80--\gevcc{380},
and the value yielding the maximum weight is selected as \mtnwa. 
The two most
energetic jets in an event are considered to have originated from the
$b$ quarks, giving two possible jet-quark assignments.
For each assumed top quark mass and jet-quark assignment,
we integrate numerically over the possible pseudorapidities
of the two neutrinos. The distribution of the neutrino pseudorapidity
is assumed to be Gaussian around zero, with a width of 1.0 obtained from {\sc
pythia}. Given a neutrino pseudorapidity, we can solve for its
transverse momentum. Up to two solutions are possible for each of the
neutrino and antineutrino transverse momenta. For each of the four
solutions, we compare the total momentum in the $x$ and $y$ directions
carried by the neutrinos to the measured $x$ and $y$ components of the
\met. We calculate a weight that is the product of two Gaussians, 
one each for the $x$ and $y$ directions, of the difference between the
measured \met in that direction and the sum of the momentum components
of the two neutrinos. We use a Gaussian width of \gev{19}, which is
optimized using MC \ttbar events. The four weights are added to form
the integrand. Note that we do not account for resolution effects in
measurements of the jets and leptons. The two integrals corresponding
to the two jet-$b$ quark assignments are added to form the top-mass-dependent
weight. In the calculation of the transverse momenta of the
neutrinos, jets are corrected using the generic jet corrections;
applying the top-specific corrections of \secref{sec:ev_reco_tscorrs}
was not found to improve the
resolution on the reconstructed top quark mass in this
channel. \Fig{f_diltemplates} (a) and (c) shows the output of the
algorithm from fully simulated MC events with different input masses.

The momenta of the decay products of the \ttbar pair are directly
correlated to the invariant masses of the top quarks. We therefore use 
another variable, the $\Ht$ (\textit{cf.} \Secref{sec:dil_sel_bgest}), to improve the precision of the measurement.
% that is strongly dependent on the top
quark mass. %; \Ht is defined as the scalar sum of \met and the
%transverse momenta of the leptons and all jets with
%$\et>\gev{15}$. 
\Fig{f_diltemplates} (b) and (d) shows \Ht
distributions for different top quark masses. The correlation
coefficient between \mtnwa and \Ht is $\sim 40\%$ in signal events and
$\sim 60\%$ in background events.

\begin{cfigure1c}
\begin{tabular}{cc}
\includegraphics[width=0.49\textwidth]{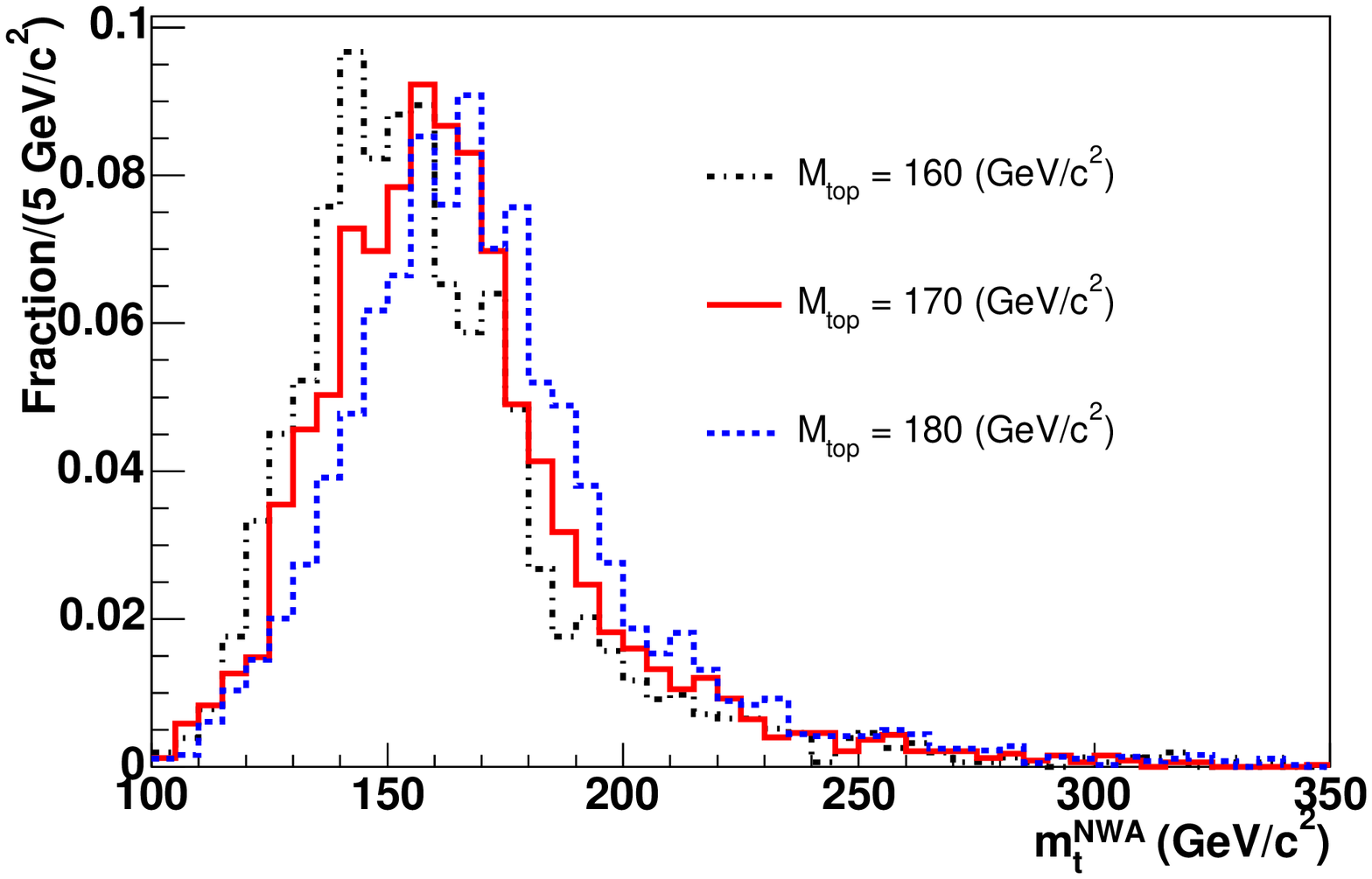}&
\includegraphics[width=0.49\textwidth]{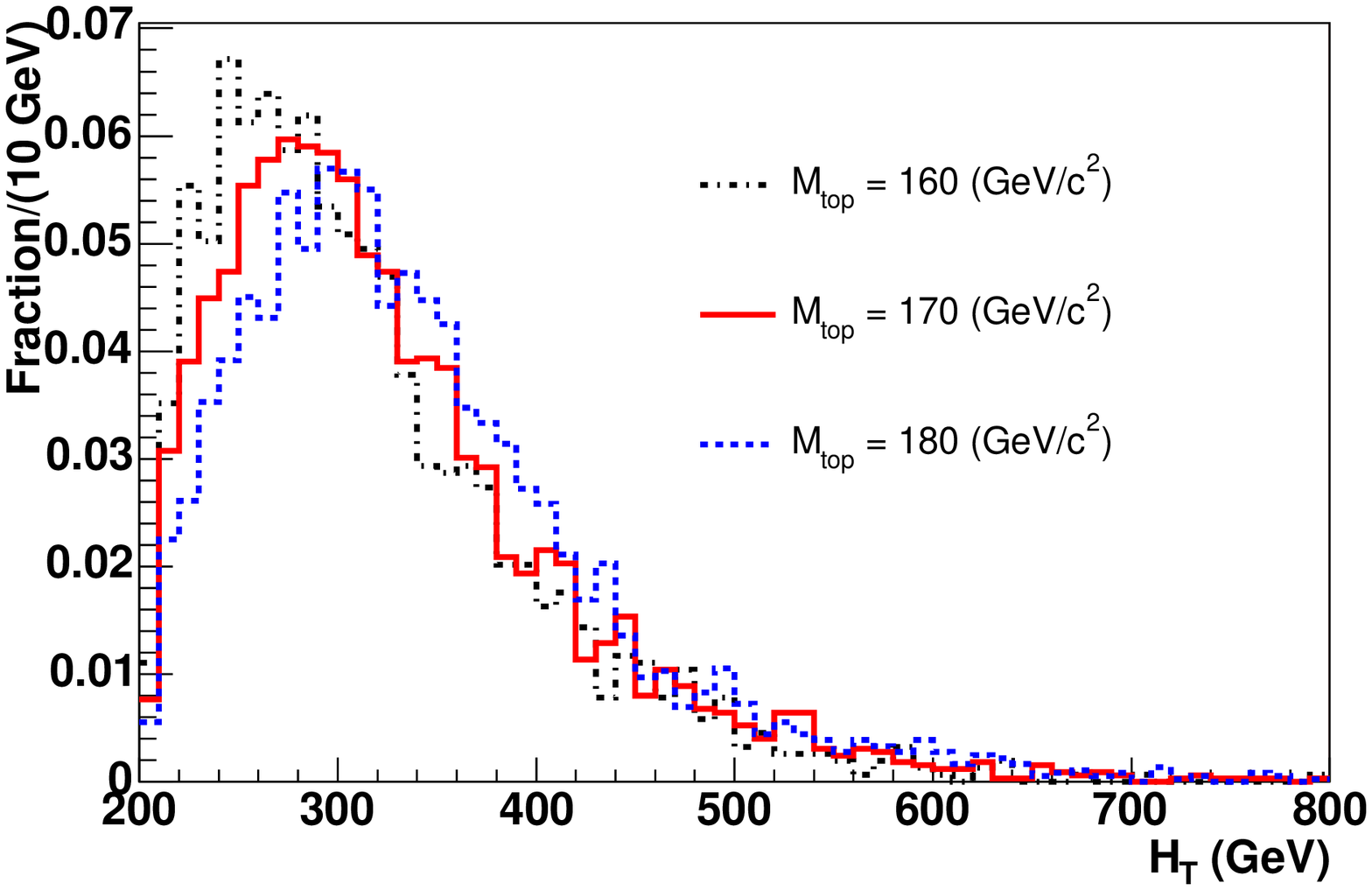}\\
(a) \zerotag \mtnwa & (b) \zerotag \Ht \\
\includegraphics[width=0.49\textwidth]{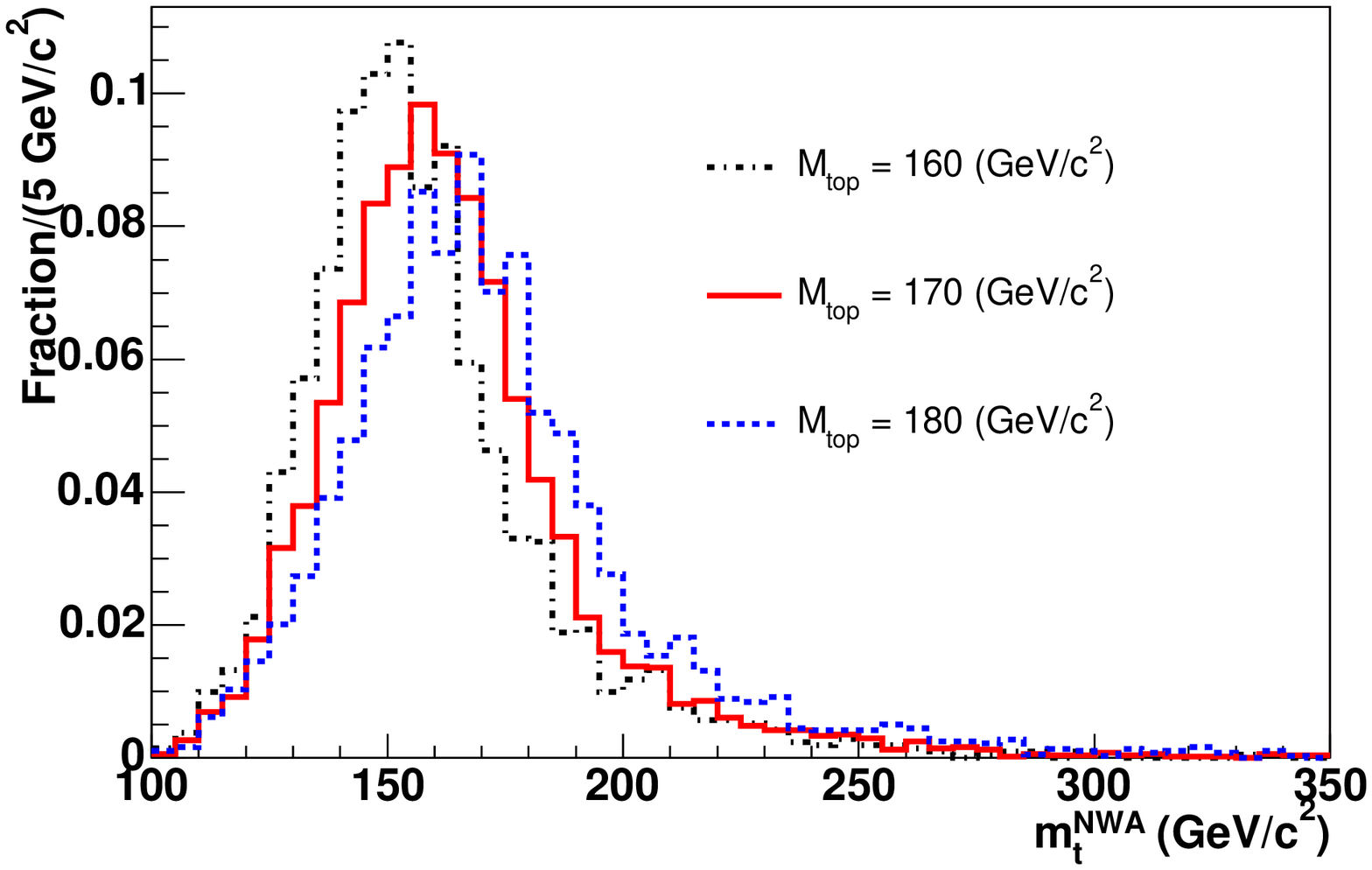}&
\includegraphics[width=0.49\textwidth]{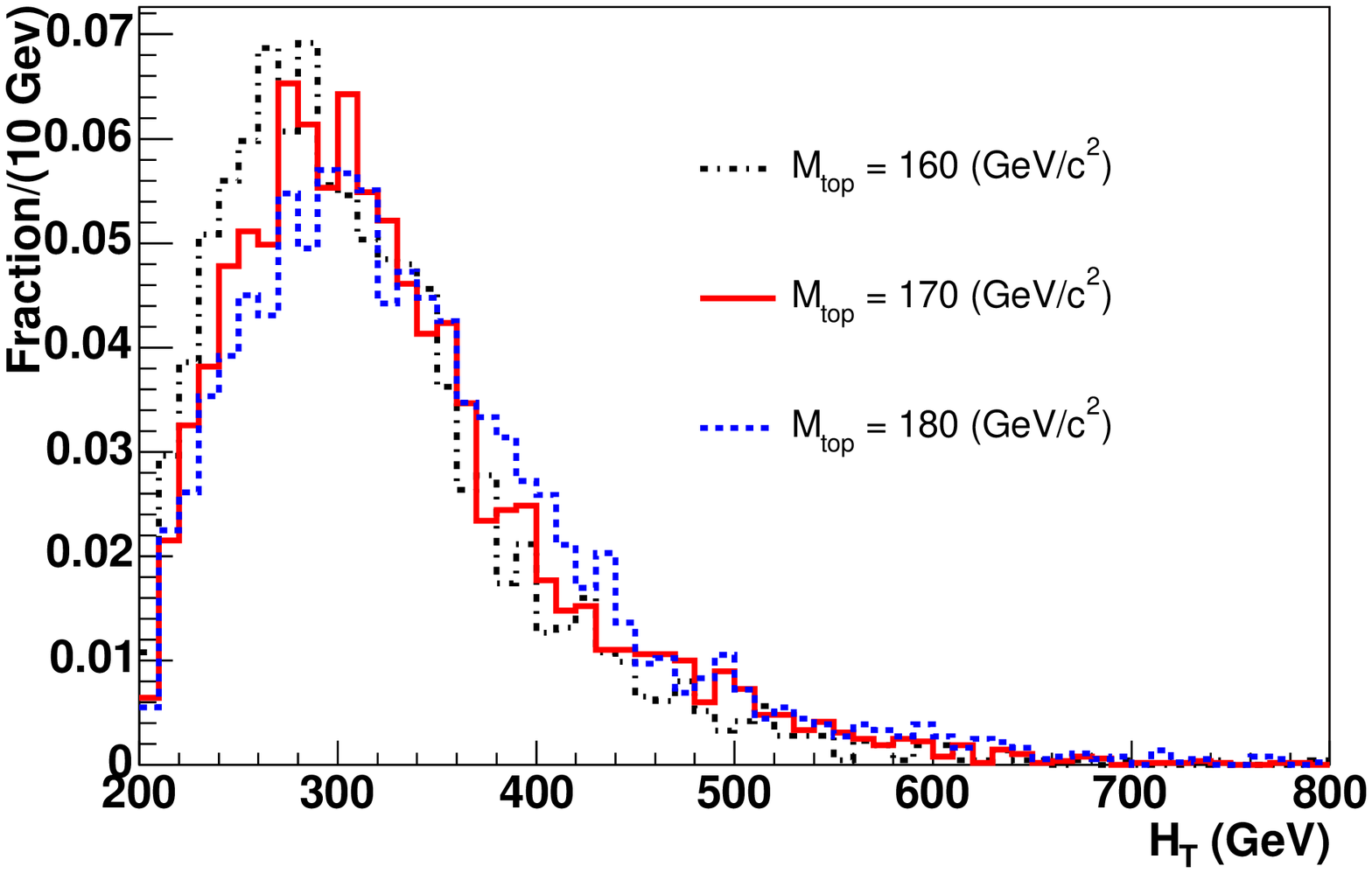}\\
(c) Tagged \mtnwa & (d) Tagged \Ht \\
\end{tabular}
\caption
{Template distributions for MC events passing the \dil selection. 
 Shown are the \zerotag \mtnwa (a) and \Ht (b)
distributions, and the tagged \mtnwa (c) and \Ht (d) distributions.
 The distributions are plotted using events with three values of \mtop and with the nominal \sigcunit{\djes=0.0}.
}
\label{f_diltemplates}
\end{cfigure1c}

\section{Mass fitting}
\label{sec:massfit}

The distributions of the observables defined in \secref{sec:massreco}
are used to determine simultaneously the two parameters \mtop and
\djes. For this, we need to know $P(x,y;\mtop,\djes)$: the probability
of observing a particular pair of values of the mass-sensitive parameters $(x,y)$, given some \mtop and
\djes. The observables $x$ and $y$ correspond to \mreco and \mjj for
\ljets events and to \mtnwa and \Ht for \dil events.  When \mtop and
\djes are fixed, the resulting $P(x,y)$ should be a normalized
probability density function (PDF) over the two-dimensional space of
the observables. The PDFs must be determined separately for signal and
background events in each subsample (e.g.\ \onetag \ljets events).
The background probabilities do not depend on \mtop.

Inaccuracies in these families of PDFs lead to biases 
in the final measurement that can be difficult to uncover or to
characterize.  Therefore, in order to achieve a precision measurement
of \mtop, it is essential to make a robust determination of $P$ for
each class of events. We accomplish this in two steps: First, at
discrete values of \mtop and \djes, we estimate the two-dimensional
PDFs for the observables from large samples of MC events
using kernel density estimation (KDE), described in \secref{sec:kde}.
Then we smooth and interpolate to find PDFs for arbitrary values of
\mtop and \djes using local polynomial smoothing (LPS), described in
\secref{sec:lps}.

The resulting probabilities are used in a combined likelihood fit
(\secref{sec:Lfit}) to measure the top quark mass. We run a rigorous
set of checks to validate the analysis machinery and calibrate the
final result using events from MC simulation; these checks are
described in \secref{sec:bias_check}.

\subsection{Kernel Density Estimation}
\label{sec:kde}

Previous template-based measurements of the top quark
mass~\cite{Abulencia:2005aj,dilmassprd} used arbitrary functional
forms to fit parameterized PDFs from histograms of the observables.
It is difficult to extend such parameterizations to two dimensions in
observables while properly accounting for correlations between \mtr and
\mjj, or \mtnwa and \Ht. In the \ljets channel, these correlations can
lead to a bias of several hundred MeV, if not properly taken into account;
 in the \dil channel, the
correlations are larger and can make such a measurement with two
observables impossible. Using a functional form with a large number of
parameters can also result in fits that are unstable with respect to
small changes in the histograms or even in the parameter
initialization.  

This measurement takes a different approach based on
KDE to form PDFs in two observables without any assumption about the
functional form. Some useful introductions to KDE can be found in
Refs.~\cite{Scott, Silverman, KDEHEP}. In KDE, the probability for an
event with observable $x$ is given by a linear sum of contributions
from all entries in the MC sample.  For a one-dimensional
distribution, this probability is given by:
\begin{equation}
\label{kernelequation}
\hat{f}(x) = \frac{1}{nh}\sum_{i=1}^{n}K\left(\frac{x-x_i}{h}\right)
\end{equation}
\noindent where $\hat{f}(x)$ is the probability to observe $x$
given, as an example, a \ttbar MC sample with known \mtop and \djes.
The sample has $n$ entries, with values of the observable given by
$x_i$.  The kernel function $K$ is a normalized function that adds less
probability to a measurement at $x$ as its distance from $x_i$
increases. The smoothing parameter $h$ (sometimes called the
bandwidth) is a number that determines the width of the kernel. Larger
values of $h$ smooth out the contribution to the kernel density estimate and
give more weight at $x$ farther from $x_i$. Smaller values of $h$
provide less bias to the kernel density estimate, but are more sensitive to
statistical fluctuations. We use an Epanechnikov kernel, defined as:
\begin{equation}
\label{epanechnikov}
K(t) = \begin{cases}
\frac{3}{4}(1-t^2) & \text{for $|t| < 1$}, \\
K(t) = 0 & \text{otherwise}.
\end{cases}
\end{equation}
so that only events with $\left|x-x_i\right| < h$ contribute to
$\hat{f}(x)$.

We use an adaptive KDE method in which the value of $h$ is replaced by
$ h_i$ so that the amount of smoothing applied to the $i$th event
depends on the value of $\hat{f}(x_i)$~\cite{adaptive}.  We run a
first pass of kernel density estimation with constant $h$. This pilot kernel
 density
estimate is then used in a second round of KDE to determine the
individual $h_i$, with $h_i \propto \hat{f}(x_i)^{-0.5}$. In the peak
of the distributions, where there are more events, we use small values
of $h_i$ to capture as much shape information as possible. In the
tails of the distribution, where there are fewer events and the kernel density
estimates are sensitive to statistical fluctuations, a larger value of
$h_i$ is used. The overall scale of $h$ is set by the root mean square
(RMS) of the distribution and by the number of entries in the MC
sample; larger (smaller) smoothing is used when fewer (more) events
are available~\cite{Scott, Silverman, OSmulti}.  If the smoothing
parameters get too large in the tails of the distribution, the kernel density
estimates can become non-local, and a point at $x_i$ can contribute
weight to an estimate at a distant $x$.  Following \refref{adaptive},
we guard against this by not allowing $h_i$ to get too large:
\begin{equation}
\label{clip}
h_i = \mbox{min}(h_i, \sqrt{10}\cdot h_0)
\end{equation}
where $h_0$ is the minimum adaptive bandwidth, which occurs in
the peak of the distribution.

KDE is extended to two dimensions by multiplying two kernels
together~\cite{2dref, 2dadaptive}:
\begin{equation}
\label{kernelequation2}
\hat{f}(x,y) =
\frac{1}{n}\sum_{i=1}^{n}\frac{1}{h_{x,i} h_{y,i}}
\left[ K\left(\frac{x-x_i}{h_{x,i}}\right)
\times K\left(\frac{y-y_i}{h_{y,i}}\right) \right]
\end{equation}
\noindent 
Note that the smoothing parameters for the two variables do not have
to be identical.  Typical values of $h$ for kernel density estimates in the
signal are 10--\gevcc{12} for \mtr, 3--\gevcc{6} for \mjj,
15--\gevcc{20} for \mtnwa and 45--\gev{55} for \Ht. For background
kernel density estimates, these numbers are slightly larger, as the number of
events passing all the cuts is smaller and the templates are wider.

The kernels in \eqnref{kernelequation2} know nothing about the
boundaries of the templates. Mathematically, the density functions can
take on any real numbers given large enough smoothing, even though
kinematic requirements and energy conservation limit
possible values of the observables. When the probability density
extends beyond such a limit, where the data are not found, the
normalization condition of the kernels does not hold. To enforce unit
normalization, we explicitly force hard boundaries and reject events in
the tails of the distribution both from the MC templates and the data,
typically removing 1-2\% of signal events and a slightly larger
fraction of background events. When kernel density estimates are calculated,
we check that each of the individual kernels is normalized within the
boundaries. If a kernel is not normalized and leaks probability
outside the boundaries, it is renormalized such that it contributes
unit weight inside the boundaries.

Using $\hat{f}(x,y)$ from \eqref{kernelequation2} as $P(x,y)$,
we can scan values of $x$ and $y$ to visualize the two-dimensional PDF
for fixed \mtop and \djes.  \Fig{ref:2dkdesig} shows two-dimensional
kernel density estimates for \ljets and \dil signal events, given
$\mtop=\gevcc{170}$ and $\djes=0.0$.  \Fig{ref:2dkdebkgd} shows the
estimates for background events at $\djes=0.0$.  The background
kernel density estimates are derived separately for the individual
contributions to the background model, taking into account the sample
sizes and RMS values, and are then combined with the appropriate
weights.  Note that since jet energy scale shifts would affect all
data events in a similar way, data-derived background templates do not
depend on \djes. The \dil tagged background contains multiple peaks
that come from the fake background. The data used to model the fakes,
which comprise half of the total tagged background in the \dil
channel, contain very few events, and also peak at different locations
than the other backgrounds.

\begin{cfigure1c}
\begin{tabular}{cc}
\includegraphics[width=0.48\textwidth,type=eps,ext=.eps,read=.eps]{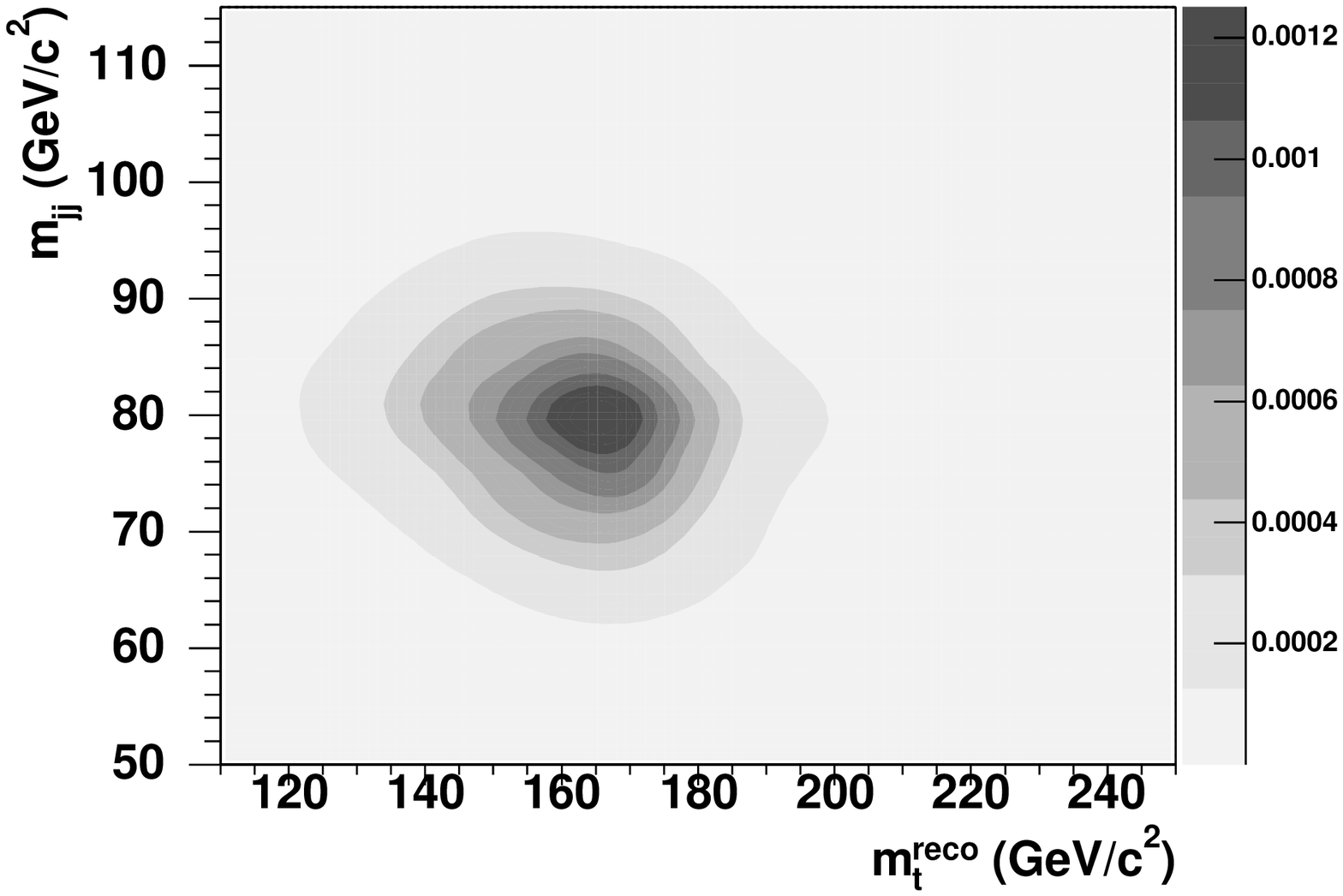} &
\includegraphics[width=0.48\textwidth,type=eps,ext=.eps,read=.eps]{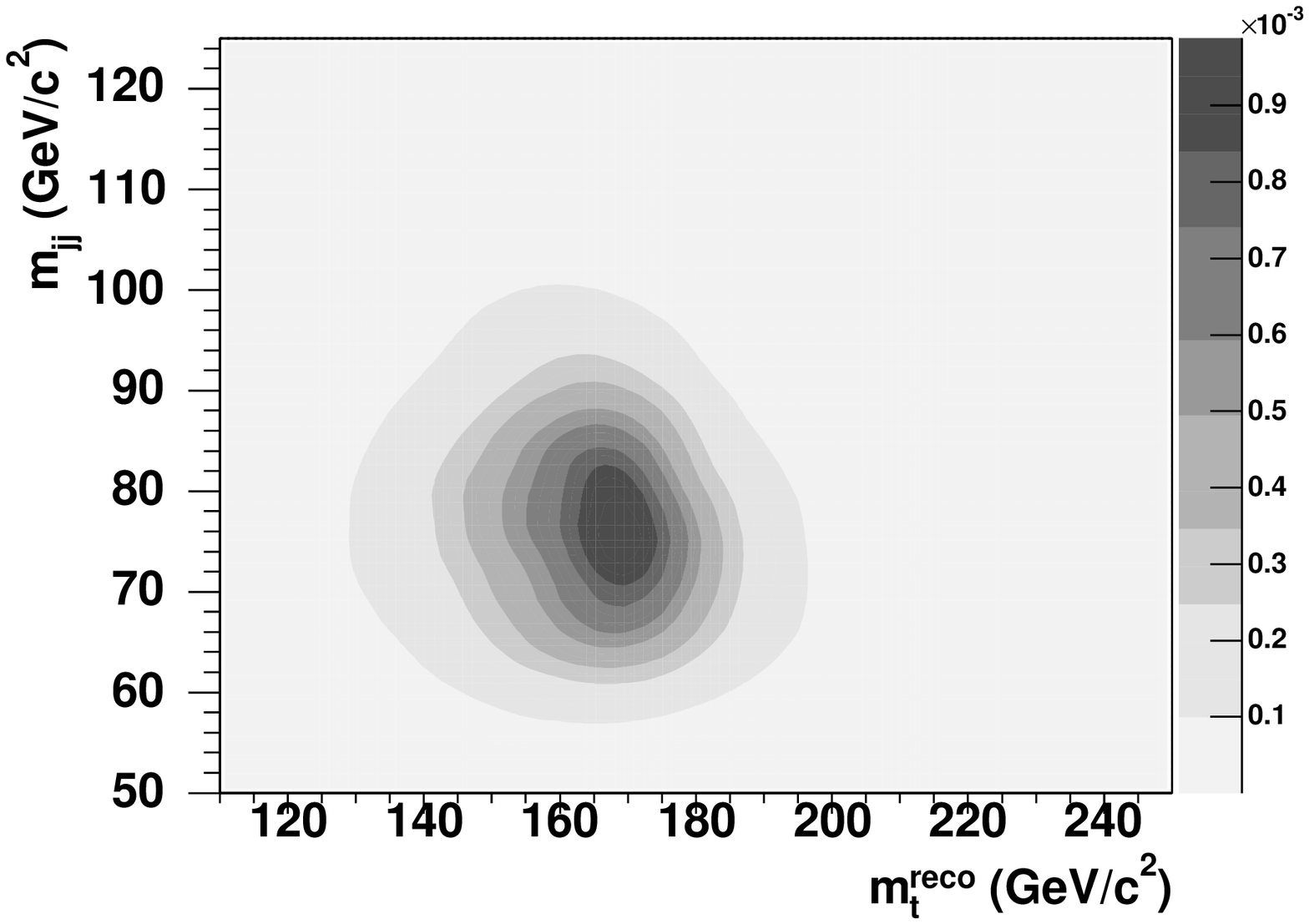} \\
(a) Lepton + jets \onetag & (b) Lepton + jets \twotag \\
\includegraphics[width=0.48\textwidth,type=eps,ext=.eps,read=.eps]{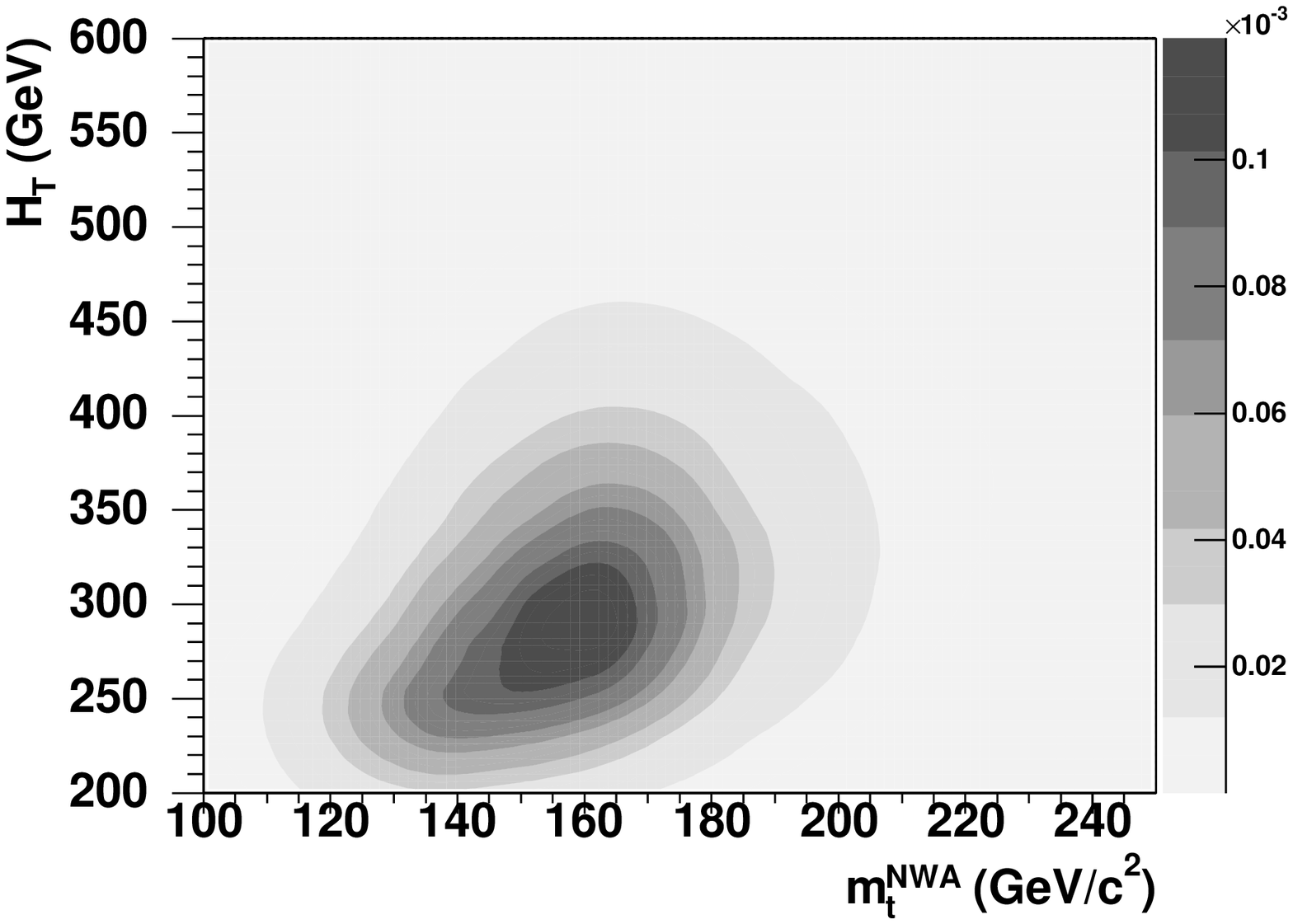} &
\includegraphics[width=0.48\textwidth,type=eps,ext=.eps,read=.eps]{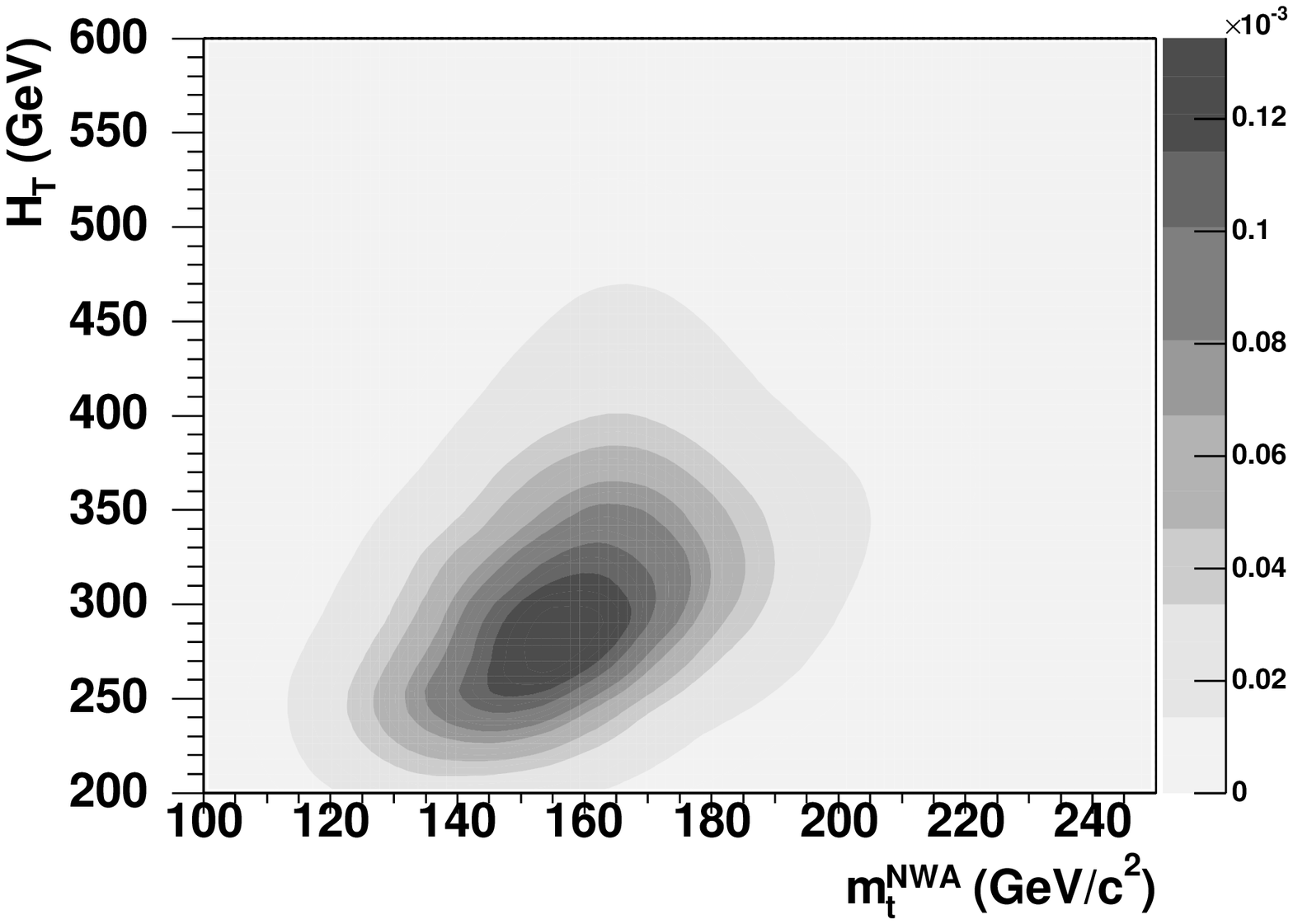} \\
(c) Dilepton 0-tag & (d) Dilepton tagged \\
\end{tabular}
\caption
{Kernel density estimates at $\mtop=\gevcc{170}$ and $\djes = 0.0$ for
\ljets \onetag (a) and \twotag (b) events, and for
\dil untagged (c) and tagged (d) events.}
\label{ref:2dkdesig}
\end{cfigure1c}

\begin{cfigure1c}
\begin{tabular}{cc}
\includegraphics[width=0.48\textwidth,type=eps,ext=.eps,read=.eps]{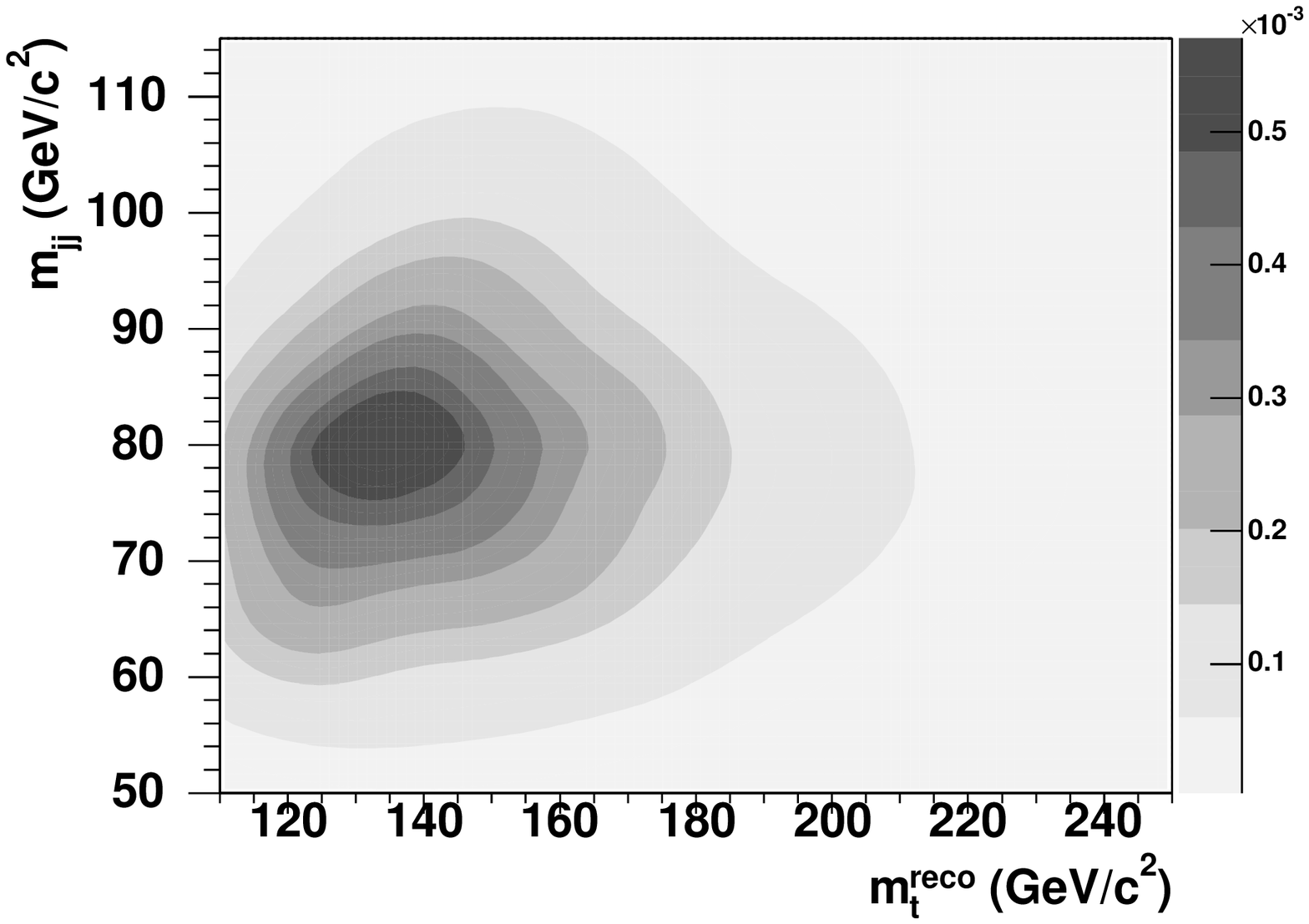} &
\includegraphics[width=0.48\textwidth,type=eps,ext=.eps,read=.eps]{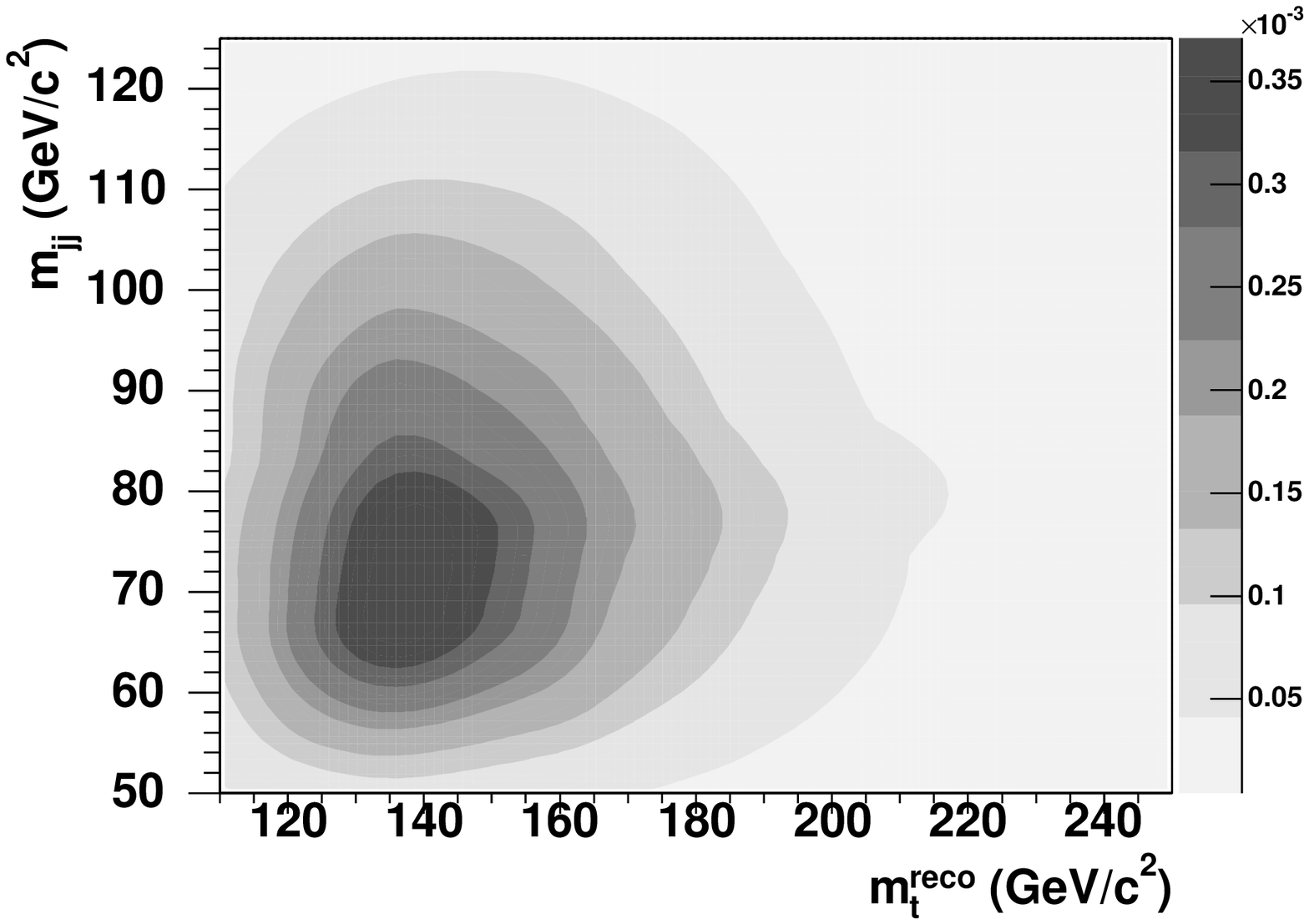} \\
(a) Lepton + jets \onetag & (b) Lepton + jets \twotag \\
\includegraphics[width=0.48\textwidth,type=eps,ext=.eps,read=.eps]{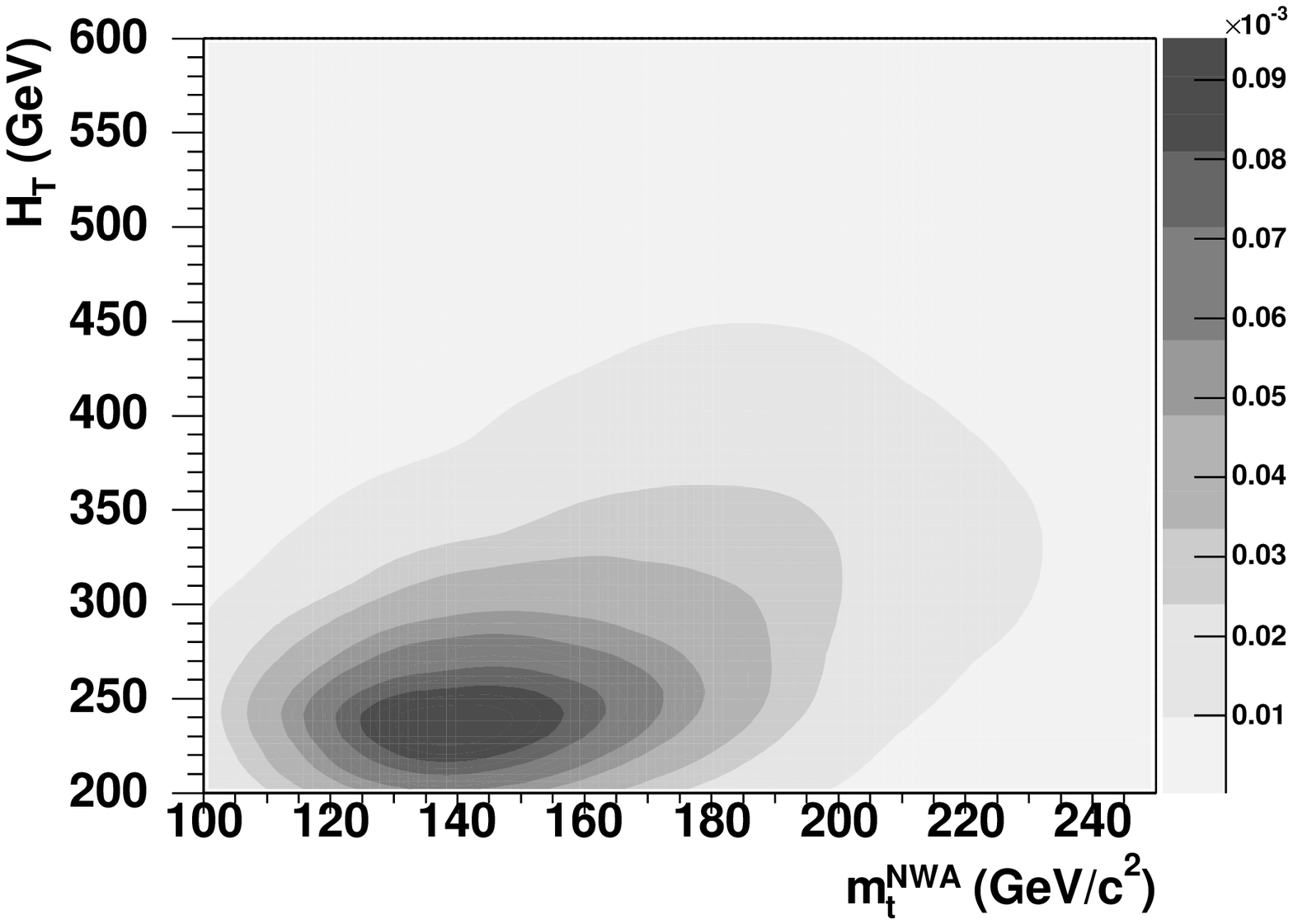} &
\includegraphics[width=0.48\textwidth,type=eps,ext=.eps,read=.eps]{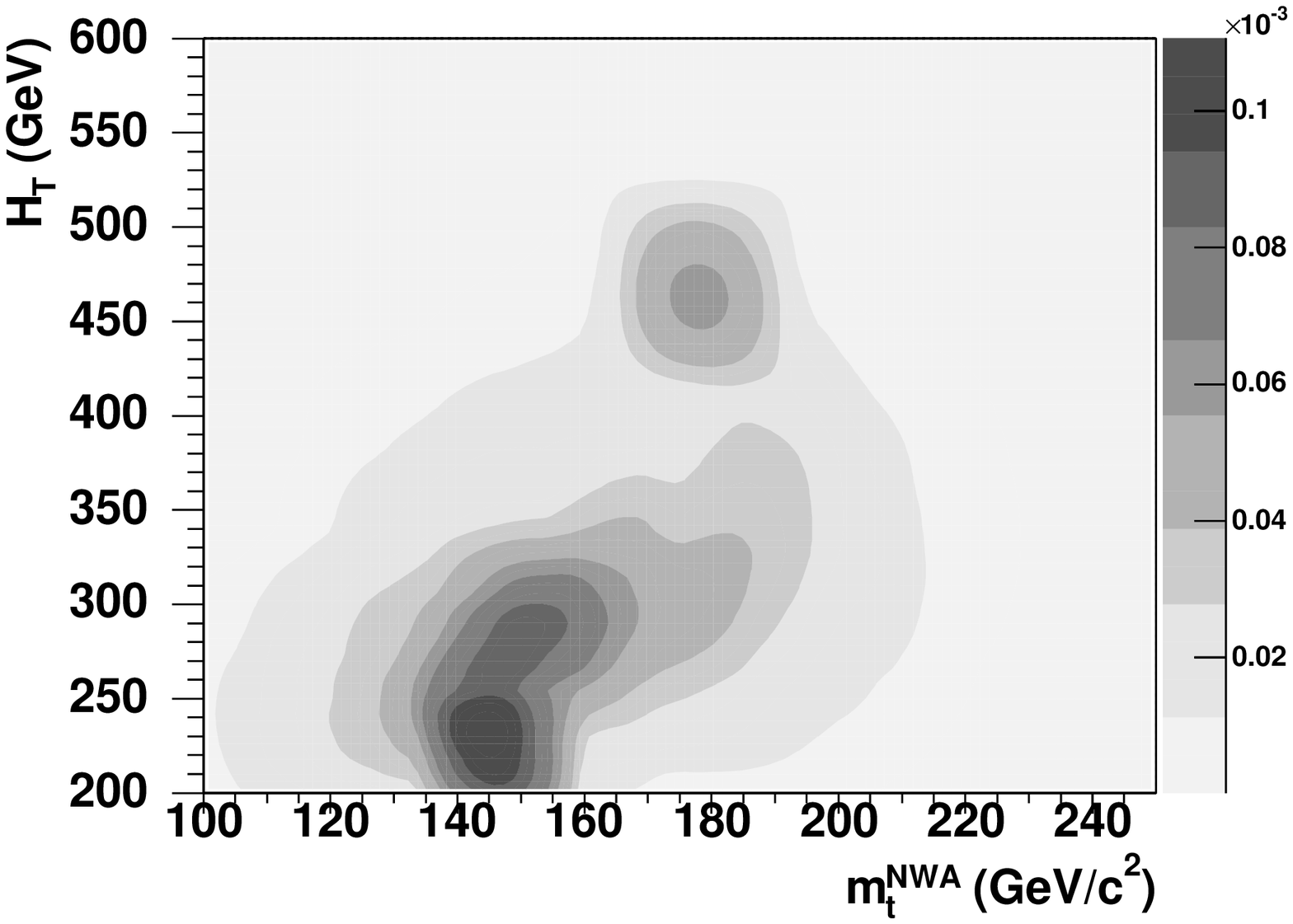} \\
(c) Dilepton 0-tag & (d) Dilepton tagged \\
\end{tabular}
\caption
{Kernel density estimates at $\djes=0.0$ for \ljets \onetag (a) and
\twotag (b) background events, and for \dil untagged (c) and tagged (d) events.}
\label{ref:2dkdebkgd}
\end{cfigure1c}

\subsection{Local Polynomial Smoothing}
\label{sec:lps}

The PDF families $P(x,y;\mtop,\djes)$ are defined on the
continuous parameters \mtop and \djes. But the signal MC samples are
produced at discrete values of \mtop and \djes, and the background MC
samples are produced at discrete values of \djes, so KDE is capable of
producing PDFs only at discrete points of the parameter space.
In addition, statistical
fluctuations in the kernel density estimates are correlated for events
with similar observables,
so it is useful to smooth out the PDFs before the likelihood fit.
To obtain PDFs that are smoothly and 
continuously varying as a function of \mtop and \djes
without assuming Gaussian likelihoods, we employ a
technique known as local polynomial smoothing (LPS)~\cite{LPSbook},
described briefly below.

LPS locally approximates the value of the PDF with a
second-order polynomial. 
The expansion uses the estimates from KDE, but gives more
weight to MC samples in a nearby region of \djes (and \mtop, if we are
smoothing out the signal probabilities).
We look for an estimate $\hat{P}(\bv{\alpha})$ for the true value of the
function $P(\bv{\alpha})$, where we have omitted the values of
the observables $(x,y)$ from the arguments of the function.  The
quantity $\bv{\alpha}$ is a two-dimensional vector $(\mtop,\djes)$ in the
case of the signal probability function or the scalar \djes for the
background. Kernel density estimation provides estimates $Y_{k}$ for
the values of $P(\bv{\alpha_{k}})$ at a number of points $\bv{\alpha_{k}}$.
We assume that $Y_{k}$ are unbiased estimators of the true probability 
values $P(\bv{\alpha_{k}})$ with the same variance.

A second order expansion of the function $P$ for points $\bv{t}$
in the neighborhood of $\bv{\alpha}$ can be written as:
\begin{equation}
P(\bv{t})=\langle \bv{c}, \bv{F}(\bv{t}-\bv{\alpha}) \rangle
\label{expansion}
\end{equation}
\noindent where the angle bracket denotes an inner product.
The coefficients of the expansion are
given by the components of the vector $\bv{c}$. The
quantity $\bv{F}$ is a vector of basis functions for second
order polynomials. For a two-dimensional $\bv{v}$,
$\bv{F}(\bv{v})$ is:
\begin{equation}
\label{avec}
\bv{F}(\bv{v})=\left(
\begin{array}{c}
1 \\
\bv{v}_{0} \\
\bv{v}_{1} \\
\bv{v}_{0} \bv{v}_{1}  \\
\frac{1}{2} \bv{v}_{0}^2 \\
\frac{1}{2} \bv{v}_{1}^2 \\
\end{array}
\right)
\end{equation}
If $\bv{v}$ is a scalar, $\bv{F}(\bv{v})$ reduces to a 3-component vector.

To evaluate $\hat{P}(\bv{\alpha})$, we minimize the criterion given by
\eqnref{mincrt} with respect to $\bv{c}$.
% Was: ...we find a vector of coefficients
% $\bv{c}=\bv{\hat{c}}$ for which the criterion given by
%\eqnref{mincrt} is minimized. 
In other words, we find the second
order expansion of $P$ around $\bv{\alpha}$ that best
matches the estimates $Y_{k}$ at points
$\bv{\alpha_{k}}$:
\begin{equation}
\sum_{k} w(\bv{\alpha_{k}})(Y_{k} - \langle \bv{c}, \bv{F}(\bv{\alpha_{k}} - \bv{\alpha})\rangle)^{2}
\label{mincrt}
\end{equation}
where the weight of each estimate is given by the factor $w(\bv{\alpha_{k}}) = W(u_{k})$, with:
\begin{equation}
u_{k}=\sqrt{\sum_{d=1}^{N_{d}} \left(\frac{\bv{\alpha}_{d}-\bv{\alpha}_{\textbf{\textit{k}},d}}{h_{d}}\right)^{2}}
\label{weightfactor}
\end{equation}

The sum in \eqnref{weightfactor} runs over the components of the
vectors, and $N_{d}$ is the dimensionality of the parameter
space. We use $W=W(u)=(1-|u|^{3})^3$ for
$|u|<1$ and $W(u)=0$ otherwise. This gives a smoothly
decreasing weight to the estimates $Y_{k}$ obtained at points
far away from the evaluation point $\bv{\alpha}$. The constants
$h_{d}$ control the amount of smoothing, larger values of
$h_{d}$ give more weight to the estimates farther away from
the point $\bv{\alpha}$.

For this analysis, signal MC samples are generated at 76 mass points
with \mtop ranging from 120 to \gevcc{240}.  The spacing between mass
points is small (\gevcc{0.5}) in the region of interest
(165--\gevcc{185}), and gets larger in the tails of the grid. Each
signal MC sample and MC-based background is processed using 29
different values of \djes from \sigcunit{-3.0} to
\sigcunit{+3.0}. Near the range of interest of nominal \djes, the
spacing is \sigcunit{0.2}.  We set the constants $h_{d}$ of
\eqnref{weightfactor} based on the performance of the analysis in
terms of expected precision and biases.  We choose
$h_{\mtop}=\gevcc{10.0}$ for signal PDF smoothing in \ljets events and
$h_{\mtop}=\gevcc{15.0}$ in the \dil events. For both categories,
$h_{\djes}=\sigcunit{0.8}$. The background MC samples have
smaller statistics and increased jitter, as many 
events are selected near jet energy thresholds and move in and out of 
the sample as \djes varies. To compensate for these effects, we choose the
larger $h_{\djes}=\sigcunit{3.0}$ for background smoothing.

%Applying LPS to obtain the probability density fuctions allows the
%likelihood to assume any form; we do not need to make the assumption
%that the likelihood is Gaussian in \mtop. \note{Is the part after the
%semicolon attacking an argument that hasn't beeen made?} Since the
%technique is applied for every event in the likelihood, correlated
%fluctuations of the individual KDE estimates will be smoothed
%out. \note{Could probably use a better rewrite--I removed a reference
%to an equation in a later section (after my reorg).}

\subsection{Likelihood fit}
\label{sec:Lfit}

We compare the two dimensional distributions of the observables in the
data with the signal and background PDFs in an unbinned extended
maximum likelihood fit~\cite{Barlow:1990vc}.
The most important parameters of the fit are
the mass of the top quark (\mtop) and the deviation from the nominal
jet energy scale (\djes). Each subsample gives two additional
parameters: the expected number of signal events (\ns) and the
expected number of background events (\nb) in the subsample. The
likelihood form is given by:
\begin{eqnarray}
\label{ltotal}
\mathcal L & = & e^{-\frac{\djes^{2}}{2}} \nonumber \\
      & \times & \lljonetag  \nonumber \\
      & \times & \lljtwotag  \nonumber \\ 
      & \times & \ldilnontag \nonumber \\ 
      & \times & \ldiltag
\end{eqnarray}
\noindent where the first term in the product constrains the measurement of the JES to
its nominal value of 0 within the uncertainty of \sigcunit{1}. Each of the
subsequent terms corresponds to one subsample, and is given by:
\begin{equation}
\label{lsample}
\lsample=\lshape \times \lbg
\end{equation}

The term in the likelihood most critical to the mass measurement is the
extended maximum likelihood shape term:
\begin{equation}
\label{lshape}
\begin{split}
\lshape&=\frac{e^{-(\ns+\nb)} (\ns+\nb)^N}{N!}\\
&\times \prod_{i=1}^{N} \frac{\ns P_{s}(x_i,y_i;\mtop,\djes) + \nb P_{b}(x_i,y_i;\djes)}{\ns+\nb}
\end{split}
\end{equation}
\noindent where the product runs over all events in a given
subsample. The observables $(x_i,y_i)$
in the $i$th event are $(\mreco,\mjj)$ in the \ljets
channel and $(\mnwa,\Ht)$ in the \dil channel. The 
quantities $P_{s}$ and $P_{b}$ designate the signal
and background PDFs as determined by KDE and LPS.
%The \lnev term is given by \eqnref{lnev} and arises because we
%fit for the expected number of signal and background events and not
%simply a background fraction.
%\begin{equation}
%\label{lnev}
%\lnev=\frac{e^{-(\ns+\nb)} (\ns+\nb)^N}{N!}.
%\end{equation}
To improve the precision of the measurement, we apply a Gaussian
constraint to the expected number of background events:
\begin{equation}
\label{lbg}
\lbg=e^{-\frac{(\nb-\nbzero)^{2}}{2\sigma_{\nbzero}^{2}}}
\end{equation}
\noindent where \nbzero is the \textit{a priori} estimate for the
expected number of background events, and $\sigma_{\nbzero}$
is the uncertainty on the estimate. Both sets of numbers are given 
in \tab{tablefinalbackgroundlj} and \tab{tablefinalbackgrounddil}.

We minimize the negative logarithm of the likelihood with respect to
all 10 parameters using {\sc minuit}. The uncertainty on
\mtop and \djes is found by searching for the points where the negative
logarithm of the likelihood minimized with respect to all other
parameters deviates by $\frac{1}{2}$ from the minimum. The uncertainty
on the top quark measurement obtained this way includes the
statistical uncertainty as well as the systematic uncertainty due to
allowed variations in the jet energy scale and the background
estimates. We scale the uncertainty on the top quark mass by the pull
width as obtained in \secref{sec:bias_check}. For the \dil-only
cross-check, we fix the value of the \djes parameter to \sigcunit{0}
and perform a fit only for
\mtop, as the \dil channel has no power to resolve \mtop and \djes
simultaneously.

\subsection{Method check}
\label{sec:bias_check}

We test the likelihood procedure using large numbers of MC simulated experiments, 
each of which is generated for a specific value of \mtop and \djes. 
In each experiment we select the number of background events from a
Poisson distribution with a mean equal to the expected number of background
events in the sample. The number of signal events is selected from a
Poisson distribution with a mean equal to the expected number of
signal events assuming a \ttbar pair production cross section of
\pb{6.7}. The signal events are drawn at random from a MC sample
generated at a given \mtop and \djes. The background events are drawn
from the entire background sample with probabilities corresponding to
the different background sources and the weights of individual events
given by the model. Individual event rates can vary based 
for example on mistag probability of jets in a given event.
An event (signal or background) can be drawn arbitrary number 
of times from a given MC sample so that different
MC experiments can share simulated data. We refer to this method of drawing 
events from MC samples as ``drawing with replacement''. 
Once the simulated data is
constructed we perform a maximum likelihood fit as described
in the previous sections. We fluctuate the constraint on JES and the
constraint on the expected
number of background events in each subsample to reflect the possible
deviation of those parameters from their \emph{a priori} expected values.
The fluctuations on the constraints are applied to estimate the
effect of our limited knowledge about the nuisance parameters on the
top quark mass measurement. The jet energy scale 
constraint in \eqnref{ltotal} is
replaced by:
\begin{equation}
\label{fluctjes}
e^{-\frac{(\djes-\djesf)^{2}}{2}}
\end{equation}
where \djesf is randomly selected in each MC simulated experiment
from a Gaussian with mean
corresponding to the \djes being tested and unit width. Similarly, the
quantity \nbzero in \eqnref{lbg} is replaced in each MC simulated experiment by a
value drawn from a Gaussian with mean of \nbzero and width
$\sigma_{\nbzero}$.

The likelihood fit should, on average, return the value of the top
quark mass used to generate the
MC simulated experiments. \Fig{biasresiduals} shows the average
residual (deviation from expectation for an unbiased measurement) from
3000 MC simulated experiments for a range of \mtop.
%Was: ...
% and \djes.
The fit to a constant shows no bias for the combined
and \ljets-only fits, and a small positive bias for the
\dil-only fit. This bias does not warrant a correction since it is
small in comparison to the expected uncertainty, and has a probability
of 9\% for a purely statistical fluctuation to generate the observed
shift. 
%The \ttbar MC samples generated at different values of
%\djes but the same value of \mtop share almost all events when \djes 
%is shifted. This leads to large correlations in the
%pseudoexperiment output, so we use only the pseudoexperiments
%generated at $\djes = \sigcunit{0}$ in the fits.

\begin{cfigure1c}
\begin{tabular}{c}
\includegraphics[height=0.25\textheight]{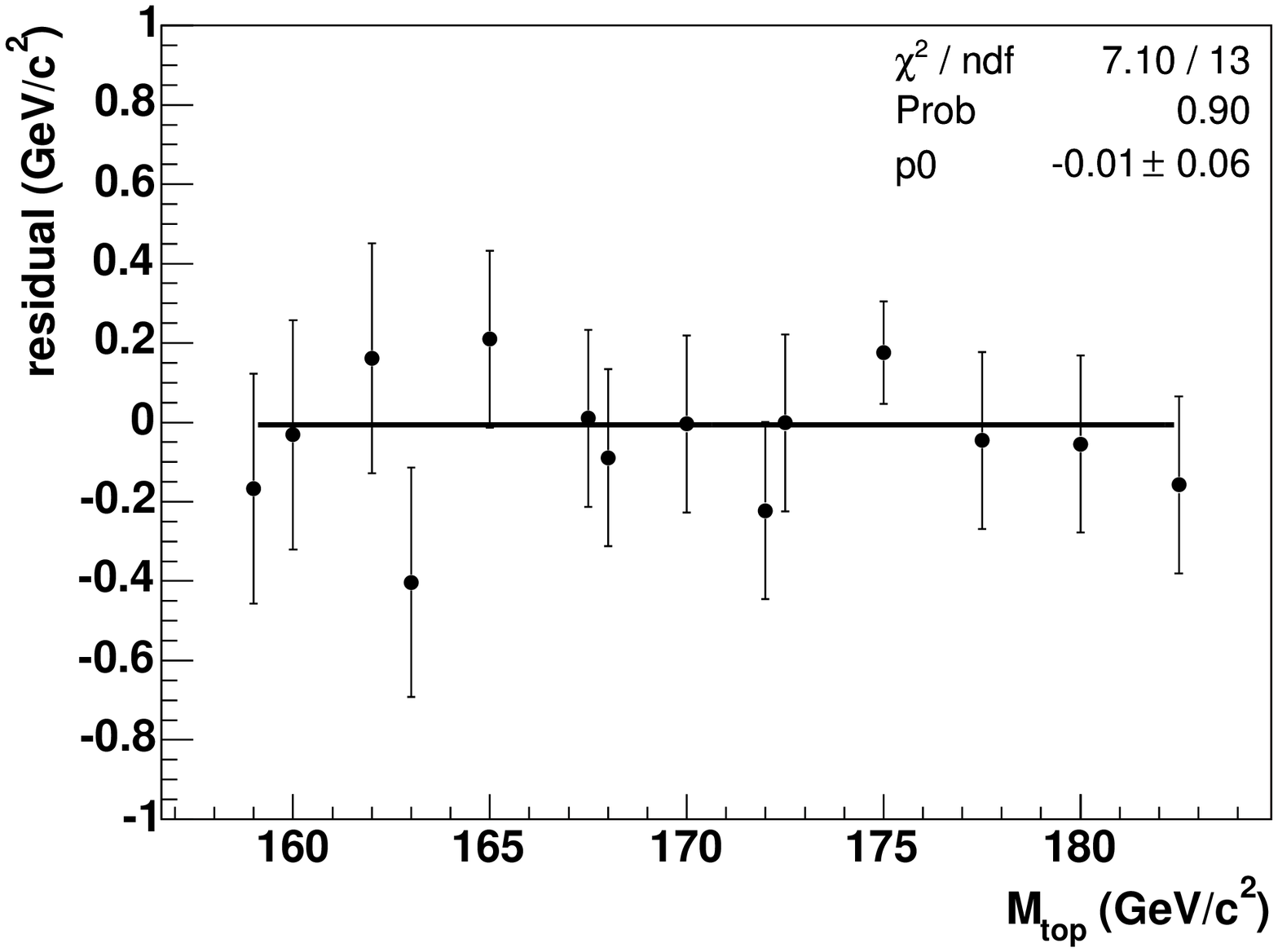} \\[-1.6pc]
(a) Lepton + jets only fit
\end{tabular}
\begin{tabular}{c}
\includegraphics[height=0.25\textheight]{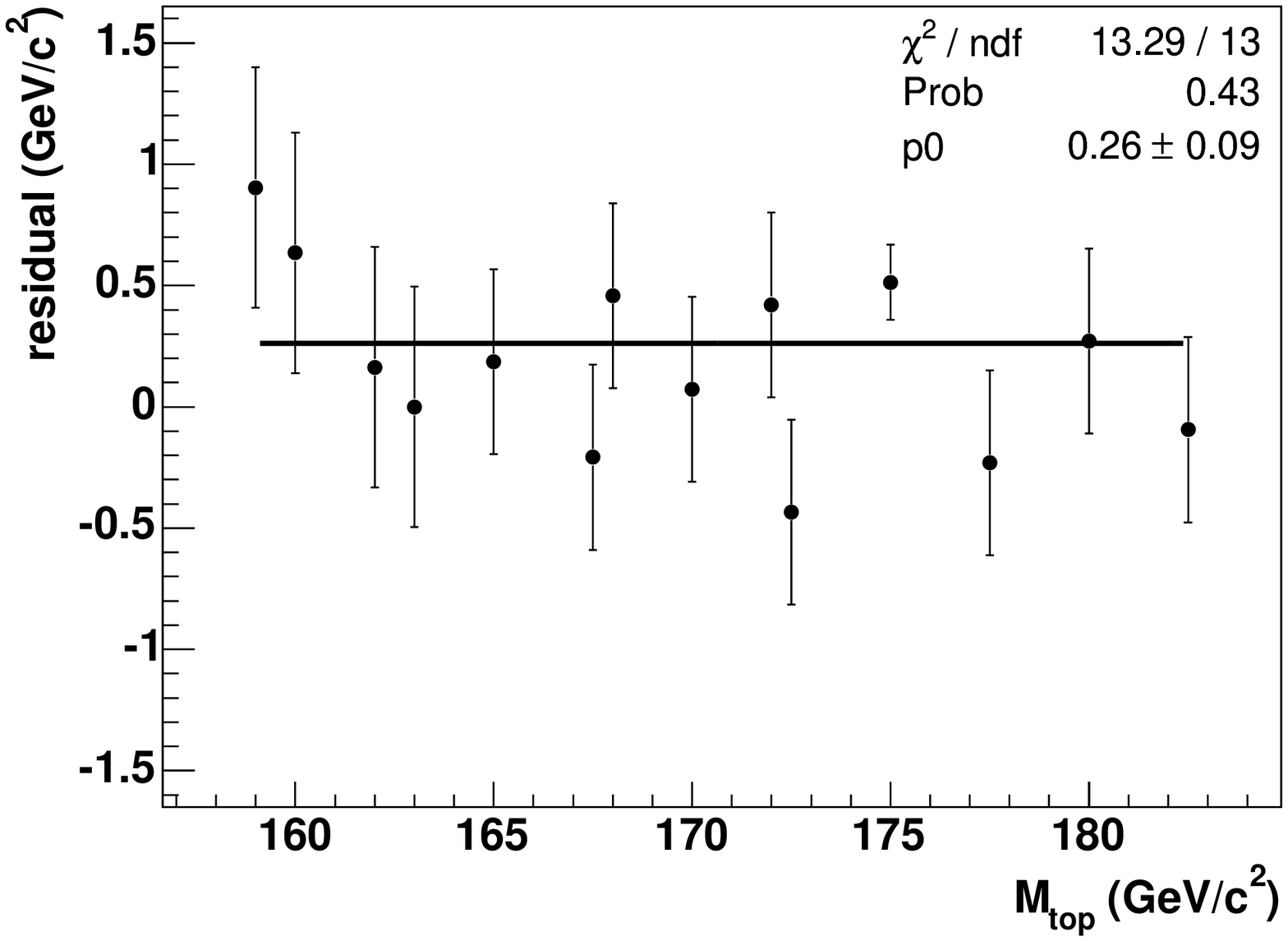} \\[-1.6pc]
(b) Dilepton only fit
\end{tabular}
\begin{tabular}{c}
\includegraphics[height=0.25\textheight]{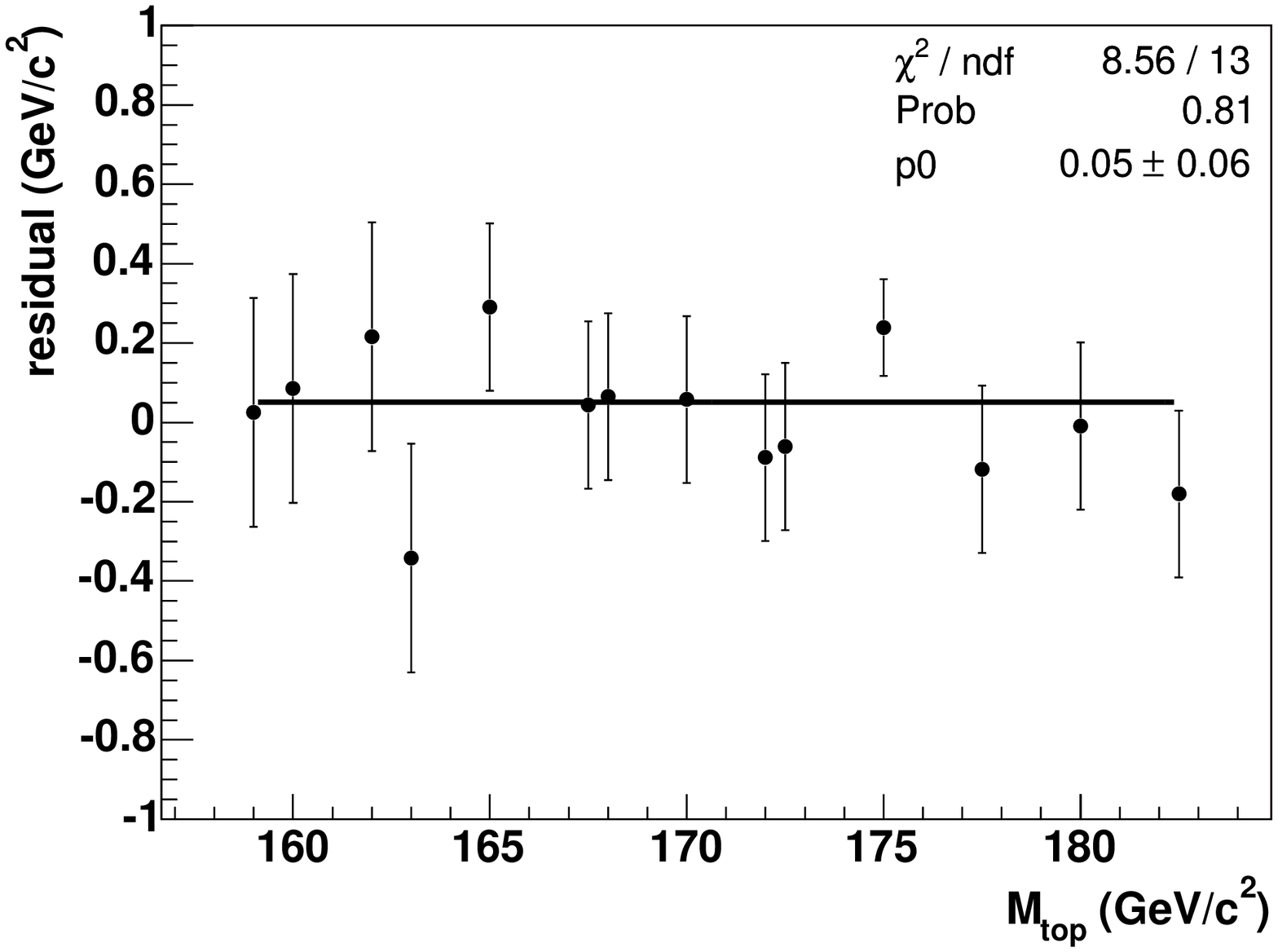} \\[-1.6pc]
(c) Combined fit
\end{tabular}
\caption[Residual bias checks]{Checks for bias in the fitted top quark mass for a) \ljets only fit, b) \dil only fit and c) combined fit.}
\label{biasresiduals}
\end{cfigure1c}

The statistical uncertainty on the measurement is extracted from the
data. To test whether the error estimate is sound, for each
MC simulated experiment we calculate the pull, defined as a ratio of the
residual to the uncertainty reported by {\sc minuit}. If the residual
is positive (negative) we use the negative (positive) error in the
ratio. The width of the pull distribution for \mtop is shown in
\fig{masspulls}. The average pull width is larger than 1.0 for
the combined and \ljets-only measurements due to the finite
number of events in the two-dimensional fits. The pull
width correction is 3\% for these measurements, and thus we increase the 
reported uncertainty in the data by this amount.

\begin{cfigure1c}
\begin{tabular}{c}
\includegraphics[height=0.25\textheight]{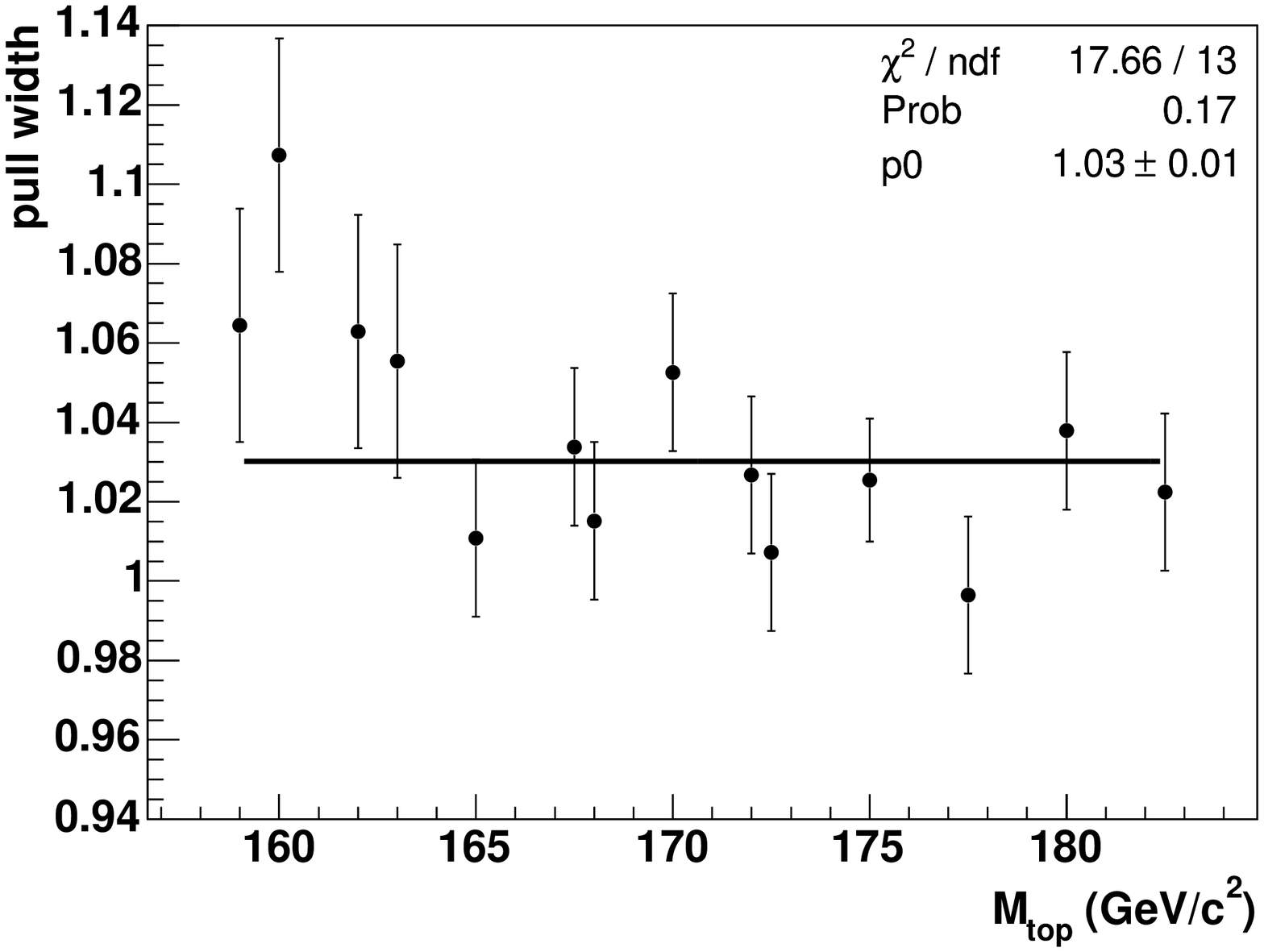} \\[-1.6pc]
(a) Lepton+ jets only fit
\end{tabular}
\begin{tabular}{c}
\includegraphics[height=0.25\textheight]{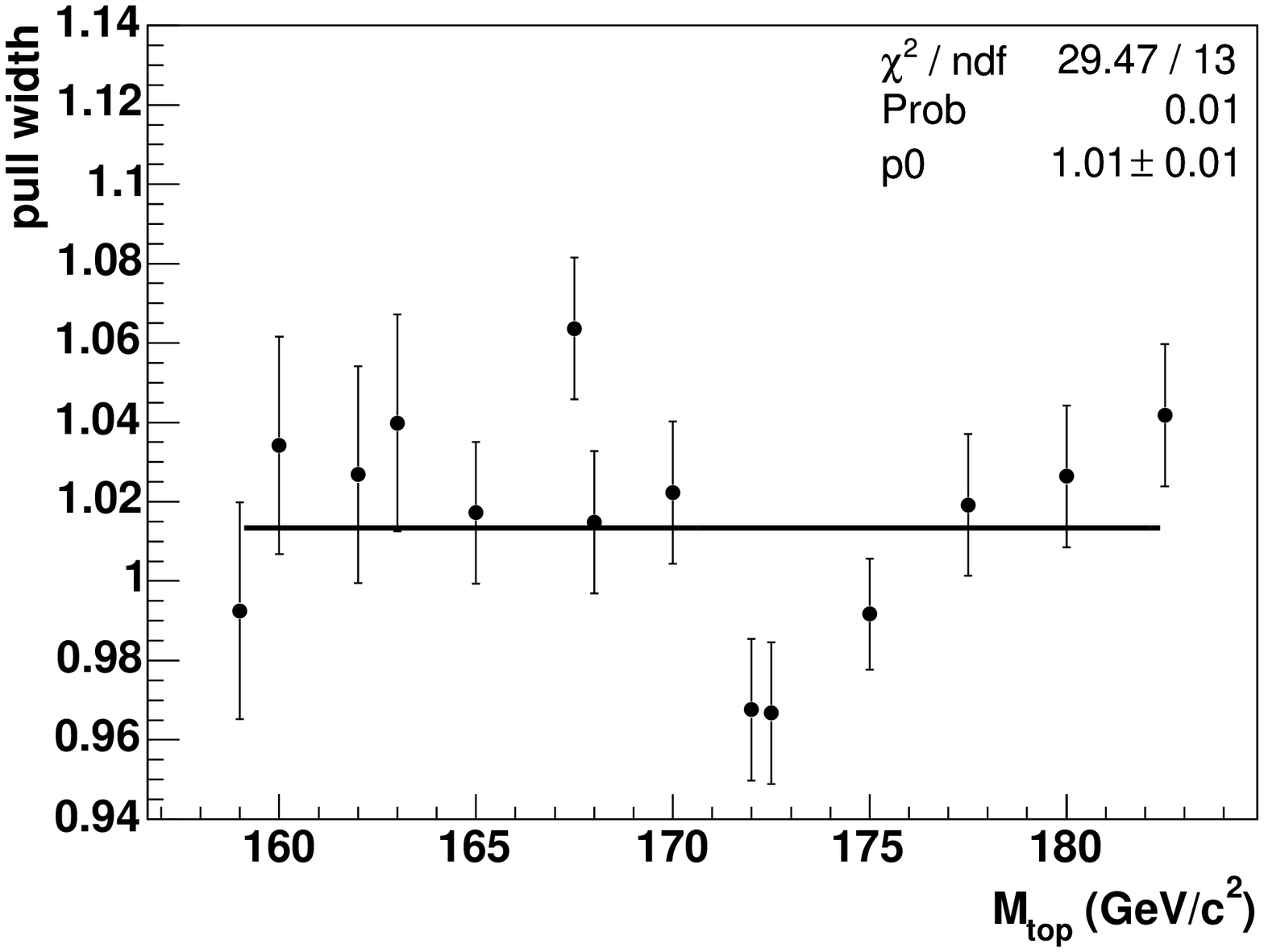} \\[-1.6pc]
(b) Dilepton only fit
\end{tabular}
\begin{tabular}{c}
\includegraphics[height=0.25\textheight]{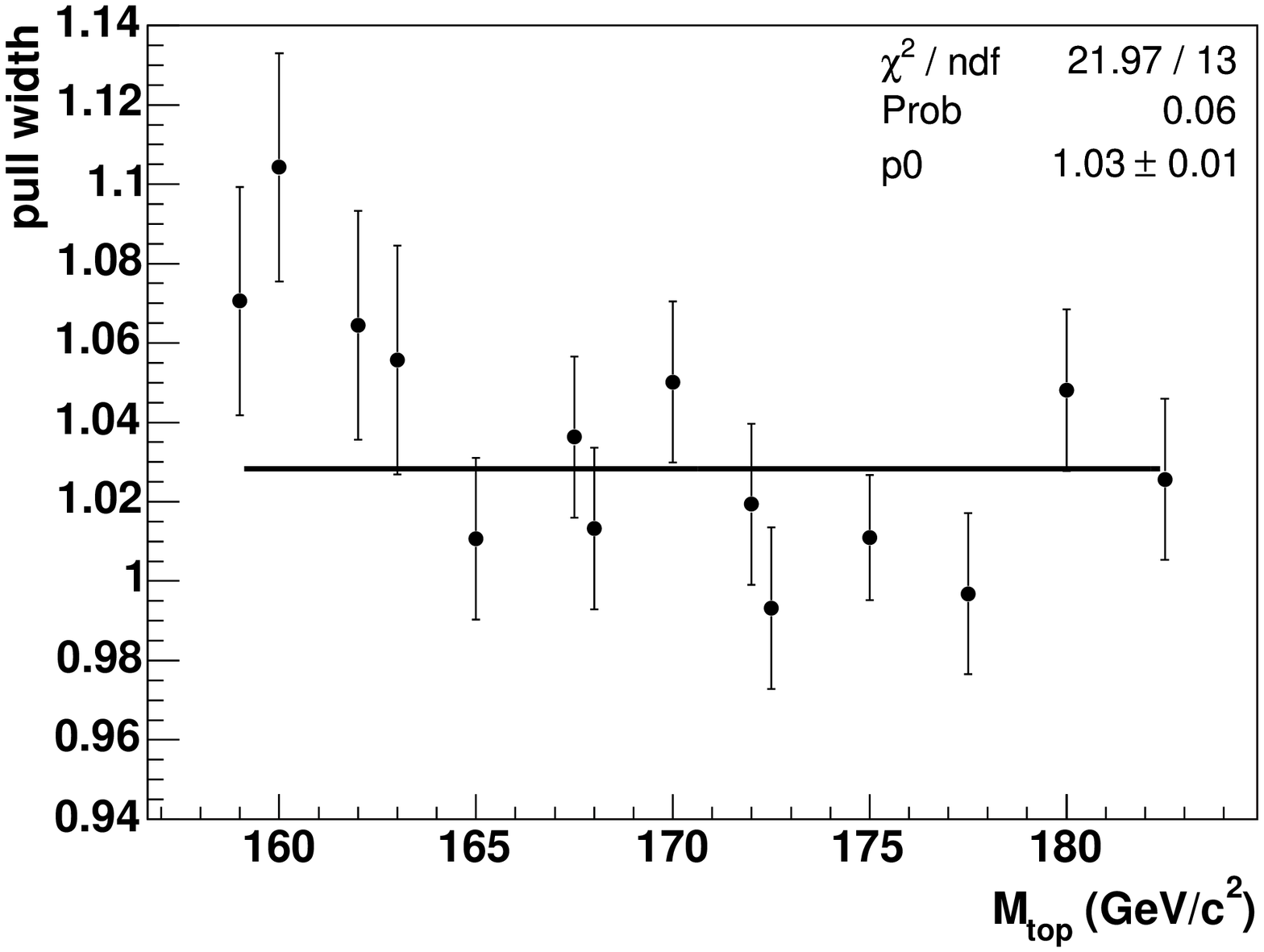} \\[-1.6pc]
(c) Combined fit
\end{tabular}
\caption[Mass pull width]{Width of the pull distribution for the fitted \mtop for the a) \ljets only fit, b) \dil only fit and c) combined fit.}
\label{masspulls}
\end{cfigure1c}

The residual and pull width for the \djes parameter are also
investigated using the MC simulated experiment ensembles.  Both the combined
fit and the \ljets-only fit show a negative bias of
$\lesssim\sigcunit{0.02}$ (with marginal statistical significance) and
a pull width of $\sim 1.04$.  Since \djes is a nuisance parameter,
whose precise value is not as important as its effect on the top quark
mass measurement, we do not correct the \djes value measured in data 
for this bias.

As noted above, when ensembles of simulated data are constructed the
events are drawn with replacement. If we were to draw the events from
MC samples without replacement such that no two MC simulated experiments were
to share events, we would have only $\sim 100$ MC simulated experiments for
each \ttbar sample. Drawing events with replacement allows us to
perform an arbitrarily large number of MC simulated experiments, fully
exploring the possible combinations of events in order to check our
machinery for possible biases. To evaluate the uncertainties on
statistics such as the residuals and pull widths, we employ the
bootstrap technique~\cite{efron, efron2}. In each bootstrap ensemble,
we draw events from the signal MC sample with replacement until
we reach the same number of events as in the original sample. We then
run 3000 MC simulated experiments using this bootstrapped sample in place
of the original sample. We repeat the above procedure 60 times. For
each of the bootstrap ensembles, we calculate the desired statistic.
The RMS of the statistic in question from
the 60 bootstrap ensembles is taken as the uncertainty on the
statistic. For example the uncertainty on the residual of fitted top quark
mass in a typical MC sample
is \gevcc{0.4} for the \dil fit and \gev{0.2} for the \ljets and
combined fits

\section{Results on data}
\label{sec:results}

The likelihood fit when applied to the data yields $\mtop =
\gevcc{\measStatJES{171.9}{1.7}}$. The \ljets-only fit yields
$\mtop = \gevcc{\measStatJES{171.8}{1.9}}$. The \dil-only fit,
which does not include an \textit{in situ} \djes measurement
but instead fixes $\djes$ to \sigcunit{0.0}, yields
$\mtop = \gevcc{\measAStat{171.2}{3.6}{3.4}}$.  The combined fit
returns $\djes = \sigcunit{\measStatMT{-0.17}{0.35}}$, and the \ljets fit
returns $\djes = \sigcunit{\measStatMT{-0.12}{0.37}}$. The results above have
been corrected for the pull width and high instantaneous luminosity effects.
Results from the combined fit, including fitted
numbers of signal and background
events for each subsample, are summarized in \Tab{t:fitResults}. The \dil-only
fit and \ljets-only fit both return \mtop value lower than the \mtop 
measured in both channels simultaneously.
This is due to the \textit{in situ} JES calibration extracted from the \ljets 
channel events being applied to the \dil channel data in the combined fit.

The log-likelihood contours for the combined measurement are shown in
\fig{ref:contourcomb}. The one-dimensional log-likelihood for the
\dil-only measurement and the log-likelihood contours for the
\ljets-only measurement are shown in \fig{ref:likedilonly} and
\fig{ref:likeljonly}, respectively. 
\Fig{templateoverlayLJ} shows the one-dimensional \ljets data with
the best-fit one-dimensional signal and background distributions
overlaid on top. \Fig{templateoverlayDIL} shows the distributions for the
\dil data.
Using the observed number of
events in data and the background expectations, 10\% of MC
experiments have a smaller error than the value measured in the
combined fit. The $p$-value for the \ljets-only fit is 21\%; the value
for the \dil-only fit is 14\%.

\begin{table}
\caption[Combined fit results.]
{The input constraints and fitted values are given for all free
parameters in the combined likelihood fit. LJ refers to \ljets
subsamples and DIL refers to \dil subsamples. Con. refers to the
constraint used in the likelihood.}
\label{t:fitResults}
\begin{ruledtabular}
\begin{tabular}{llcccc}
\multicolumn{2}{l}{Category} & LJ \twotag & LJ \onetag & DIL tagged & DIL 0-tag\\
\hline
\mtop & con. & \multicolumn{4}{c}{None} \\
     & fit &
\multicolumn{4}{c}{\mathversion{bold}\gevcc{\measErr{171.9}{1.7}}}\\
\hline
\jes  & con. & \multicolumn{4}{c}{\sigcunit{\measErr{0.0}{1.0}}} \\
     & fit & \multicolumn{4}{c}{\sigcunit{\measErr{-0.17}{0.35}}}\\
\hline
$n_s$ & con. & \multicolumn{4}{c}{None} \\
     & fit & \measAErr{96.4}{10.4}{9.7} & \measAErr{184.1}{17.7}{17.1} &
             \measAErr{56.9}{8.2}{7.5} & \measAErr{43.8}{10.4}{9.8} \\
\hline
$n_b$ & con. & \measErr{4.2}{1.9} & \measErr{42.7}{12.5} &
                    \measErr{3.9}{1.0} & \measErr{44.3}{7.0} \\
     & fit & \measErr{3.4}{1.9} & \measAErr{47.4}{10.2}{10.0} &
             \measErr{3.9}{1.0} & \measAErr{41.5}{6.5}{6.4} \\
\end{tabular}
\end{ruledtabular}
\end{table}

\begin{cfigure}
\includegraphics[width=\columnwidth]{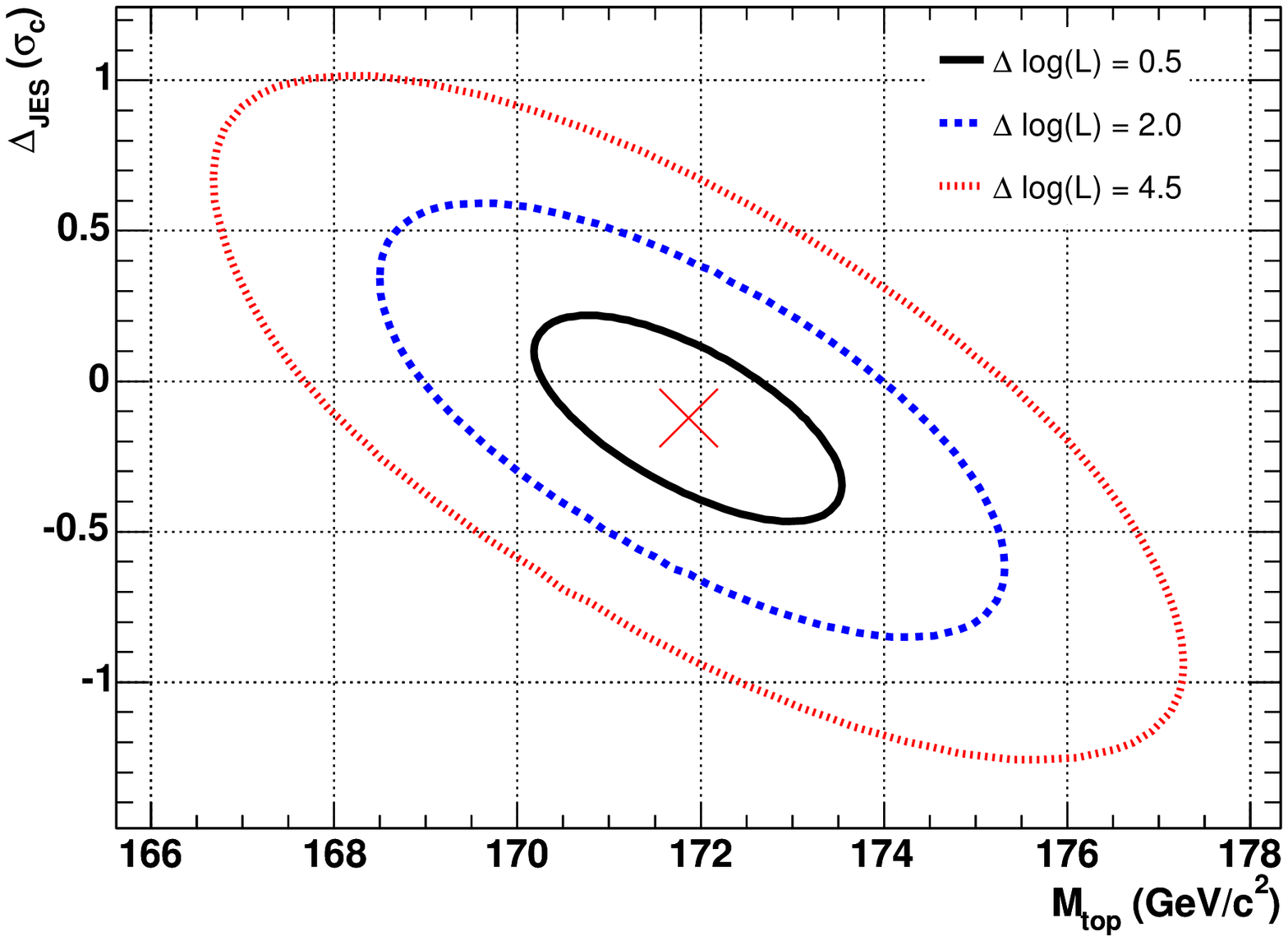}
\caption{Negative log-likelihood contours for the combined fit. The minimum is
indicated by the 'x' and corresponds to the most probable
top quark mass and $\djes$, given the data. The contours are
drawn at values of 0.5, 2.0 and 4.5 of the increase of the
log-likelihood from the minimum value. These curves correspond to the
1,2 and \sigunit{3} uncertainty on the measurement of the top quark mass.}
\label{ref:contourcomb}
\end{cfigure}

\begin{cfigure}
\includegraphics[height=0.4\textheight]{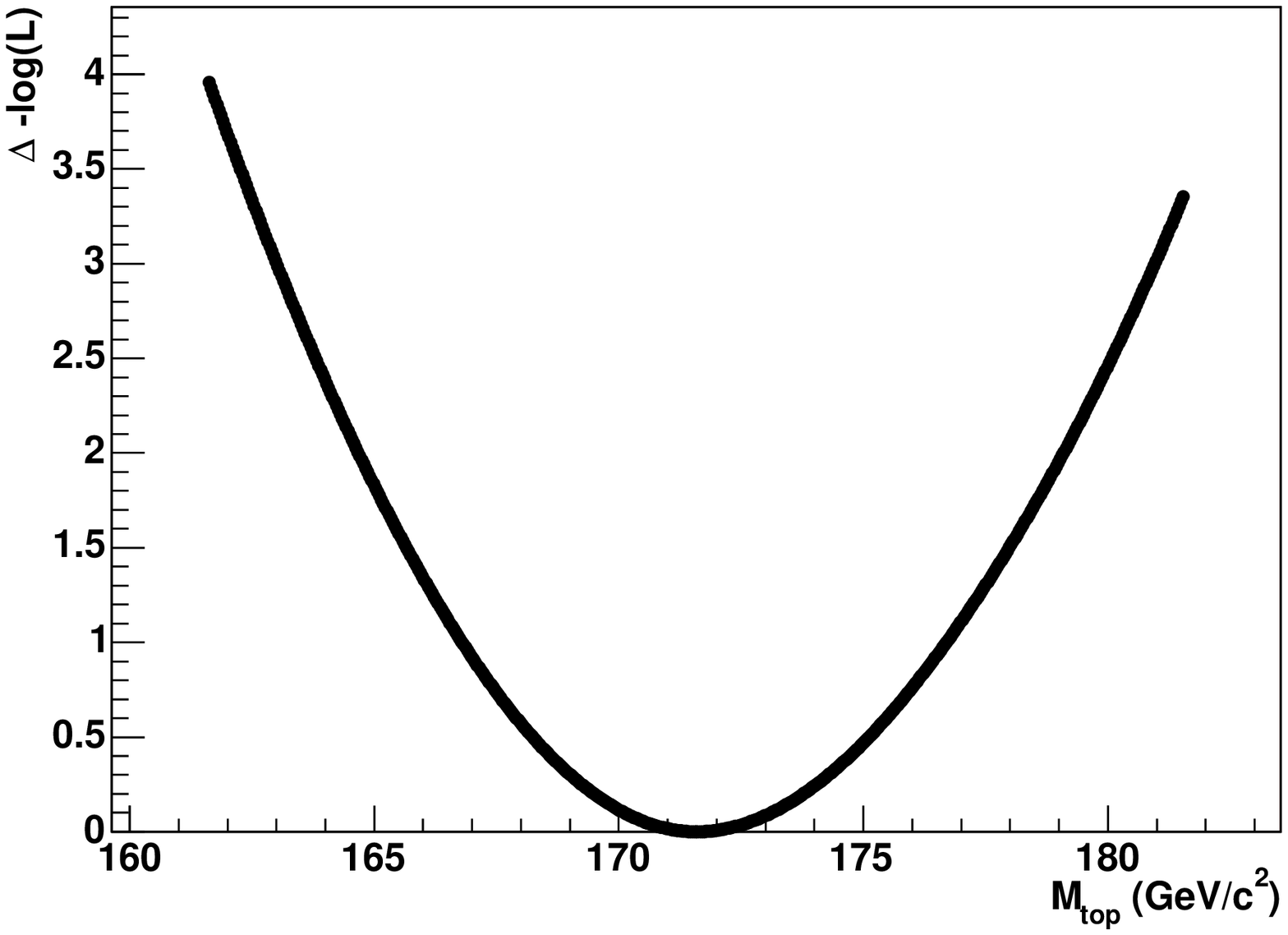}
\caption
{One-dimensional log-likelihood for the \dil-only fit.}
\label{ref:likedilonly}
\end{cfigure}

\begin{cfigure}
\includegraphics[height=0.4\textheight]{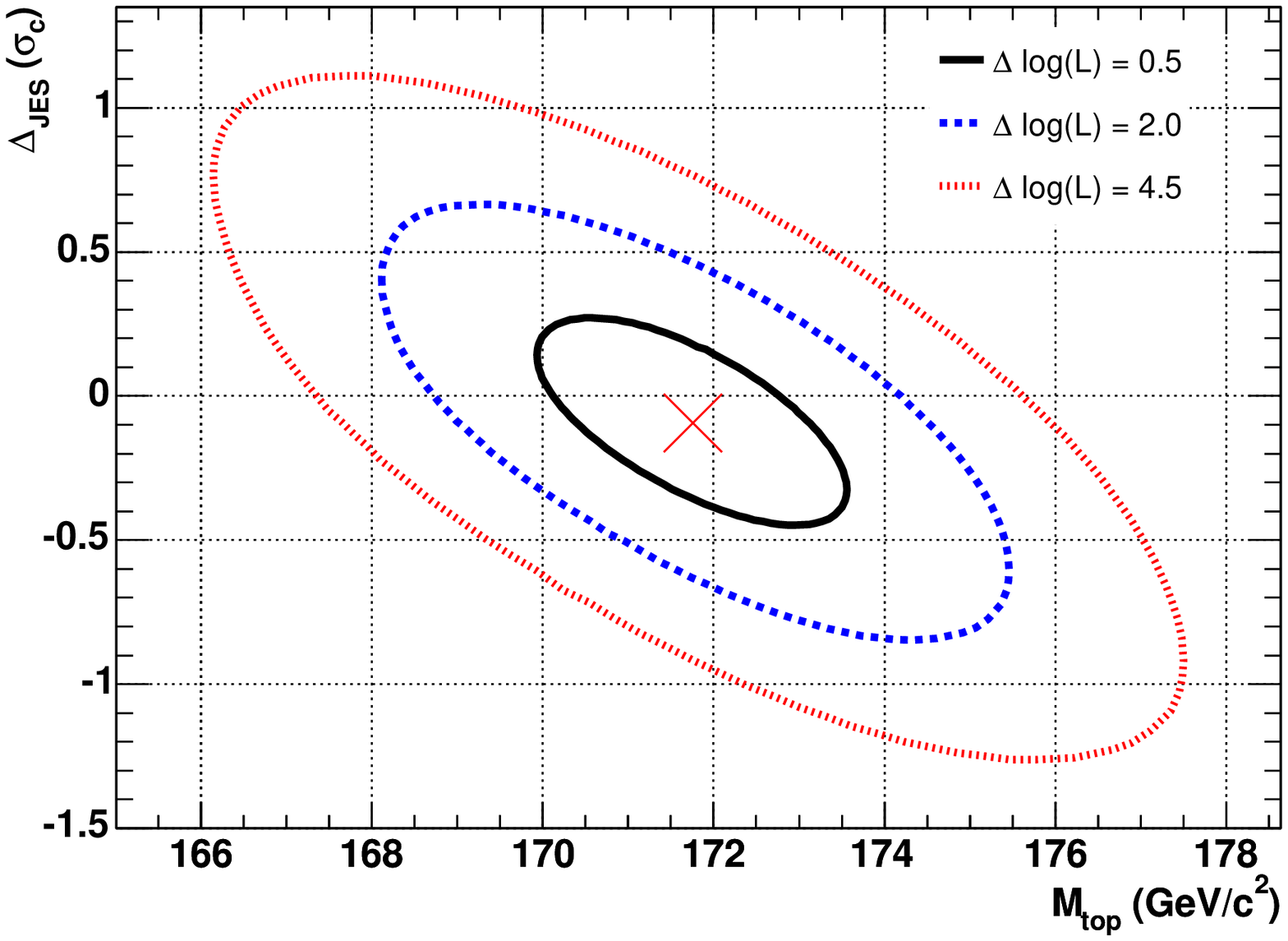}
\caption
{Negative log-likelihood contours for the \ljets-only fit. The minimum is
indicated by the 'x' and corresponds to the most probable
top quark mass and $\djes$, given the data. The contours are
drawn at values of 0.5, 2.0 and 4.5 of the increase of the
log-likelihood from the minimum value. These curves correspond to the
1,2 and \sigunit{3} uncertainty on the measurement of the top quark mass.}
\label{ref:likeljonly}
\end{cfigure}

%We fit without the \emph{a priori} Gaussian constraints on \djes and
%\nb, and the result is unchanged after rounding, showing that these
%priors do not significantly affect our result.
We fit without the \emph{a priori} JES and background
constraints and measure the same $\mtop =
\gevcc{\measStatJES{171.9}{1.7}}$, showing that these priors do not
significantly affect our result. We also fit separately in the 
several individual subsamples: the first \invfb{1} and last \invfb{0.9}
of data, electron and muon events in the \ljets-only fit, and different
lepton pair type events in the \dil-only fit. 
In addition we quote the top quark mass fitted in \ljets and \dil
subsamples separated by $b$ tagging multiplicity.
The results are consistent across cross-checks and are summarized in 
\Tab{xchecks1}. The results are not corrected for pull width effects and 
bias due to incorrect instantaneous luminosity profile of the MC 
samples. With
the exception of the fit without the \emph{a priori} JES constraint, all
cross-checks include the JES prior. The fits in the four subsamples separated
by $b$ tagging multiplicity  include the \emph{a priori} background
constraints; all
other cross-checks do not include background constraints.

\begin{table}
\caption[Cross checks]
{Cross-checks on the data. LJ refers to the \ljets-only fit, DIL
refers to the \dil-only fit, and Combo refers to the combined
fit. All numbers are uncorrected for pull width effects and 
bias due to incorrect instantaneous luminosity profile of the MC
samples. For the dilepton-only fits, \djes is fixed to \sigcunit{0.0}.}
\begin{center}
\begin{ruledtabular}
\begin{tabular}{cccc}
\label{xchecks1}
Fit type&Sample&\mtop (GeV/c$^2$)&\djes (\sigcnoarg)\\
\hline

 \multirow{3}{*}{Nominal}&Combo&\measErr{171.9}{1.7}&\measErr{-0.12}{0.34}\\
% \cline{2-4}
 &LJ&\measErr{171.8}{1.8}&\measErr{-0.09}{0.36}\\
% \cline{2-4}
 &DIL&\measAErr{171.6}{3.5}{3.3}&-\\
 \hline
 \multirow{3}{*}{No JES prior}&Combo&\measErr{171.9}{1.7}&\measAErr{-0.14}{0.36}{0.37}\\
% \cline{2-4}
 &LJ&\measErr{171.8}{1.9}&\measAErr{-0.11}{0.39}{0.38}\\
% \cline{2-4}
 &DIL&\measAErr{171.6}{3.5}{3.3}&-\\
 \hline
 \multirow{3}{*}{No bkgd prior}&Combo&\measErr{171.9}{1.7}&\measAErr{-0.11}{0.35}{0.34}\\
% \cline{2-4}
 &LJ&\measErr{171.8}{1.8}&\measErr{-0.06}{0.36}\\
% \cline{2-4}
 &DIL&\measErr{171.5}{3.4}&-\\
 \hline
 1-tag LJ&LJ&\measAErr{169.1}{3.1}{2.6}&\measAErr{-0.17}{0.48}{0.57}\\
 \hline
 2-tag LJ&LJ&\measAErr{173.6}{2.6}{2.3}&\measAErr{0.20}{0.47}{0.50}\\
 \hline
 0-tag DIL&DIL&\measAErr{170.1}{6.4}{7.6}&-\\
 \hline
 Tagged DIL&DIL&\measAErr{172.2}{4.4}{4.0}&-\\
 \hline
 $e$ only LJ&LJ&\measErr{172.2}{2.7}&\measErr{-0.09}{0.51}\\
 \hline
 $\mu$ only LJ&LJ&\measAErr{171.3}{2.4}{2.3}&\measAErr{-0.04}{0.46}{0.47}\\
 \hline
$e e$ only DIL&DIL&{\measErr{169.0}{8.0} }&-\\
\hline
$e \mu$ only DIL&DIL&{\measAErr{173.6}{5.2}{4.0} }&-\\
\hline
$\mu \mu$ only DIL&DIL&{\measAErr{167.9}{7.5}{6.8} }&-\\
\hline
 \multirow{3}{*}{First \invfb{1}}&Combo&\measAErr{171.7}{2.3}{2.4}&\measAErr{0.45}{0.55}{0.50}\\
% \cline{2-4}
 &LJ&\measAErr{172.2}{2.5}{2.4}&\measAErr{0.59}{0.52}{0.55}\\
% \cline{2-4}
 &DIL&\measErr{166.1}{5.0}&-\\
 \hline
 \multirow{3}{*}{Last \invfb{0.9}}&Combo&\measErr{171.7}{2.7}&\measAErr{-0.70}{0.53}{0.59}\\
% \cline{2-4}
 &LJ&\measAErr{170.2}{3.1}{3.0}&\measAErr{-0.61}{0.54}{0.64}\\
% \cline{2-4}
 &DIL&\measAErr{175.2}{5.3}{4.7}&-\\

\end{tabular}
\end{ruledtabular}
\end{center}
\end{table}

\begin{cfigure1c}
\begin{tabular}{cc}
\includegraphics[width=0.49\textwidth]{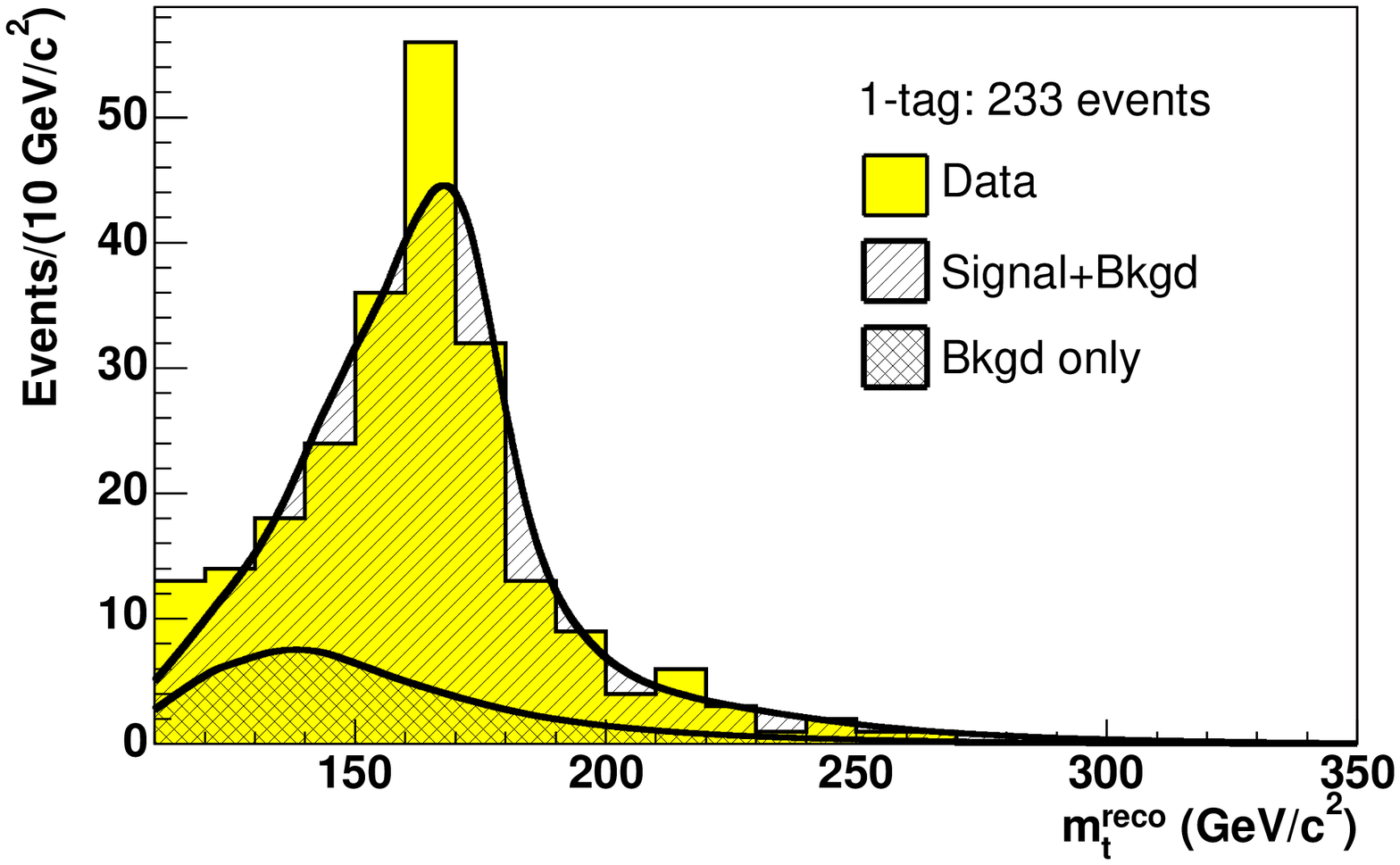} &
\includegraphics[width=0.49\textwidth]{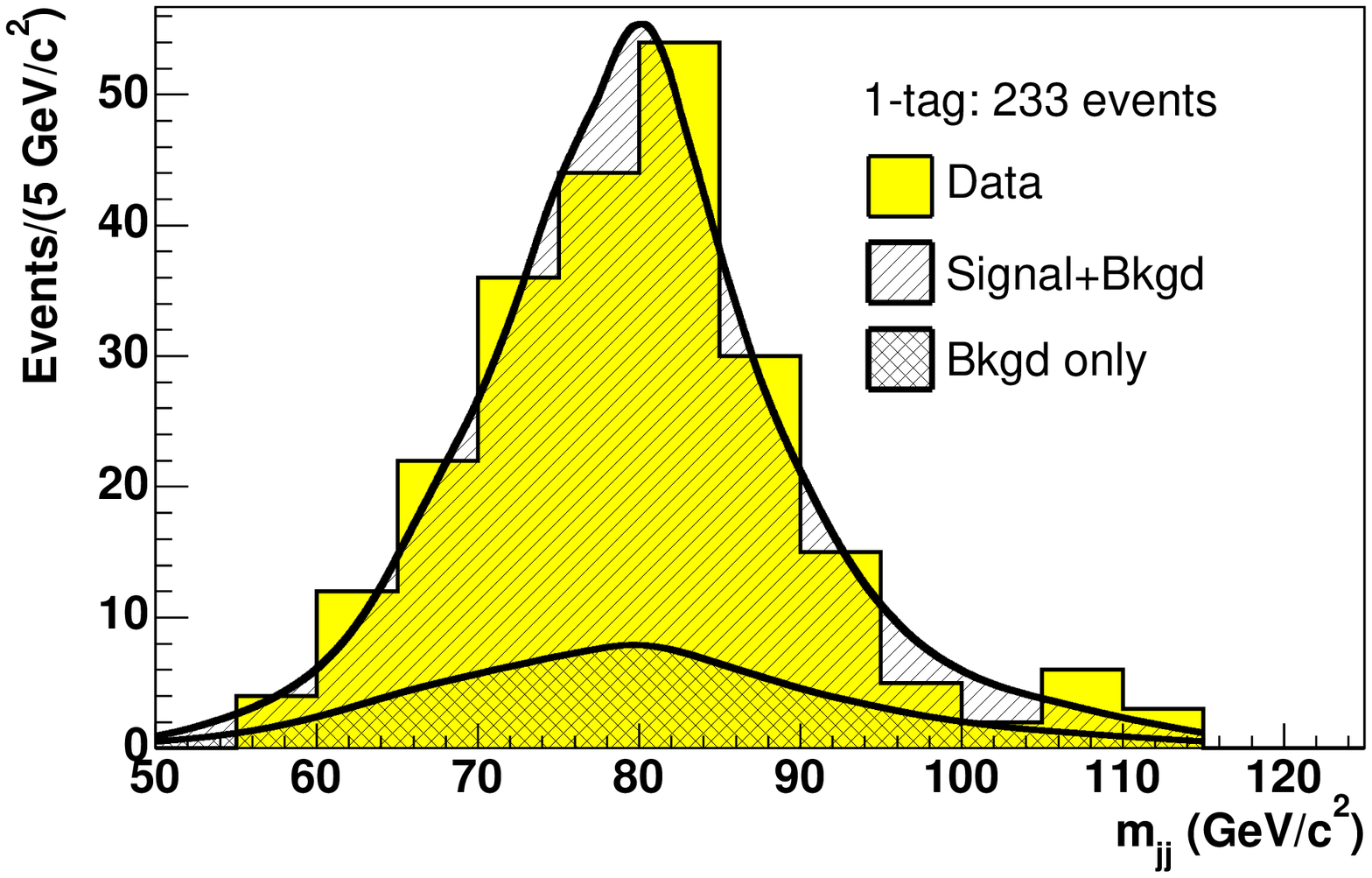} \\
(a) \onetag \mreco & (b) \onetag \mjj \\
\includegraphics[width=0.49\textwidth]{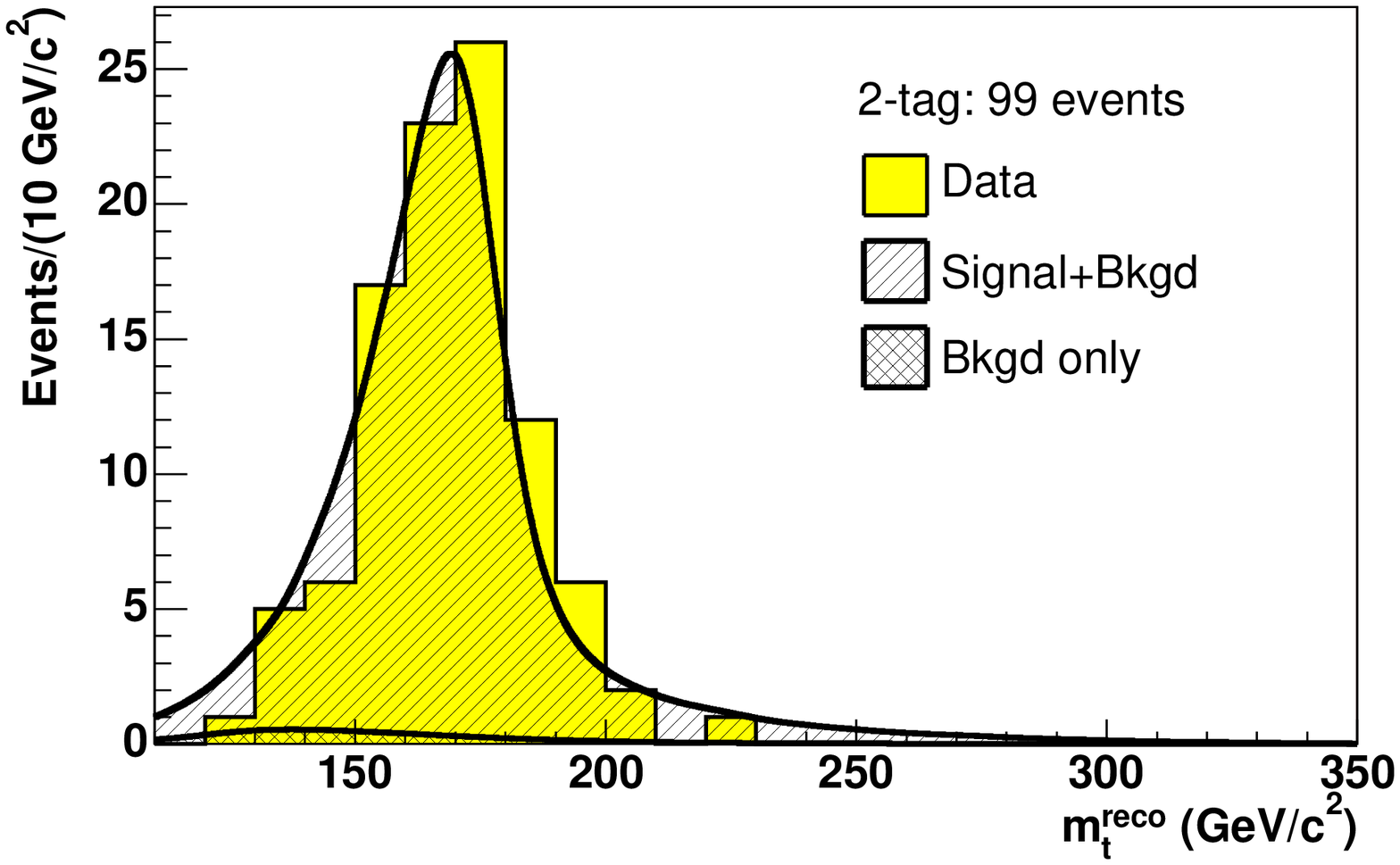} &
\includegraphics[width=0.49\textwidth]{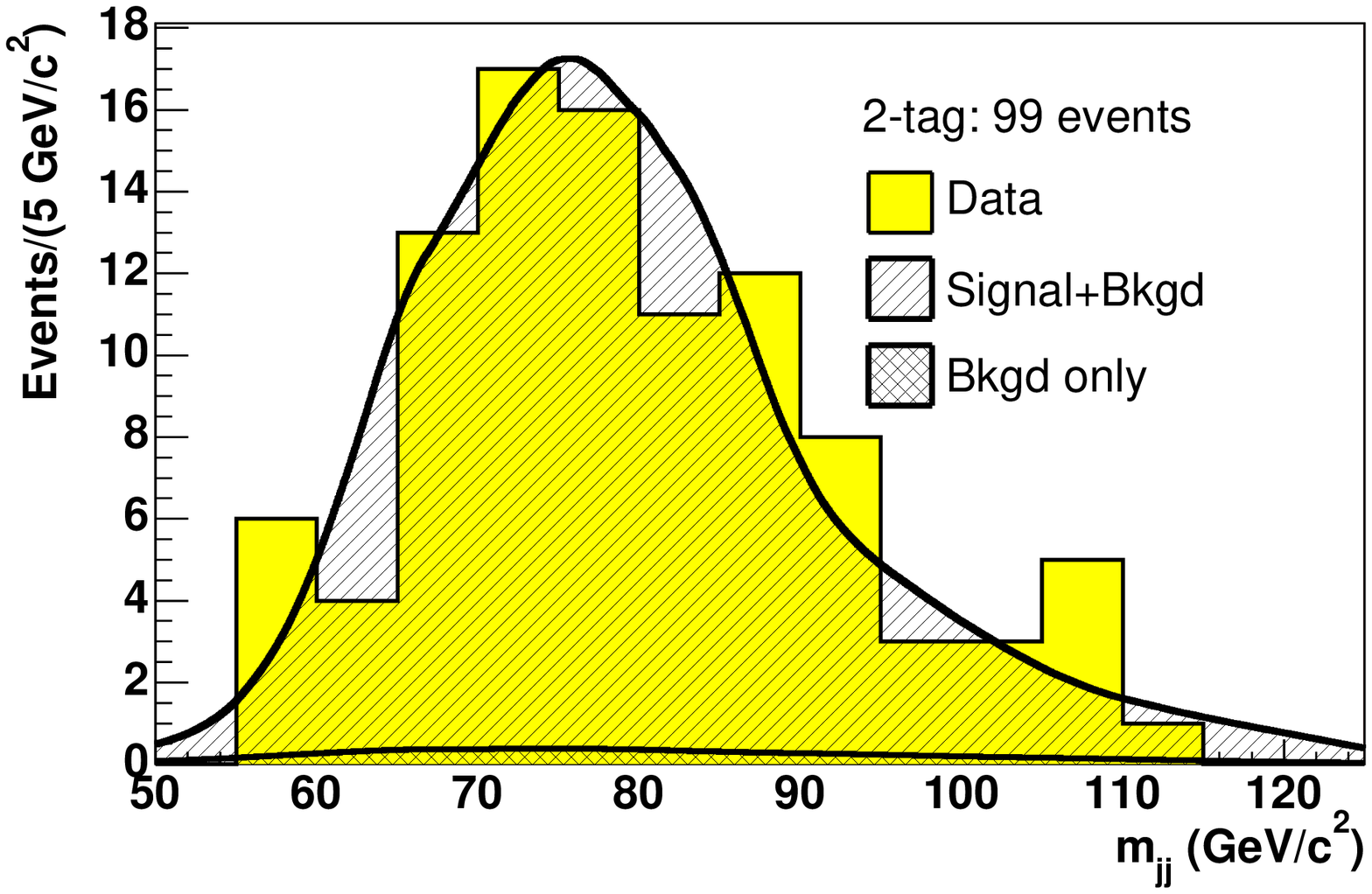} \\
(c) \twotag \mreco & (d) \twotag \mjj 
\end{tabular}
\caption{
One-dimensional \ljets data with density estimates overlaid using
$\mtop = \gevcc{172.0}$, $\djes=0.0$, and a full background model.
The expected numbers of events are set to the values from the
constrained fit.  Shown are the \onetag \mreco (a) and \mjj
(b) distributions, and the \twotag \mreco (c) and
\mjj (d) distributions.}
\label{templateoverlayLJ}
\end{cfigure1c}

\begin{cfigure1c}
\begin{tabular}{cc}
\includegraphics[width=0.49\textwidth]{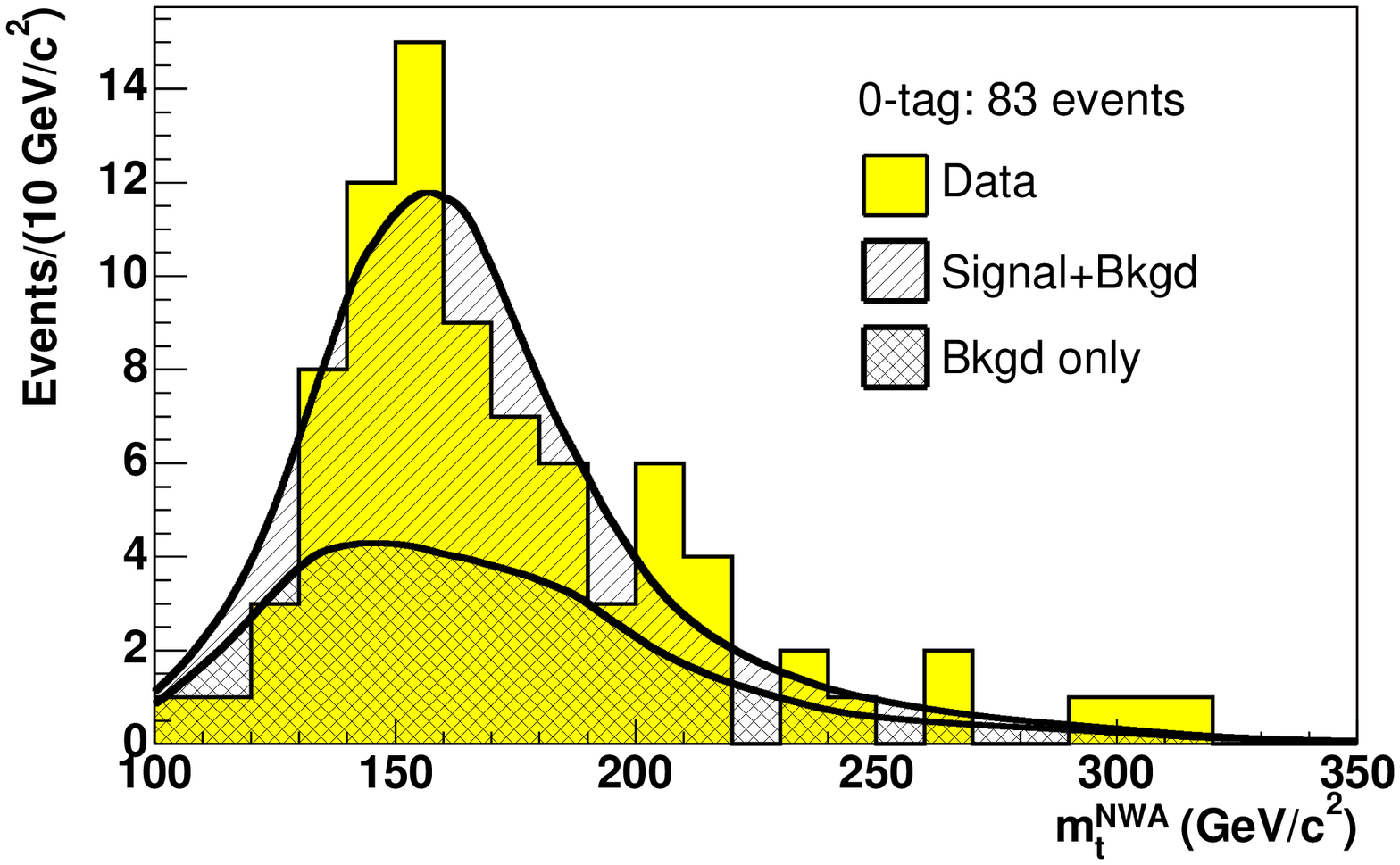} &
\includegraphics[width=0.49\textwidth]{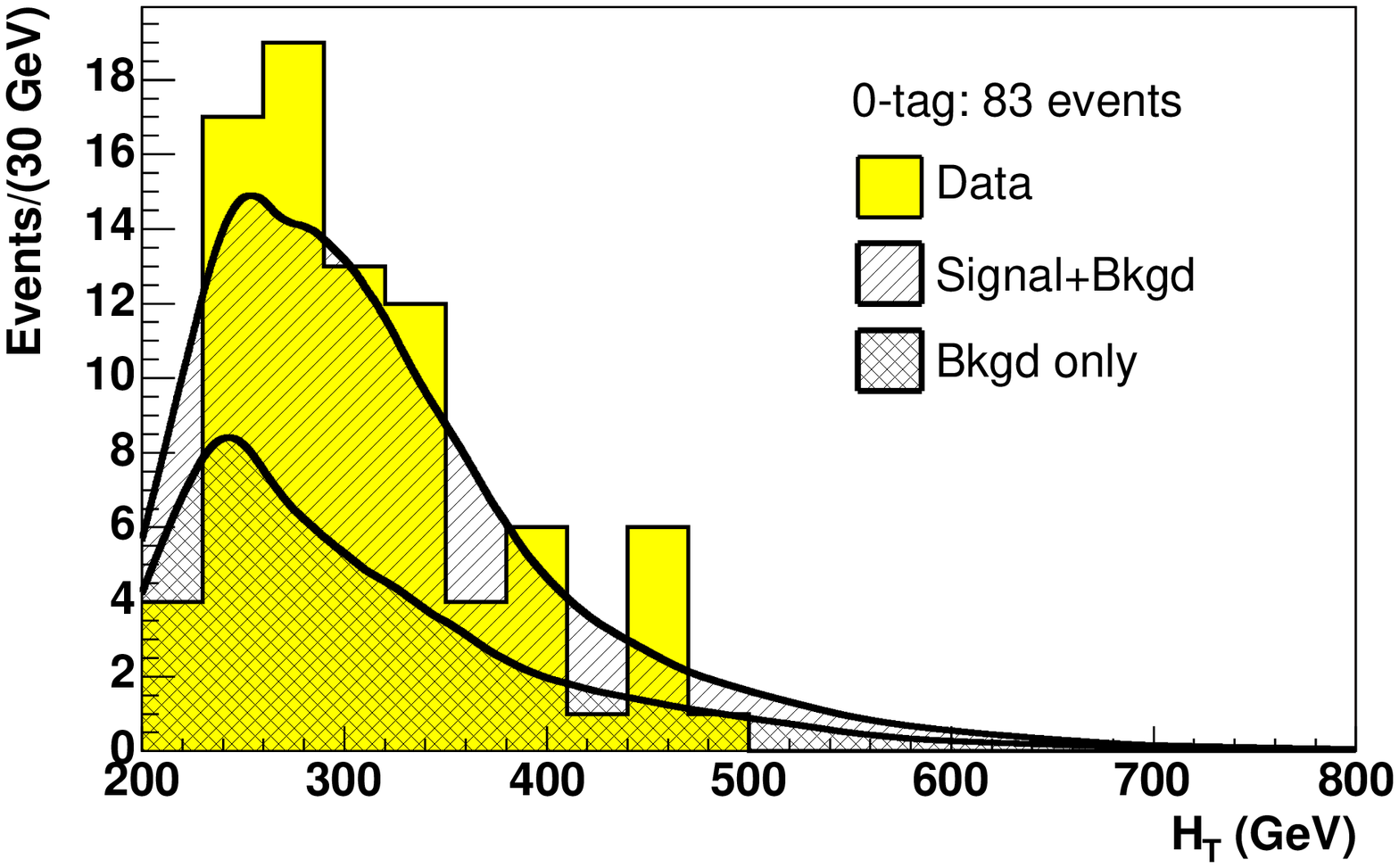} \\
(a) \zerotag \mtnwa & (b) \zerotag \Ht \\
\includegraphics[width=0.49\textwidth]{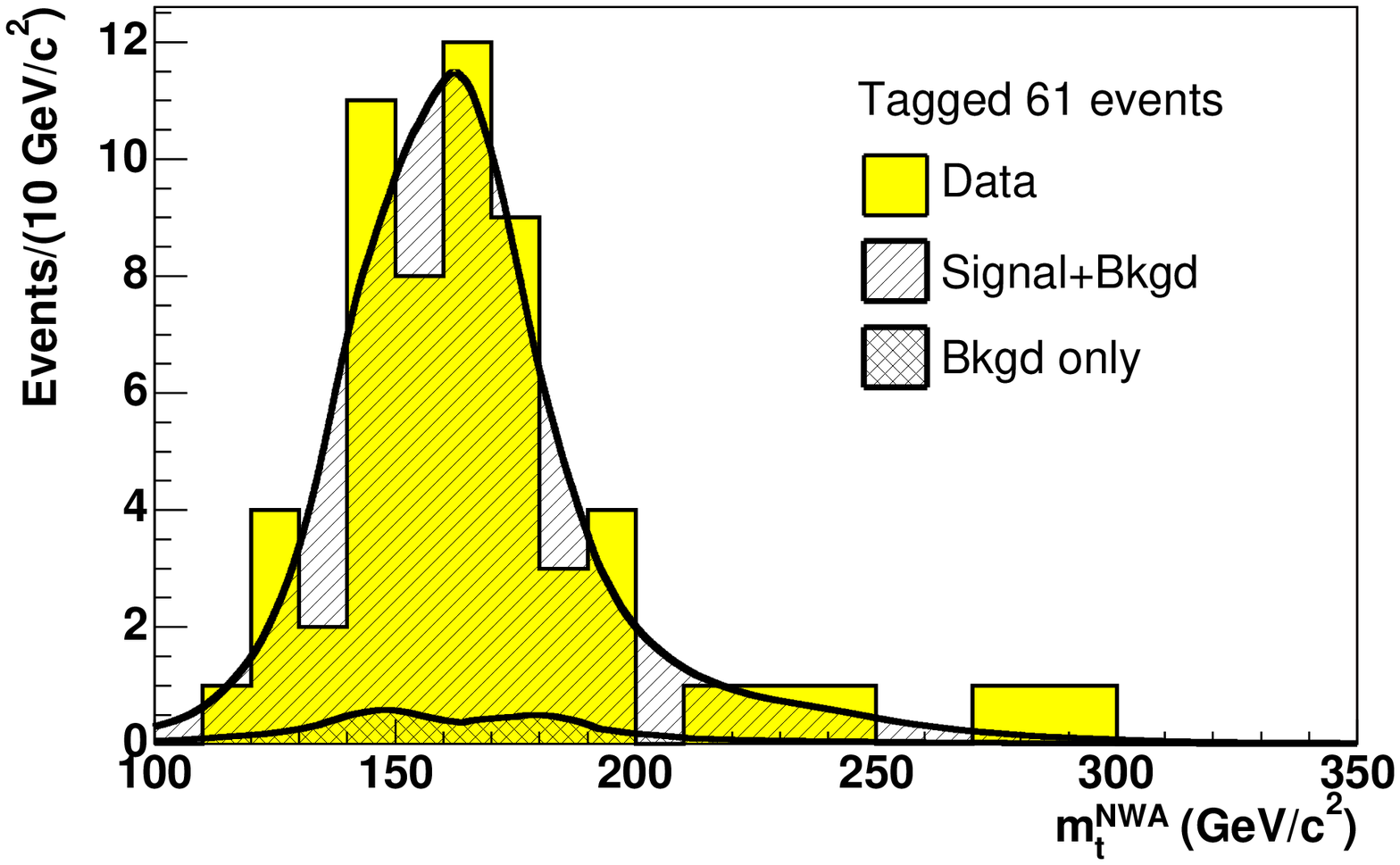} &
\includegraphics[width=0.49\textwidth]{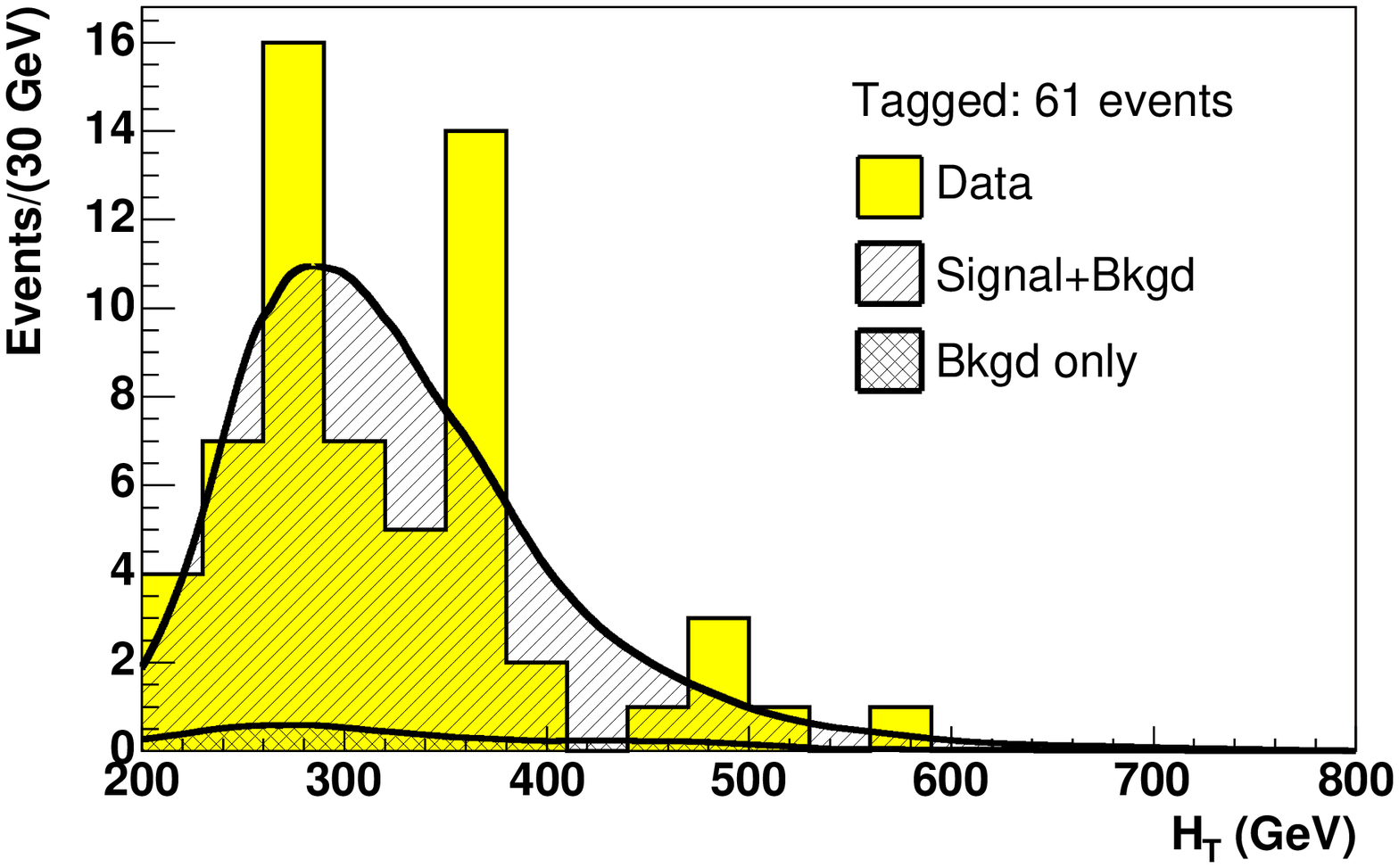} \\
(c) Tagged \mtnwa & (d) Tagged \Ht \\
\end{tabular}
\caption{
One-dimensional \dil data with density estimates overlaid using $\mtop
= \gevcc{172.0}$, $\djes=0.0$, and full background model.  The
expected numbers of events are set to the values from the constrained
fit.  Shown are the \zerotag \mtnwa (a) and \Ht (b)
distributions, and the tagged \mtnwa (c) and \Ht (d) distributions.}
\label{templateoverlayDIL}
\end{cfigure1c}

\section{Systematic uncertainties}
\label{sec:systematics}

We examine a variety of effects that could affect our measurement
by comparing MC simulated experiments in which we change systematic parameters
within their uncertainties.  As a single nuisance parameter,
the measured \djes does not fully capture the complexities
of jet energy scale uncertainties, particularly those with 
different $\eta$ and $\pt$ dependence. Fitting for the global
JES removes most of these effects, but not all of them.  In order
to estimate the total residual JES uncertainty, we vary JES parameters
within their uncertainties in
both signal and background MC generated data and measure resulting shifts 
in $\mtop$.  We also conduct MC simulated experiments where we
assume JES uncertainties are not fully correlated between jets
of different momenta.  So as not to bias the results, we remove
the JES prior for these experiments.  For the dilepton-only
measurement, which has no \emph{in situ} calibration,
these systematics dominate.
To form a \emph{b} jet energy scale systematic we replace the 
default parameters 
of the Bowler fragmentation function~\cite{Bowler} in the {\sc pythia}
simulation with the
parameters obtained by the D\O~collaboration in a {\sc pythia} tune to
the LEP and SLD data~\cite{Wojteksthesis}.
We also vary the semileptonic
branching fractions of \emph{b} and \emph{c} quarks within their uncertainties
given by~\cite{zresonance} and \cite{PDBook}. The calorimeter response to $b$
 jets is varied to capture differences in absolute jet energy scale
uncertainties for light flavor and $b$ quarks.
Effects due to uncertain modeling of initial-state gluon radiation
(ISR) and final-state gluon radiation (FSR) are studied by
extrapolating uncertainties in the $\pt$ of Drell-Yan events to the
$\ttbar$ mass region, resulting in a systematic on ISR-FSR
modeling~\cite{Abulencia:2005aj}. Note that unlike in \refref{Abulencia:2005aj}, we
coherently shift parameters affecting both ISR and FSR, as the
uncertainties on the two effects should be correlated. 
We measure the uncertainty due to
generator choice by comparing MC simulated experiments generated with {\sc herwig} and {\sc pythia}.
A systematic on different parton distribution
functions is obtained by varying the independent eigenvectors of the
{\sc cteq6m} set~\cite{refcteq6m}, comparing parton distribution
functions with different values of $\Lambda_{QCD}$, and comparing {\sc
cteq5l}~\cite{refcteq5l} with {\sc mrst72}~\cite{refmrst}.
The gluon fusion fraction uncertainty is calculated by reweighting the MC
samples to increase the fraction of
$\ttbar$ events initiated by gluons instead of quarks from the 6\% in
the leading order MC samples used for the measurement to 20\%, which is 
given as 
the $1\sigma$ upper bound on the gluon fusion fraction
in~\cite{theory_csection}.
Systematic uncertainties due to
lepton energy scale are estimated by propagating shifts on
electron and muon energies within their uncertainties.
Background shape systematic
uncertainties are obtained by varying the fraction of the different
types of backgrounds in MC simulated experiments. For lepton+jets
backgrounds, we generate further changes in the shapes by varying
 the $Q^2$ used in the calculation of
hard scattering and shower evolution in the
range $M_W^2/4$--$4 M_W^2$. 
%results in a background shape systematic uncertainty.
We also substitute the anti-electron QCD model for the 
non-isolated lepton model in MC simulated experiments. 
% gives an additional QCD modeling systematic uncertainty. 
For dilepton backgrounds changing the shape of the Drell-Yan sample according
to the difference in the missing energy distribution observed in data and simulation
gives one systematic effect.
We also shift the fake model in ways expected to maximally correlate with the
reconstructed mass.
The systematic uncertainty due to limited signal MC statistics is taken
as the uncertainty on the fit to a constant of the residuals obtained in
MC experiments (\fig{biasresiduals}). We study the effects of limited
background MC statistics using the bootstrap technique, where multiple
background MC data ensembles are generated. 
The ``pileup'' systematic is induced by the possible mismodelling of the
Minimum Bias events in the MC samples. These events are found to deposit
more energy in jets in simulation than expected from data.
It has been suggested that color effects may cause a systematic bias of order
\gevcc{0.5} which is not accounted in our studies~\cite{Wicke:2008iz}.

The systematic uncertainties are summarized in
\tab{systtablesummary}. The total systematic uncertainty is \gevcc{1.1} for
both the combined and the lepton+jets measurement, and \gevcc{3.8} for
the dilepton-only measurement.

\begin{table}
\begin{center}
\caption{Summary of systematics uncertainties. All numbers have units of
GeV/c$^2$. Comb refers to the combined fit, LJ refers to the lepton+jets-only fit and DIL refers to the dilepton-only fit.}
\begin{ruledtabular}
\begin{tabular}{cccccc}
\label{systtablesummary}
Systematic & Comb & LJ & DIL \\
\hline
Residual JES				&0.7	&0.7	&3.5\\
Generator				&0.7	&0.8	&1.3\\
Parton distribution functions		&0.3	&0.3	&0.5\\
$b$ jet energy				&0.2	&0.2	&0.2\\
Background shape			&0.2	&0.2	&0.3\\
Gluon fusion fraction				&0.2	&0.2	&0.2\\
Initial and final state radiation	&0.1	&0.2	&0.2\\
MC statistics				&0.1	&0.1	&0.5\\
Lepton energy scale			&0.1	&0.1	&0.3\\
Pileup					&0.1	&0.1	&0.1\\
\hline
Combined				&1.1	&1.1	&3.8\\
\end{tabular}
\end{ruledtabular}
\end{center}
\end{table}

\section{Conclusions}
\label{sec:conclusions}

We present the first measurement of the top quark mass across 
multiple decay topologies using a joint likelihood fit. Our procedure
includes a full treatment of correlations between systematics, and
does not assume Gaussian likelihoods or symmetric errors in
the channels being combined. In
\invfb{1.9} of data, we measure:
\begin{align*}
\label{eqn:conclusion_comb}
\mtop = \gevcc{\measStatJESSystALIGNED{171.9}{1.7}{1.1}} \\
= \gevcc{\measErr{171.9&}{2.0}}
\end{align*}
with cross-checks using events from the lepton+jets and dilepton
channels separately:
\begin{align*}
\mtop = \gevcc{\measStatJESSystALIGNED{171.8}{1.9}{1.1}}  \\
= \gevcc{\measErr{171.8&}{2.2}}\\
&\makebox[2.0in][r]{(lepton+jets only);}\\
%\begin{flushright}\text{(lepton+jets only);}\end{flushright}\\
\mtop = \gevcc{\measAStatSystALIGNED{171.2}{3.6}{3.4}{3.8}}\\
= \gevcc{\measAErr{171.2&}{5.3}{5.1}}\\
&\makebox[2.0in][r]{(dilepton only).}\\
%\begin{flushright}\text{(dilepton only).} \end{flushright}
%\shoveright{\text{(dilepton only)}.}
\end{align*}

This measurement increases our understanding of physics in the top
quark sector, and contributes to tests of the mechanism of electroweak
symmetry breaking.  In addition, the analysis methods and tools
described in this article will be applicable to other measurements at
the Tevatron experiments, and soon at CERN's Large Hadron Collider.

The precision of top quark mass measurements at the Tevatron is
approaching limits set by our understanding of non-perturbative QCD
phenomena.  Nevertheless, some further improvements are expected as
CDF accumulates a factor of 3--4 times more data during \runii and as
studies of important systematic effects provide additional constraints
on those uncertainties.

\begin{acknowledgments}
We thank the Fermilab staff and the technical staffs of the
participating institutions for their vital contributions. This work
was supported by the U.S. Department of Energy and National Science
Foundation; the Italian Istituto Nazionale di Fisica Nucleare; the
Ministry of Education, Culture, Sports, Science and Technology of
Japan; the Natural Sciences and Engineering Research Council of
Canada; the National Science Council of the Republic of China; the
Swiss National Science Foundation; the A.P. Sloan Foundation; the
Bundesministerium f\"ur Bildung und Forschung, Germany; the Korean
Science and Engineering Foundation and the Korean Research Foundation;
the Science and Technology Facilities Council and the Royal Society,
UK; the Institut National de Physique Nucleaire et Physique des
Particules/CNRS; the Russian Foundation for Basic Research; the
Ministerio de Ciencia e Innovaci\'{o}n, Spain; the Slovak R\&D Agency;
and the Academy of Finland.
\end{acknowledgments}

\bibliographystyle{apsrev}
\bibliography{PRD}

\end{document}